\documentclass{aa}  

\renewcommand{\rm}{\mathrm}

\def\({\left(}
\def\r){\right)}

\newcommand*{\nt}{\textrm}
\newcommand*{\dif}{\textrm{d}}
\usepackage{physics}
\usepackage{ulem}
\usepackage{cancel}
\usepackage{bm}
\usepackage{verbatim}
\usepackage{subcaption}
\usepackage{graphicx}
\usepackage{txfonts}
\usepackage{subcaption}
\usepackage{acronym}
\usepackage[dvipsnames]{xcolor}
\usepackage[margin=10pt,font=small,labelfont=bf]{caption}
\usepackage[colorlinks = true,
            linkcolor = blue,
            urlcolor  = blue,
            citecolor = blue,
            anchorcolor = blue]{hyperref}

\usepackage[noabbrev]{cleveref}
\usepackage{float}
\usepackage{todonotes}

\begin{document} 

\title{KiDS-1000 cosmology:\\ Constraints from density split statistics}
   \author{Pierre A. Burger\inst{1}
   \and Oliver Friedrich\inst{2}
   \and Joachim Harnois-D\'eraps\inst{3}
   \and Peter Schneider\inst{1}
   \and Marika Asgari\inst{4}
   \and Maciej Bilicki\inst{5}
   \and Hendrik Hildebrandt\inst{6}
   \and Angus H. Wright\inst{6}
   \and Tiago Castro\inst{7,8,9}
   \and Klaus Dolag\inst{2,10}
   \and Catherine Heymans \inst{6,11}
   \and Benjamin Joachimi \inst{12}
   \and Konrad Kuijken \inst{13}
   \and Nicolas Martinet\inst{14}
   \and HuanYuan Shan \inst{15,16}
   \and Tilman Tröster\inst{17}
    }
   \institute{Argelander-Institut f\"ur Astronomie, Auf dem H\"ugel 71, 53121 Bonn, Germany 
   \and Universitäts-Sternwarte, Fakultät für Physik, Ludwig-Maximilians-Universität München, Scheinerstr.1, 81679 München
   \and School of Mathematics, Statistics and Physics, Newcastle University, Newcastle upon Tyne, NE1 7RU, UK
   \and E.A Milne Centre, University of Hull, Cottingham Road, Hull, HU6 7RX, United Kingdom
  \and Center for Theoretical Physics, Polish Academy of Sciences, al. Lotników 32/46, 02-668 Warsaw, Poland
  \and Ruhr University Bochum, Faculty of Physics and Astronomy, Astronomical Institute (AIRUB), German Centre for Cosmological Lensing, 44780 Bochum, Germany
   \and INAF -- Istituto Nazionale di Astrofisica,
    Osservatorio Astronomico di Trieste, via Tiepolo 11, 34143 Trieste, Italy
    \and INFN -- Istituto Nazionale di Fisica Nucleare, Sezione di Trieste,  via  Valerio  2,  34127 Trieste,  Italy
    \and Institute for Fundamental Physics of the Universe, Via Beirut 2, 34151 Trieste, Italy
    \and Germany Max-Planck-Institut für Astrophysik, Karl-Schwarzschild-Straße 1, 85741 Garching, Germany
    \and Institute for Astronomy, University of Edinburgh, Royal Observatory, Blackford Hill, Edinburgh, EH9 3HJ, UK
    \and Department of Physics and Astronomy, University College London, Gower Street, London WC1E 6BT, UK
    \and Leiden Observatory, Leiden University, P.O.Box 9513, 2300RA Leiden, The Netherlands
    \and Laboratoire d'Astrophysique de Marseille, Aix-Marseille Univ., Rue Frédéric Joliot-Curie 38, 13388 Marseille, France
    \and Shanghai Astronomical Observatory (SHAO), Nandan Road 80, Shanghai 200030, China 
    \and University of Chinese Academy of Sciences, Yuquan Road 19A, Beijing 100049, China
    \and Institute for Particle Physics and Astrophysics, ETH Zürich, Wolfgang-Pauli-Strasse 27, 8093 Zürich, Switzerland
\\ \email{pburger@astro.uni-bonn.de}
             }

   \date{Received 3 August 2022 / Accepted 28 October 2022}

  \abstract
  {Weak lensing and clustering statistics beyond two-point functions can capture non-Gaussian information about the matter density field, thereby improving the constraints on cosmological parameters relative to the mainstream methods based on correlation functions and power spectra.}
  {This paper presents a cosmological analysis of the fourth data release of the Kilo Degree Survey based on the density split statistics, which measures the mean shear profiles around regions classified according to foreground densities. The latter is constructed from a bright galaxy sample, which we further split into red and blue samples, allowing us to probe their respective connection to the underlying dark matter density.}
  {We used the state-of-the-art model of the density splitting statistics and validated its robustness against mock data infused with known systematic effects such as intrinsic galaxy alignment and baryonic feedback.}
  {After marginalising over the photometric redshift uncertainty and the residual shear calibration bias, we measured for the full KiDS-bright sample a structure growth parameter of $S_8\equiv \sigma_8 \sqrt{\Omega_{\rm m}/0.3}=0.73^{+0.03}_{-0.02}$ that is competitive and consistent with two-point cosmic shear results, a matter density of $\Omega_{\rm m}=0.27\pm0.02$, and a constant galaxy bias of $b=1.37\pm0.10$. }
{}

   \maketitle

%
\section{Introduction}
Gravitational lensing, the theory that describes the deflection of light by massive objects, reveals a wealth of information about the evolution of matter structure in the Universe \citep[see, e.g.][for recent cosmic shear analyses]{Hamana:2020,Asgari2021,Amon2022}. 
Due to the accurate theoretical description and control over systematic inaccuracies, the most commonly used methods focus on two-point statistics, namely the two-point correlation functions and their Fourier counterparts called power spectra. These statistics are excellent for capturing the Gaussian information contained in the data and are complete if the data are Gaussian-distributed, such as the cosmic microwave background \citep[CMB; e.g.][]{Aghanim:2020}. In the late Universe, however, non-linear gravitational instabilities generate a significant amount of non-Gaussian features, whose information can only be accessed with higher-order statistics.
Furthermore, since higher-order statistics scale differently with cosmology and are affected differently by residual systematic effects, the constraining power on cosmological parameters increases by jointly investigating second- and higher-order statistics \citep[see, e.g.][]{Kilbinger2005,Berge2010,Pires:2012,Fu:2014,Pyne2021}.

As the current analysis of the estimation of cosmological parameters reaches the per cent level, tensions arise between observations of the early and late or local Universe. A famous tension is the one for the Hubble parameter $H_0$ \citep{diValentino:2021h} but it is not the subject of this work. More interesting for us is the tension in the matter clustering parameter $S_8=\sigma_8\sqrt{\Omega_\mathrm{m}/0.3}$, where it seems that the local Universe is less clustered than observations of the CMB suggest \citep{Hildebrandt:2017, joudaki2020, Heymans:2021,diValentino:2021s}.

A recent development in analysis methods has enabled the joint investigation of weak lensing and galaxy clustering data \citep{vanUitert2018, Joudaki2018,DES:2018, DES2021}, yielding significantly better constraints, especially along the $\sigma_8$-$\Omega_\mathrm{m}$ degeneracy axis. Even though foreground clustering data introduces largely uncertain astrophysical parameters, such as the galaxy bias, that complicate the analysis, these joint analyses inform us better about the correlation between galaxies and the underlying matter distribution \citep{Sanchez2017}. Here again, two-point statistics have been favoured so far for the reasons mentioned above, such that the combination of all measurements (cosmic shear, galaxy clustering, and galaxy-galaxy lensing) are generally referred to as `3$\times$2pts' statistics.

To access the additional information contained in the non-Gaussian features, a competitive statistic to the 3$\times$2pts method was recently proposed in \citet[][]{Gruen:2015}, coined the `density-split statistics' (DSS hereafter). This technique measures the tangential shear on the full pixelated survey footprint and bins the resulting shear profiles as a function of the foreground mass density. For example, high galaxy density regions generally trace large matter over-density regions, in which the tangential shear is expected to be larger, and this varies with cosmology. The DSS, therefore, captures information both from the shape and amplitude of the shear profiles and from the number of foreground galaxies in each density bin, with the latter helping to measure the galaxy bias significantly.

The first ingredient needed is a prediction model to interpret the measurements and constrain cosmological and astrophysical parameters. This can be constructed either from simulations \citep[see, e.g.,][for examples of simulation-based inference using the lensing peak count]{Harnois-Deraps2021, Zuercher2022} or from analytical calculations, where for instance \citet{Reimberg2018} and \citet{Barthelemy2020} made use of large deviation theory (LDT) to model the reduced-shear correction to the aperture mass probability distribution function (PDF). Another approach to access higher-order moments is discussed in \citet{Halder2021}, \citet{Halder2022}, and \citet{Heydenreich2022b}, where third-order cosmic shear statistics were modelled directly from the bispectrum. Although, the simulation-based approach has advantages regarding the numerical incorporation of critical systematic effects such as the intrinsic alignment (IA) of galaxies \citep[see, e.g.][hereafter HD22]{Harnois-Deraps2022} and baryonic processes extracted from hydrodynamical simulations. However, it typically requires large simulation suites that jointly vary all the parameters under consideration. On the other hand, analytical modelling of the DSS can better dissect the basic underlying properties of the LSS, and it can be computed sufficiently fast enough at any point in the cosmological parameter space. Such a model was derived in \citet[][hereafter F18]{Friedrich:Gruen:2018}, based on non-perturbative modelling of the matter density PDF. For a given cosmology, mean foreground galaxy density, and redshift distributions of the foreground and background galaxies, the F18 model predicts the mean tangential shear profiles and the PDF of the galaxy counts in each mass density bin. In \citet[][hereafter G18]{Gruen:Friedrich:2018}, the F18 model was used to constrain cosmological parameters from measurements of the Dark Energy Survey (DES) First Year and Sloan Digital Sky Survey (SDSS) data, yielding results competitive with the main DES 3$\times$2\,pt analysis \citep{DES:2018}.

To date, no cosmological constraints from DSS exist, except that of G18. However, the methods have been improved significantly. In particular, \citet{Brouwer2018} have presented a contemporary measurement of the DSS extracted from the third data release of the Kilo-Degree Survey data (KiDS), wherein the foreground galaxies were selected to mimic the spectroscopic Galaxy And Mass Assembly survey \citep[hereafter GAMA;][]{Driver:2011}. They developed an optimal methodology in their work, notably showing how the resulting signal-to-noise ratio (S/N) depends on the smoothing scale for the density map of foreground galaxies.

\citet[hereafter B22]{Burger2022} modified the analytical model by F18 for an application to galaxy density fields smoothed with general filters. As discussed in \citet[][]{Burger:2020}, compensated filter functions outperform the previously used top-hat filter functions in terms of the overall S/N of the shear signals and in recovering the correlation between the galaxy and matter density contrast. B22 mention another advantage of compensated filter functions: they are more compact in Fourier space and, therefore, can better suppress large-$\ell$ modes where baryonic effects play an important role, as studied in \citet{Asgari2020}. On the downside, compensated filters complicate the LDT-like calculations \citep{Barthelemy2020}. Nevertheless, B22 show that the density split statistics with compensated filters can still be accurately modelled in a computationally tractable manner after calibrating residual inaccuracies at large and small scales on the simulations of \citet{Takahashi2017}.

The current paper presents the first cosmological inference based on a DSS analysis of the KiDS data. We exploited the model advances presented in B22, using the dense sample of bright galaxies presented in \citet{Bilicki2021}, to construct our foreground density maps and compute the tangential shear from the lensing catalogue constructed from the fourth KiDS data release. Our inference includes a marginalisation over several residual systematic uncertainties. We verified with numerical $N$-body and hydrodynamical simulations that our measurements are robust against the IA of galaxies and baryonic feedback.

This work is structured as follows. In Sect.~\ref{Aperture_stat} we review the basics of the DSS and introduce small modifications made to our model compared to the one from B22. In Sect.~\ref{Sect:Obs_Data} we present the observed data used in our analysis, and then we describe in Sect.\,\ref{Sect:Simu_Data} the simulations needed for the validation of our inference pipeline which is described in Sect.~\ref{sec:inference_description}. In Sect.\,\ref{sec:validation} we perform our validation of the model together with an investigation on IA and baryonic physics which could potentially contaminate our results. In Sect.~\ref{sec:results} we finally present our main results and conclude with a discussion and summary in Sect.~\ref{sec:Conclusion}.

\section{Theoretical background}
\label{Aperture_stat}
The DSS essentially measures the tangential shear around sub-areas of the sky that are assigned according to the galaxy foreground density. It is therefore closely related to aperture statistics, which we introduce here first.
Given a convergence field $\kappa(\boldsymbol{\theta})$, the aperture mass map is defined as
\begin{equation}
    M _{\nt{ap}}\left(\boldsymbol{\theta}\right) \coloneqq \int \dd^2\theta' \,\kappa(\boldsymbol{\theta}+\boldsymbol{\theta}')\,U(|\boldsymbol{\theta'}|) \, ,	
    \label{Map}
\end{equation}
where $\boldsymbol{\theta}$ is the position on the flat sky, and $ U(\vartheta)$ is a compensated, axisymmetric filter function, such that $ \int\vartheta\, U(\vartheta)~ \dif\vartheta=0$. The aperture mass, $M_{ \nt{ap}}$, can also be expressed in terms of the tangential shear $\gamma_{\nt{t}}$ \citep{Schneider:1996} and a second filter function $Q$ as
\begin{equation}
     M_{ \nt{ap}}(\boldsymbol{\theta}) = \int\dd^2\theta'\,\gamma_{\nt{t}}(\boldsymbol{\theta}+\boldsymbol{\theta}')\,Q(|\boldsymbol{\theta}'|)\, ,
     \label{eq:MapQ}
\end{equation}
where
\begin{equation}
 Q(\vartheta) = \frac{2}{\vartheta^2} \int\limits_0^{\vartheta}\dd \vartheta'\,\vartheta'\,U(\vartheta') - U(\vartheta) \, .
 \label{NewQ}
\end{equation}
The above relation between the two filters $U$ and $Q$ can be inverted,
\begin{equation}
 U(\vartheta) = 2\int\limits_{\vartheta}^{\infty} \dd\vartheta'\, \frac{Q(\vartheta')}{\vartheta'} - Q(\vartheta) \, ,
\label{NewU}
\end{equation}
allowing us to work either with convergence maps or shear catalogues.
Replacing the convergence by the foreground galaxy number counts $n(\boldsymbol{\theta})$ in Eq.~(\ref{Map}), we define the aperture number counts, or simply aperture number, as \citep[][]{Schneider:1998}
\begin{equation}
    N_{ \nt{ap}}(\boldsymbol{\theta}) \coloneqq \int\dd^2\theta' \,n(\boldsymbol{\theta}+\boldsymbol{\theta}')\,U(|\boldsymbol{\theta}'|)\, .
    \label{Nap}
\end{equation}
This definition is equivalent to the `Counts-in-Cell' (CiC) statistics mentioned in \citet{Gruen:2015} if the filter $U$ is defined as a top-hat. In that case, however, $U$ is not compensated; hence, one can not relate the filters $U$ and $Q$.

The general idea of the DSS is to divide the survey area into quantiles $\mathcal{Q}$ according to the aperture number $N_\mathrm{ap}$ and then measure the mean tangential shear in the corresponding quantiles $\langle \gamma_\textrm{t} |  \mathcal{Q} \rangle$. We detail in Sect.~\ref{subsec:DSS_estimator} how we achieve this in the data and in Sect.~\ref{subsec:DSS_model} how we predict it analytically.

\subsection{Measuring the DSS vector}
\label{subsec:DSS_estimator}
We followed several ordered pipeline steps to extract the DSS data vector for the cosmological inference. First, we distributed the foreground (lens) galaxies onto a \texttt{HEALPix} \citep{HEALPix2005} grid $n(\Vec{\theta})$ of $\texttt{nside}=4096$, which resulted in a pixel area of $A_\mathrm{HP}\approx 0.74\,\mathrm{arcmin}^2$. Second, we determined the aperture number field $N_\mathrm{ap}$ with a filter function $U$, with a finite filter radius $\Theta$ and maximal one transition from positive to negative values at $\theta_\mathrm{tr}$. This was achieved with the \texttt{healpy} function \texttt{smoothing}, with a beam window function that is the $U$-filter in the spherical harmonic space determined with \texttt{healpy} function \texttt{beam$2$bl}. Since Eq.~\ref{Nap} assumes full knowledge of $n(\boldsymbol{\theta})$ on the sky which has to be modified in the presence of a mask $m(\boldsymbol{\theta})$ as
    \begin{align}
        N_\mathrm{ap}(\Vec{\theta}) &= \frac{\int_0^{\theta_\mathrm{tr}}U(\Vec{\theta'}) \, \dd^2 \theta' }{\int_0^{\theta_\mathrm{tr}} m(\Vec{\theta}+\Vec{\theta'})U(\Vec{\theta'}) \, \dd^2 \theta'} \int_0^{\theta_\mathrm{tr}} n(\Vec{\theta}+\Vec{\theta'})U(\Vec{\theta'}) \,\dd^2 \theta' \nonumber \\&+
        \frac{\int_{\theta_\mathrm{tr}}^{\Theta} U(\Vec{\theta'})\,\dd^2 \theta'}{\int_{\theta_\mathrm{tr}}^{\Theta} m(\Vec{\theta}+\Vec{\theta'})U(\Vec{\theta'})\,\dd^2 \theta'} \int_{\theta_\mathrm{tr}}^{\Theta} n(\Vec{\theta}+\Vec{\theta'})U(\Vec{\theta'})  \,\dd^2 \theta'\, ,
    \end{align}
    where in this work $m(\Vec{\theta})$ was the KiDS-1000 mask. For example, the second part of this equation vanishes for a top-hat filter. We divided the filter in this way to prevent the masked area from entering the positive part to decrease $N_\mathrm{ap}$ systematically and artificially increase the aperture number when entering the negative filter region. By separating the compensated filter into its positive and negative parts, we could correct both individually. Furthermore, since this correction became less accurate in heavily masked regions, we included only those pixels where the number of unmasked pixels within the given filter radius (effective area) was greater than $50\%$ of the total number of pixels inside the same circle (maximal area), which we treated as our `good' pixels. This can change from pixel to pixel because our \texttt{HEALPix} map originated from a flat sky mask. To avoid a pixel being considered good, yet for more than $50\%$ of pixels to be missing in the negative part, we included for compensated filters only those pixels where the effective area for the positive and the negative part was greater than $50\%$ of their individual maximal area. With our choice of the 50\% threshold, we attempted to achieve a compromise between statistical power and falsely measured $N_\mathrm{ap}$ values. G18 considered only regions with at least $80\%$ coverage, but since the KiDS footprint is very narrow, we had to relax that threshold to avoid shot noise-dominated data vectors. Therefore, we decided to use the highest threshold that yielded shear profiles that do not deviate significantly from shear profiles with smaller threshold values. The result is shown in Fig.~\ref{fig:shear_cutoff}, where it is seen that the shear profiles with threshold values of $50\%$ or smaller are quite similar and start to deviate for higher threshold values. Next, we allocated those good pixels to five quantiles $\mathcal{Q}$ according to their $N_\mathrm{ap}$ value. The pixels from each quantile are then correlated with the tangential shear information from the source catalogues using the \texttt{treecorr} \citep{Jarvis:Bernstein:2004} software in ten log-spaced bins with angular separation $10\,\mathrm{arcmin} < \vartheta < 120\,\mathrm{arcmin}$. This resulted in measurements of the five tangential shear profiles $\langle \gamma_\textrm{t} |  \mathcal{Q} \rangle$, that is, one per quantile. For all measured profiles, the shear around random points is subtracted, which ensures that the average overall quantiles vanish by definition. Finally, we constructed our data vector, which consists of the shear profiles from the highest two and lowest two quantiles, plus the mean of the aperture number values in the same four quantiles. We must exclude the information of one quantile $\mathcal{Q}$ since the other four quantiles fully determine it by construction, for the reason explained above. The same is true for mean aperture number values in those quantiles, whose average is fixed by the total galaxy number density measured in the data. This results in ten $\theta$ bins, four quantiles and two source bins in a data vector of $80+4=84$ elements. Although the information content is independent of which quantile is removed, we exclude the middle quantile for the whole analysis since it would have the least cosmological information, if the quantiles were analysed separately.

\subsection{Modelling the DSS vector}
\label{subsec:DSS_model}
Our modelling of the DSS signal is inspired by the LDT approach and builds from the original F18 model and the subsequent improvements presented in B22. We refer the reader to these two references for the complete details of the model calculations and highlight here only the broad principles and the minor modifications we have made. Briefly, the model consists of three key ingredients: (i) the PDF of the matter density contrast, smoothed with the filter function $U$, labelled $\delta_{{\rm m},U}$; (ii) the expectation value of the convergence inside a radius $\vartheta$ given the smoothed matter density contrast defined above; (iii) the distribution of $N_\mathrm{ap}$ values given $\delta_{{\rm m},U}$. B22 shows how these are computed for arbitrary filter functions and quantile counts. We, however, focus here on the `adapted compensated' filter case, introduced in B20 and shown in figure 3 in B22, and five quantiles. As in B22, the model was calibrated using the full-sky simulations described in \citet{Takahashi2017} to suppress the residual differences of the modelled and measured data vector. The KiDS-1000 lens distribution used in the current paper peaks at a lower redshift than that in B22, and we know that the DSS model is slightly less accurate in that case, but we subsequently show that the calibration is accurate enough to yield unbiased results. 

Since real galaxies are not expected to be perfectly Poisson-distributed, we modified the distribution of the aperture number computed in the B22 model in a way that allows for super-Poissonian shot noise. Inspired by F18, we achieved this by scaling the galaxy number density $n_0$ with a free parameter $\alpha>0$, such that $n_0\,\alpha^{-1}$ can be interpreted as an effective number density of Poissonian tracers. This implies that the quantity $p(N_{\textrm{ap}}\,\alpha^{-1}|\delta_{\mathrm{m},U})$ follows a log-normal distribution, instead of $p(N_{\textrm{ap}}|\delta_{\mathrm{m},U})$ as in B22. Consequently, the characteristic function $\Psi$, determining the parameters of the log-normal distribution, must be modified to (see equation 36 of B22)
\begin{align}
\Psi(t)=&\exp\left(2\pi \frac{n_0}{\alpha} \int_0^\infty \dd \vartheta\;\vartheta\; \left(1+b\, \langle w_\vartheta|\delta_{\mathrm{m},U}\rangle\right)  \left[\mathrm{e}^{\mathrm{i} t U(\vartheta)} -1\right]\right) \, ,
\label{eq:CF_alpha}
\end{align}
where $w_\vartheta$ is the mean 2D density contrast on a circle at $\vartheta$ (see Eq.~37 in B22), and $b$ is the linear galaxy bias.
Moreover, to ensure that the mean aperture number remains constant, we further modified the calculation of $p\left(N_{\textrm{ap}}|\delta_{\mathrm{m},U}\right)$ as (see equation A39 of B22):
\begin{align}
p\left(N_{\textrm{ap}}|\delta_{\mathrm{m},U}\right) \rightarrow \frac{1}{\alpha}p\left({N_{\textrm{ap}}\alpha^{-1}}|\delta_{\mathrm{m},U}\right) \, .
\label{eq:alpha}
\end{align}
Since the expectation value $\langle N_{\textrm{ap}} | \delta_{\mathrm{m},U} \rangle \propto \alpha$ and the variance $\langle (N_{\textrm{ap}} -  \langle N_{\textrm{ap}} | \delta_{\mathrm{m},U} \rangle )^2 | \delta_{\mathrm{m},U} \rangle \propto \alpha^2$ we see that the ratio of variance to expectation value is proportional to $\alpha$ as required to describe deviations from Poissonian samples. Similar to \citet{Friedrich:Gruen:2018}, we require $\alpha>0.1$ in our parameter sampling for numerical reasons. We also compared our definition of this $\alpha$ parameter to the one implemented in \citet{Friedrich:Gruen:2018} and found no differences in the predictions. 

Compared with numerical simulations, this model has been shown in B22 to be accurate, with residual inaccuracies to be everywhere significantly smaller than the statistical noise of the KiDS-1000 data. We, therefore, do not need to include a modelling error in our uncertainty budget. Furthermore, we also tested a non-linear galaxy bias model, where we exchanged the constant galaxy bias $b$ with $b=b_\mathrm{1}+b_\mathrm{2} \delta_{\mathrm{m},U}>0$. However, $b_\mathrm{2}$ was highly correlated with other parameters such as $\Omega_\mathrm{m}$ which prevented our parameter estimation from converging and therefore had to be excluded for this analysis. We also tested if the assumption of a linear galaxy bias is satisfied (see Fig.~\ref{fig:MCMC_bias} and its description), and the results of that test can be summarised as follows: a linear galaxy bias model is sufficient if an analysis using shear and $N_\mathrm{ap}$ information gives similar cosmological results as using only shear information since the shear profiles are basically insensitive to the galaxy bias model.

\section{Observational data}
\label{Sect:Obs_Data}
In our analysis, we exploit the fourth data release of the KiDS \citep{Kuijken:2015,Kuijken:2019,deJong:2015,deJong:2017}, which is a public survey carried out at the European Southern Observatory\footnote{The KiDS data products are public and available through \url{http://kids.strw.leidenuniv.nl/DR4}}. KiDS was designed for weak lensing applications, producing high-quality images with VST-OmegaCAM camera. Thanks to the infrared data from its overlapping partner survey VIKING \citep[VISTA Kilo-degree Infrared Galaxy survey,][]{Viking2013}, galaxies are observed in nine optical and near-infrared bands, $u,g,r,i,Z,Y,J,H,K_s$, allowing for better control over redshift uncertainties \citep[][hereafter H21]{Hildebrandt2021} to earlier releases. The weak lensing data in KiDS DR4 are collectively called `KiDS-1000' as they cover $\sim 1000\,\mathrm{deg}^2$ of images; this reduces to $777.4\,\mathrm{deg}^2$ of the effective area after masking. These galaxies are further split into lens and source samples, which we discuss in more detail in the following sections, with properties summarised in Table \ref{table:data_overview}.

\subsection{Lens catalogues}
\label{Sect:Obs_Data:lenses}
Our primary lens catalogue is the `KiDS-bright' sample described in \citet[][hereafter Bi21]{Bilicki2021},  a flux-limited galaxy catalogue with accurate and precise photometric redshifts, $z_\mathrm{ph}$, derived using the nine photometric bands available in the KiDS-1000 data. This highly pure and complete\footnote{By purity, we mean very low fractions of stars and quasars (point sources) or artefacts. Completeness is evaluated with respect to GAMA.} galaxy dataset was selected to match the properties of the partly overlapping Galaxy And Mass Assembly \citep[GAMA,][]{Driver:2011} spectroscopic redshift, $z_\mathrm{sp}$, dataset. KiDS-bright is limited to $r < 20\,\mathrm{mag}$, covers $\sim 1000\, \mathrm{deg}^2$ and contains about one million galaxies after artefact masking. To obtain photometric redshift estimates, Bi21 took advantage of the large amount of spectroscopic calibration data measured by GAMA and trained a supervised machine-learning neural network algorithm implemented in the ANNz2 software \citep{Sadeh2016} to map an input space of 9-band magnitudes to an output redshift. The ANNz2 training sample consists of matched KiDS galaxies with spectroscopic redshifts from the GAMA equatorial fields, where that survey is the most complete and provides representative training data. The trained model was subsequently applied to the entire inference dataset, the photometrically selected KiDS galaxies with the same $r<20$ cut and magnitudes detected in the same nine bands. This sample spans the redshift range of $0<z\lesssim 0.6$; however, since our analytical DSS model is less accurate for very small redshifts, we further exclude all galaxies with $z_\mathrm{ph}<0.1$. This cut only slightly lowers the number of lenses and results in a projected number density of 0.325\,arcmin$^{-2}$, as summarised in the first row of Table \ref{table:data_overview}.

\begin{figure}
\centering
\includegraphics[width=\columnwidth]{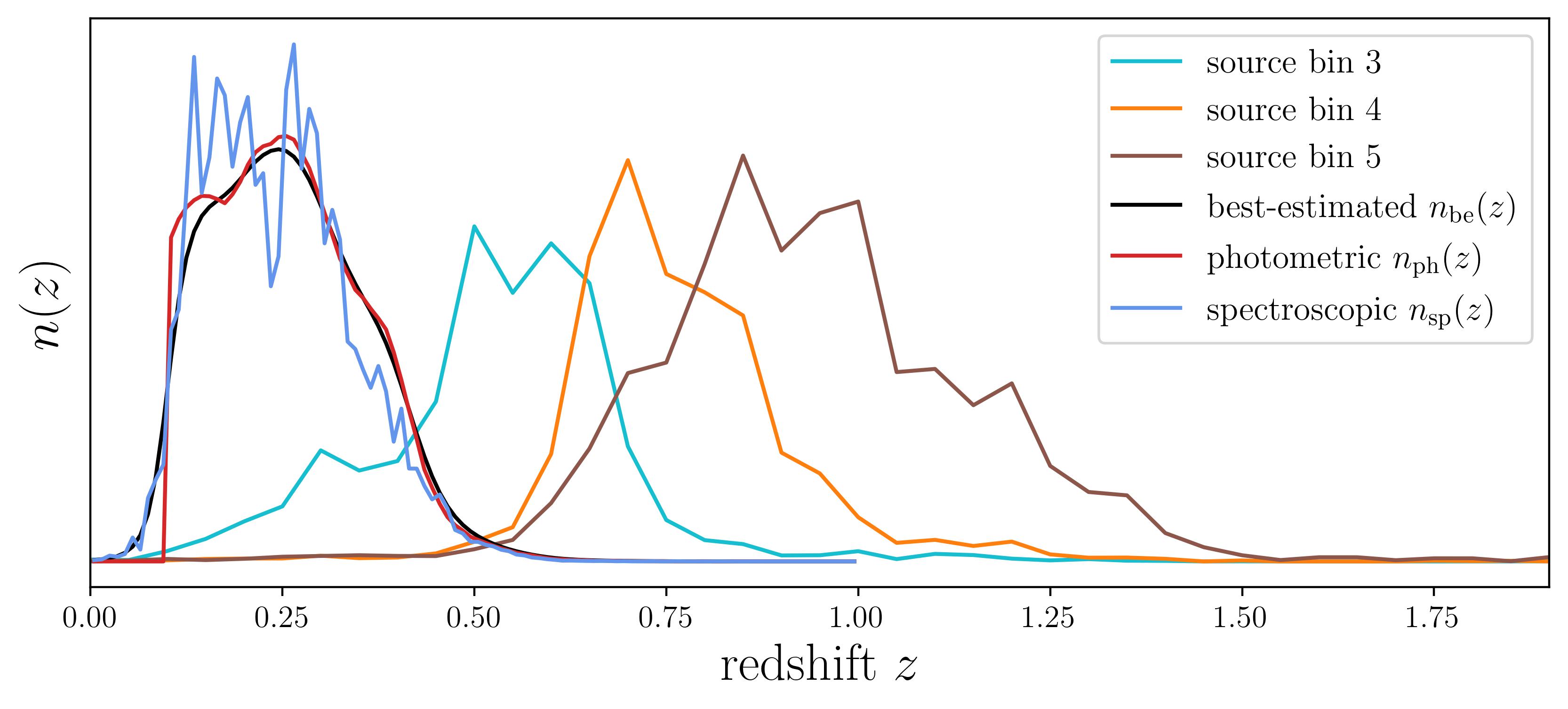}
\caption{Redshift distributions, $n(z)$, of the galaxy samples. The lens sample is obtained from the KiDS-bright galaxies described in Bi21. The blue line shows $n_\mathrm{sp}(z)$, i.e. the redshift distribution of KiDS galaxies for which we have spectra from the GAMA survey. The red line shows $n_\mathrm{ph}(z)$, the distribution of the full KiDS-bright sample as estimated by ANNz2 with a photometric redshift cut of $z_\mathrm{ph}<0.1$. The black line shows our best-estimated $n_\mathrm{be}(z)$: a smoothed version of $n_\mathrm{ph}(z)$ that better accounts for photometric redshift errors.
The cyan, orange, and brown lines show the third, fourth and fifth redshift bins of the KiDS-1000 data, as estimated in H21. As the third bin strongly overlaps the sources, it is excluded from the analysis.}
\label{fig:nofz}
\end{figure}

The main properties of interest to us are the galaxy bias, which has not been measured before for the KiDS-bright sample, the galaxy number density, and the redshift distribution, which is needed in the modelling.
As shown in Bi21, the photometric redshift distribution of the full KiDS-bright sample is measured with high precision: jackknife sub-sampling reveals negligible mean bias, with a small overall scatter of ${\sigma_z \approx 0.018(1+z)}$. However, for our theoretical model, we need to estimate our newly selected foreground sample's true $n(z)$ distribution. We take advantage of the very good match between the GAMA spectroscopic sample and the KiDS-bright dataset, allowing us to build an accurate model of the photometric redshift error distribution, as discussed in Bi21.
Following the description in \citet{PeacockBilicki2018}, the best-estimated $n_\mathrm{be}(z)$ of the true $n(z)$ can be obtained from a convolution between the normalised photometric redshift distribution of all the galaxies in our selected sample, $n_\mathrm{ph}(z)$, and a photo-$z$ error model $p_{\delta z} (\Delta z)$,
\begin{equation}
    n_\mathrm{be}(z) = \int n_\mathrm{ph}(z_\mathrm{ph})\,p_{\delta z} \left(z_\mathrm{ph}-z\right) \, \dd z_\mathrm{ph}\, .
    \label{eq:nz}
\end{equation}
Following \citet{Bilicki2014}, we adopted a `modified Lorentzian',
\begin{equation}
    p_{\delta z} (\Delta z) \propto \left(1 + \frac{ \Delta z^2}{2as^2}\right)^{-a} \, ,
    \label{eq:Lorentzian}
\end{equation}
which was shown in Bi21 to reproduce the photo-$z$ errors better than a Gaussian error model. To estimate the parameters $a$ and $s$, Eq.~\eqref{eq:Lorentzian} is fitted to the KiDS-bright galaxies that also have GAMA spectroscopic redshifts, where $\Delta z = z_\mathrm{ph}-z_\mathrm{sp}$. The best-fit $a$ and $s$ values for our selection are provided in the top row of Table \ref{table:data_overview}, and the resulting $n_\mathrm{be}(z)$ is shown in black in Fig.\,\ref{fig:nofz}.

\begin{table*}
\centering
\caption{Overview of the observational KiDS-1000 data.}
\begin{tabular}{ccccc}
\hline
\hline
name & $n_\mathrm{eff} [\mathrm{arcmin}^{-2}]$ & $\delta \langle z \rangle$ & $a$ & $s$ \\
\hline
Full KiDS-bright sample & 0.325 & $0.0 \pm 0.01$ & 2.509  & 0.018 \\
Red KiDS-bright sample & 0.131 & $0.0 \pm 0.01$ &  2.630 & 0.016  \\
Blue KiDS-bright sample & 0.165 & $0.0 \pm 0.01$ &  2.619 & 0.020   \\
 \hline
name & $n_\mathrm{eff} [\mathrm{arcmin}^{-2}]$ & $\delta \langle z \rangle$ & $\sigma_\epsilon$ & $m$-bias $\times 10^{3}$\\
\hline
Source sample bin 4 & 1.26 & $0.011 \pm 0.0087$ &  0.25 & $8\pm 12$  \\
Source sample bin 5 & 1.31 & $-0.006 \pm 0.0097$ & 0.27 & $12\pm 10$  \\
\hline
\hline
\vspace{0.1cm}
\end{tabular}
\tablefoot{The upper and lower parts describe the lens and source catalogues, respectively. The $a$ and $s$ parameters enter the Lorentzian fitting function (see Eq. \ref{eq:Lorentzian}) and capture the uncertainty on the redshift of our lens sample. The uncertainties on the mean redshift, $\Delta \langle z \rangle$, for the lenses are motivated by the uncertainty from the sources, although it is probably an upper limit of the error. We show the mean and uncertainty on the redshift bias for the source samples, taken from H21 and used in \citet{Asgari2021}. The rightmost columns display the measured ellipticity dispersion per component, $\sigma_\epsilon$ measured in \citet{Giblin:2020}, and the shear multiplicative $m$-bias correction updated in \citet{JLB2022}.}
\label{table:data_overview}
\end{table*}

In order to estimate the uncertainty on our $n(z)$ estimate, we notice that Eq.~\eqref{eq:nz} has two ingredients: the photometric redshift distribution $n_\mathrm{ph}(z)$, which is `exact' in the sense that they are directly measured, and the photo-$z$ error model $p_{\delta z}$. Putting aside possible systematic effects related to adopting this machine learning approach in this framework, we only need to quantify the uncertainty associated with our choice of the error model. To account for this, we measured how much the output $n(z)$ changes when the $a$ and $s$ parameters are determined from different sub-samples on the sky. For this, we split the KiDS-bright $\times$ GAMA matched sample into ten sub-samples along the right ascension, where each sub-sample has the same number of objects. We fitted $a$ and $s$ to each sub-sample with Eq.~\eqref{eq:Lorentzian} and convolve the resulting $p_{\delta z}$ with the full $p(z_\mathrm{ph})$ of the KiDS-bright sample. The resulting ten $n(z)$ distributions are almost indistinguishable from our best estimate, as displayed in Fig.~\ref{fig:nz_a_s}. We, therefore, conclude that we can safely neglect the error coming from the Lorentzian fit. 

It is difficult to estimate all uncertainties accurately on the $n(z)$ estimate since, for this, we would need to test the ANN2z algorithm on another spectroscopic survey with the same selection. We investigate this further and study the impact of changing the shape and the mean of the $n(z)$. Therefore, besides the best-estimated $n_\mathrm{be}(z)$, we also consider using the $n_\mathrm{ph}(z)$ directly, as well as the spectroscopic redshift distribution $n_\mathrm{sp}(z)$ itself, coming from the matched KiDS-bright $\times$ GAMA galaxies; both also shown in Fig.~\ref{fig:nofz}. We further allow the $n(z)$ to shift along the redshift direction to give the analysis some flexibility, where the shift value $\delta \langle z \rangle$ is drawn from a Gaussian with a standard deviation of 0.01 and vanishing mean, motivated by the uncertainty on the mean of the source redshift distribution.

In addition to the full lens sample described above, we take advantage of the colour information contained in the KiDS-bright data to construct colour-selected sub-samples. This allows us to constrain the bias of blue and red galaxy populations separately from the DSS signal. Following Bi21, we use an empirical split between red and blue galaxies based on their location on the absolute $r$-band magnitude $M_r$ and the rest-frame $u-g$ colour diagram. The rest-frame quantities are based on LePhare \citep{Arnouts1999} with the derivations presented in Bi21. We apply a cut through the green valley in the colour-magnitude diagram, which results in a line that delimits the red and blue samples that satisfy
\begin{equation}
    u-g = 0.825-0.025\, M_r \, .
\end{equation}
We identify those galaxies that are at least 0.05 mag above (below ) the cut line as red (blue) galaxies. We estimate their underlying redshift distributions following the same approach as for the full sample, and the resulting $n(z)$ are shown in Fig.~\ref{fig:nofz_redblue}. The effective number densities and best-fit parameters of the modified Lorentzian redshift error model are also listed in the second and third rows of Table~\ref{table:data_overview}. We finally note that, while these colour-selected sub-samples are particularly interesting from a galaxy formation perspective, our main cosmological results are obtained from the full lens sample, which has the highest signal-to-noise.

\subsection{Source catalogues}

The fiducial KiDS-1000 cosmic shear catalogue consists of five tomographic bins, whose redshifts are calibrated using the self-organising map (SOM) method\footnote{The SOM method organises galaxies into groups based on their nine-band photometry and finds matches within spectroscopic samples. Galaxies for which no matches are found are removed from the catalogue.} of \citet{Wright:2020} and presented in \citet{Hildebrandt2021}. Although the five bins can be exploited in a cosmic shear analysis as in \citet{Asgari2021}, for the DSS analyses, one must also be cautious about source-lens coupling, which arises if sources and lenses belong to the same gravitational potential. This can significantly affect our signal and bias our cosmological inference if left unmodelled. A significant redshift overlap between the source and lens distribution can result in further contamination by the IA of source galaxies that are tidally connected with the foreground lenses. We measure this effect in Sect. \ref{sect:cosmo-SLICS+IA} from IA-infused weak lensing simulations and show that, given the KiDS-bright $n(z)$, this can be avoided by excluding the first three tomographic bins of the KiDS-1000 sources from the analysis. An additional lens-source coupling complication is the so-called boost factor, which arises due to the clustering of sources with over-dense and the anti-correlation of sources with under-dense regions. This can be taken into account by modifying the source $n(z)$ depending on clustering properties \citep{Gruen:Friedrich:2018}. Since we exclude the third bin, we do not need to consider this further complication. The fourth and fifth redshift bins, shown as the orange and brown lines in Fig.\,\ref{fig:nofz}, are separate enough from the lens $n(z)$ to avoid any appreciable lens-source coupling and are therefore used in our cosmological analysis. The uncertainty on the redshift distribution and the residual systematic offsets are very small, as listed in the last two rows of Table \ref{table:data_overview}. The galaxy shear estimates are provided by the lensfit tool \citep{Miller2013,Fenech2017} and are described in more detail in \citet{Giblin:2020}, where it is shown that shear-related systematic effects do not cause more than a $0.1\sigma$ shift in $S_8 \equiv \sigma_8(\Omega_\mathrm{m}/0.3)^{0.5}$ when measured by cosmic shear two-point functions.

\section{Simulated data}
\label{Sect:Simu_Data}

Besides the real KiDS-1000 data, we validate our inference pipeline on several simulated data sets, study the impact of key systematic uncertainties and carry out the cosmological inference. 
We use the publicly available \texttt{FLASK} tool (Full-sky Log-normal Astro-fields Simulation Kit) described in \citet{FLASK2016} to estimate the covariance of errors in the DSS data vector of the KiDS-1000; the cosmo-SLICS+IA simulations, described in HD22, to quantify the impact of IA on our measurements and to validate the new $N_\mathrm{ap}$ segment in our pipeline  (see the end of Sect.~\ref{Aperture_stat}) that was not present in the B22 model; and the Magneticum lensing simulations, first introduced in \citet{Hirschmann2014}, to investigate the impact of stellar and AGN feedback. More details are provided in the following sections.

\subsection{\texttt{FLASK} log-normal simulations}
\label{sec:FLASK}
Our cosmological inference analysis requires an estimate of the error covariance of the DSS data vector. Since an analytical covariance matrix for the DSS is challenging to compute, we make use instead of an ensemble of log-normal simulations produced with the publicly available \texttt{FLASK} tool\footnote{\texttt{FLASK}: \url{http://www.astro.iag.usp.br/~flask/}} \citep{FLASK2016}. In \citet{Hilbert:2011} it is shown that log-normal random fields are a good approximation to the 1-point PDF of the weak lensing convergence and shear field, and \citet{Friedrich2020} show that they are in fact accurate enough to estimate the covariance matrix for higher-order statistics in Stage-III lensing surveys (see their figure 4).\footnote{We tested that an area-rescaled covariance matrix coming from over 600 fully independent $N$-body simulations \citep[see][for a description of the SLICS simulation suite]{Harnois-Deraps2018} results in similar constraints.} Compared to full $N$-body simulations, \texttt{FLASK} log-normal random fields are computationally cheap to create. The fact that \texttt{FLASK} outputs full-sky maps has the advantage that it can easily be masked to match the footprint of the data, making area re-scaling unnecessary. 
For the creation of our mock catalogues, we use the cosmological parameters that approximately match current cosmological analyses and fixed the matter density parameter to $\Omega_\mathrm{m}=0.3$, the normalisation of the matter power spectrum to $\sigma_8=0.74$, the dimensionless Hubble parameter to $h=0.7$, the dark energy equation-of-state parameter to $w_0=-1$ and the power spectrum power-law index to $n_s=0.97$. Furthermore, we provide \texttt{FLASK} with the angular power spectrum of the projected matter density field, the convergence power spectrum for both source bins, and the two matter-lensing cross-spectra. By assuming a flat universe throughout this paper, given the $n(z)$ shown in Fig.\,\ref{fig:nofz}, and using the PYCCL software package\footnote{Currently available here: \url{https://github.com/LSSTDESC/CCL}} \citep{pyccl} to get the 3D matter density contrast power spectrum $P_\delta(\ell/\chi,\chi)$, we calculate the angular power spectrum by use of the Limber-approximated projection \citep{Kaiser1992} as
\begin{equation}
C_{i,j}(\ell) = \int\limits_0^\infty \dd\, \chi \frac{W_i(\chi)W_j(\chi)}{\chi^2}\,P_\delta(\ell/\chi,\chi)\, ,
\label{eq:power_Cl}
\end{equation}
where $i,j$ are placeholder for either the galaxy or convergence projection, such that $W_g(\chi) = n_\mathrm{l}(z[\chi])\frac{\dd z(\chi)}{\dd \chi}$ for the lenses with redshift distribution $n_\mathrm{l}$, while for the source with redshift distribution $n_\mathrm{s}$ we have instead
\begin{equation}
    W_\mathrm{s}(\chi) = \frac{3\Omega_\mathrm{m}H_0^2}{2c^2}\int_\chi^\infty \dd \chi'\, \frac{\chi(\chi'-\chi)}{\chi'a(\chi)}\,
    n_\mathrm{s}(z[\chi'])\frac{\dd z[\chi']}{\dd \chi'} \;.
\end{equation}
Besides these angular power spectra, \texttt{FLASK} needs the log-normal shift parameters $\kappa_0$ and $\delta_0$, where $-\kappa_0$ and $-\delta_0$ defining the lower limits of the log-normal random variable of the convergence and matter density fields, respectively. Whereas the shift parameter $\kappa_0 = \{0.02,0.03\}$ for the convergence power spectra for the two source bins can be determined directly from the fitting formula equation (38) in \citet{Hilbert:2011}, we estimated the shift parameter $\delta_0 = \{ 0.57,0.59,0.56\}$ for the three lens samples (full, red, blue), as described in \citet{Gruen:2015}, by assuming that it can be approximated by the shift parameter of the smoothed density contrast, which in turn is calculated from our model for a top-hat filter function (see equation 23 in B22). Given this setup, \texttt{FLASK} returns a foreground density map $\delta_{\mathrm{m,2D}}({\boldsymbol \theta})$ and two sets of correlated shear and convergence grids $\gamma_{1,2}({\boldsymbol \theta}), \kappa({\boldsymbol \theta})$, one per tomographic source bin. We populate our mock KiDS-bright galaxies on the density map by sampling, for each pixel ${\boldsymbol \theta}$, a Poisson distribution with mean parameter $\lambda=n_\mathrm{eff}\big[1+b\,\delta_{\mathrm{m,2D}}({\boldsymbol \theta})\big]$, where $b=1.4$ is the (constant) linear galaxy bias estimated from preliminary analyses with only some realisations\footnote{Also different values would not affect the posteriors as we discuss in Sect.\ref{sec:scaling_covariance}.} and $n_\mathrm{eff}=0.325\,\mathrm{arcmin}^{-2}$ is the mean galaxy density of the KiDS-bright sample. Similarly, we populate the two source planes by Poisson-sampling for each pixel a number of source galaxies $n_\mathrm{pix}$ with parameter $\lambda = n_\mathrm{eff} A_\mathrm{pix}$, where $A_\mathrm{pix}$ is the area of the pixel under consideration\footnote{The public KiDS-1000 mask is provided on a flat sky with a resolution of $0.01\,\mathrm{arcmin}^2$. This results in a \texttt{HEALPix} mask that varies from pixel to pixel given the fact that the pixelation is different and has a size of $0.74\,\mathrm{arcmin}^2$.}, and the effective number density $n_\mathrm{eff}$ is taken from Table \ref{table:data_overview}. We finally combine the two shear components of each object with their convergence to construct reduced shear components $g_{1,2}$, and further combine these with a shape noise contribution $\epsilon^{\nt{s}}$  taken from sampling a Gaussian  distribution with vanishing mean and deviation $\sigma_\epsilon$ also taken from Table \ref{table:data_overview}. This results in catalogues containing observed ellipticities ${\boldsymbol\epsilon}^{\nt{obs}}$ transformed as \citep{Seitz1997}
\begin{equation}
\epsilon^{\nt{obs}} = \frac{{\boldsymbol \epsilon}^{\nt{s}}+{\boldsymbol g}}{1+{\boldsymbol \epsilon}^{\nt{s},*}{\boldsymbol g}} \, .
\label{eq:ebos}
\end{equation}
The quantities in bold here are all complex numbers, and the asterisk `$*$' indicates complex conjugation. This procedure ensures that we match the number of foreground and background galaxies in the data and the associated shape noise level.

\subsection{cosmo-SLICS+IA}
\label{sect:cosmo-SLICS+IA}

As mentioned earlier, the second suite of simulations is used to validate the inference pipeline and study the impact of IA on our DSS measurements.
We use for this the fiducial suite of the cosmo-SLICS presented in \citet{Harnois-Deraps2019}, which consists of a set of 50 simulated light cones of 100 deg$^2$ each, run in a $\Lambda$CDM universe with $\Omega_{\nt{m}}=0.2905$, $\Omega_{\Lambda}=0.7095$,
$\Omega_{\nt{b}}=0.0473$, $h=0.6898$, $\sigma_8=0.836$ and $n_{\rm s}=0.969$. The mocks follow the non-linear evolution of $1536^3$ particles up to $z=0$, computed by the {\sc cubep$^3$m} $N$-body code \citep{Harnois-Deraps2013}. For Fourier modes of comoving wave number $k<2.0\,h\,\nt{Mpc}^{-1}$, the cosmo-SLICS three-dimensional dark matter power spectrum $P(k)$ agrees within 2$\%$ with the predictions from the Extended Cosmic Emulator \citep{Heitmann:Lawrence:2014}, followed by a progressive deviation for higher $k$-modes \citep{Harnois-Deraps2019}, offering a sufficient resolution to model Stage-III galaxy surveys. The particle data were assigned onto mass sheets at 18 redshifts and then post-processed into $10\times10$ deg$^2$ light cones. Lensing maps were produced at 18 source redshift planes for each cosmo-SLICS light cone and used to interpolate lensing information onto galaxy catalogues.

Similar to B22, we construct cosmo-SLICS mock source samples that reproduce a number of key data properties, including the tomographic $n(z)$, the galaxy number density $n_\mathrm{eff}$ and the shape noise levels. As for the \texttt{FLASK} simulations, we use Eq.~\eqref{eq:ebos} to add shape noise to the reduced shear signal. Source galaxies are placed at random positions on the light cones, and the shear quantities ($\gamma_{1/2}, \kappa$) are interpolated at these positions from the enclosing lensing maps. 

We also construct mock KiDS-bright samples by populating the light cone mass maps with galaxies that trace the underlying dark matter field linearly, following the method presented in \citet{Harnois-Deraps2018}. We here again fix the galaxy bias to 1.4 and an effective number density of $n_\mathrm{eff}= 0.325\,\mathrm{arcmin}^{-2}$.

\subsubsection{IA infusion}
The impact of galaxy IA is a known secondary signal to the cosmic shear measurements that have been neglected in past DSS studies. In this paper, we verify the validity of this assumption by measuring our statistics in simulated source data that are infused with IA. We find that IA influences our data vector only if the lenses' $n(z)$ overlap with that of the sources. Following the methods described in HD22, the IA properties of these galaxies are computed as
\begin{eqnarray}
\epsilon_1^{\mathrm{IA}} = - \frac{A_{\mathrm{IA}}\bar{C_1}\bar{\rho}(z)}{D_+(z)} (s_{xx} - s_{yy}) \;,\quad    \epsilon_2^{\mathrm{IA}} = - \frac{2 A_{\mathrm IA}\bar{C_1}\bar{\rho}(z)}{D_+(z)} s_{xy}\, ,
\label{eq:tidal_th}
\end{eqnarray}
where $s_{ij} = \partial_{ij} \phi$ are the Cartesian components of the projected tidal field tensors interpolated at their positions, with $\phi$ being the gravitational potential. In the above expression, $A_{\mathrm{IA}}$ captures the strength of the coupling between the ellipticities and the tidal field, $\bar{\rho}(z)$ is the matter density, $D_+(z)$ is the linear growth factor,  $\bar{C_1}= 5\times 10^{-14} M_{\odot}^{-1} h^{-2}$  Mpc$^3$, as calibrated in \citet{Brown2002}. These intrinsic ellipticity components $\epsilon_{1/2}^{\rm IA}$ are then combined with the cosmic shear signal by Eq.~\eqref{eq:ebos}, resulting in an IA-contaminated weak lensing sample that is consistent with the NLA model of \citet{Bridle2007}. We refer the reader to HD22 for full details about the IA infusion method. We test several values of $A_{\mathrm{IA}}$, more precisely, we infused $A_{\mathrm{IA}}=\{1,1.5,2\}$, and inspect in each case the impact on the DSS data vector.

\subsection{Magneticum}
\label{sec:magneticum}
Baryon feedback is also known to affect the distribution of the large-scale structure significantly, as the sustained outflows of energy arising from stellar winds, supernovae and AGN reduce the clustering on intra-cluster scales by up to tens of per cent \citep{vanDaalen2011}. The exact strength of this suppression is still largely uncertain, with different hydro-dynamical simulations predicting  different redshift and scale dependencies \citep[see, e.g.][for a review of recent results]{Chisari2015}. Without consensus, we opted to measure the DSS in one of these hydro-dynamical simulations for which the impact is quite high and inspect how an extreme baryon impact would affect our data vector.

The Magneticum lensing simulations were first introduced in \citet{Hirschmann2014} and used to mock up KiDS-450 and Stage-IV cosmic shear data \citep{Martinet2021b},
and subsequently in \citet{Harnois-Deraps2021} to study the impact of baryons in the peak count analysis of the Dark Energy Survey Y1 data. The underlying matter field is constructed from the {\it Magneticum Pathfinder} simulations,\footnote{\url{www.magneticum.org}} more specifically by the {\it Run-2} and {\it Run-2b} data described in \citet{Hirschmann2014} and \citet{Ragagnin2017}. These are based on the {\sc Gadget3} smoothed particle hydrodynamical code \citep{Springel2005} and are able to reproduce a large number of observations \citep[see][for more details]{Castro2021}. These both co-evolve dark matter particles of mass $6.9 \times 10^8 h^{-1}M_\odot$ and gas particles with mass $1.4 \times 10^8 h^{-1}M_\odot$, in comoving volumes of side 352 and 640 $h^{-1} \,{\mathrm{Mpc}}$, respectively. Included key mechanisms are radiative cooling, star formation, supernovae, AGN, and their associated feedback on the matter density field. 
From sequences of projected mass planes, we use the procedure outlined above for the cosmo-SLICS simulations to generate KiDS-1000 sources and KiDS-bright lenses for ten pseudo-independent light cones, each covering 100 deg$^2$. We repeat the same procedure on dark matter-only light cones, such that any difference is caused by the presence of baryons. 

The cosmo-SLICS+IA and Magneticum light cones are square-shaped, a geometry that accentuates the edge effects when the aperture filter overlaps with the light cone boundaries. One could, in principle, weight the outer rims for each $N_\mathrm{ap}$ map, such that the whole map can be used; although this would increase our statistical power, it could also introduce a systematic offset. We opted instead to exclude the outer rim for each realisation, resulting in an effective area of $36\,\mathrm{deg}^2$, where a $2\,\mathrm{deg}$ band has been removed, matching the size of the adapted filter. This procedure also ensures that roughly the same number of background galaxies are used to calculate the shear profile around each pixel.

\section{Cosmological parameter inference}
\label{sec:inference_description}
Before performing several Markov chain Monte Carlo (MCMC) samplings in the following two sections, we describe here the pipeline of our Monte-Carlo sampler. In our different MCMC runs, the model vector, the data vector and the covariance matrix are varied, but the overall pipeline stays the same.
\begin{table}[h!]
\centering
\caption{All varied parameters and their prior knowledge.}
\begin{tabular}{cc}
\hline
\hline
parameter & prior \\
\hline
$\Omega_\mathrm{m}$ & $\mathcal{U}(0.20,0.50)$  \\
$\sigma_8$ & $\mathcal{U}(0.45,1.00)$  \\
bias $b$ & $\mathcal{U}(0.5,2.5)$  \\
$\alpha$ & $\mathcal{U}(0.1,8)$  \\
\hline
$\delta \langle z \rangle$ full KiDS-bright sample & $\mathcal{N}(0.0,0.01)$  \\
$\delta \langle z \rangle$ red KiDS-bright sample & $\mathcal{N}(0.0,0.01)$  \\
$\delta \langle z \rangle$ blue KiDS-bright sample & $\mathcal{N}(0.0,0.01)$  \\
$\delta \langle z \rangle$ source bin 4,5 & $\mathcal{N}([0.011,-0.006],C_{\delta \langle z \rangle})$  \\
$m$-bias source bin 4 & $\mathcal{N}(0.002,0.012)$  \\
$m$-bias source bin 5 & $\mathcal{N}(0.007,0.010)$  \\
\hline
\hline
\end{tabular}
\tablefoot{Uniformly $\mathcal{U}$ and normally distributed $\mathcal{N}$ priors on the parameters used in our cosmological inferences. The normally distributed priors on the multiplicative shear $m$-bias and photometric redshift errors $\delta \langle z \rangle$ are used only for the real data analysis, not for the simulations where we set them to zero. The $\delta \langle z \rangle$ for the sources follow a joint normal distribution with covariance matrix $C_{\delta \langle z \rangle}$ shown in figure 6 of H21.}
\label{table:parameter_overview}
\end{table}

Our statistical analysis has the following two free cosmological parameters that we fitted for: the matter density parameter $\Omega_\mathrm{m}$ and the normalisation of the power spectrum $\sigma_8$. We additionally vary the galaxy bias term $b$ and the super-Poisonnian shot-noise parameter $\alpha$ (see Eq.\,\ref{eq:CF_alpha}). We detail the prior ranges of all parameters in Table \ref{table:parameter_overview}, where we also show the Gaussian priors for the nuisance parameters used in the data analysis (but not in the simulation-based validation runs).

For the estimated covariance matrix $\Tilde{C}$, which itself is a random variable, \cite{Percival2021} suggested a procedure that uses a more general joint prior of the mean and covariance matrix as the Jeffreys prior proposed in equation 6 in \cite{Sellentin2016}. The method by \cite{Percival2021} leads to credible intervals that can also be interpreted as confidence intervals with approximately the same coverage probability. From a data vector $\vb{d}$ and a covariance matrix $\Tilde{C}$ measured from $n_\mathrm{r}$ simulated survey realisations, the posterior distribution of a model vector $\vb{m}$ that depends on $n_\theta$ parameters $\boldsymbol{\Theta}$ is
\begin{equation}
    \boldsymbol{P}\left(\vb{m}(\boldsymbol{\Theta})|\vb{d},\Tilde{C}\right) \propto |\Tilde{C}|^{-\frac{1}{2}} \left( 1 + \frac{\chi^2}{n_{\rm r}-1}\right)^{-m/2}\, ,
    \label{eq:t_distribution}
\end{equation}
where
\begin{equation}
\chi^2 =  \left[\vb{m}(\boldsymbol{\Theta})-\vb{d}\right]^{\rm T} \Tilde{C}^{-1} \left[\vb{m}(\boldsymbol{\Theta})-\vb{d}\right] \, .
\label{eq:chi2}
\end{equation}
The power-law index $m$ is 
\begin{equation}
    m = n_\theta+2+\frac{n_\mathrm{r}-1+B(n_\mathrm{d}-n_\theta)}{1+B(n_\mathrm{d}-n_\theta)}
\end{equation}
with $n_{\rm d}$ being the number of data points and
\begin{equation}
    B = \frac{n_\mathrm{r}-n_\mathrm{d}-2}{(n_\mathrm{r}-n_\mathrm{d}-1)(n_\mathrm{r}-n_\mathrm{d}-4)} \, .
    \label{eq:B}
\end{equation}
By setting $m=n_\mathrm{r}$ the formalism of \cite{Sellentin2016} is recovered. 

Finally, since the model prediction is too slow for our MCMC, we use the emulation tool contained in \texttt{CosmoPower} \citep{COSMOPOWER2022}, which was first developed to emulate power spectra but can easily be adapted for arbitrary vectors. We trained the emulator on 4000 model points in the parameter space $\{\Omega_{\rm m}, \sigma_8, b, \alpha \}$ distributed in a Latin hypercube, where we also included $\delta \langle z \rangle$ Gaussian distributed values with the mean as shown in Table \ref{table:parameter_overview} but twice the standard deviation.\footnote{For the validation analysis we used only 2000 nodes as for that analysis the nuisance parameters were not modelled.} To quantify the accuracy of the emulator, we calculated the model at 500 independent points in the same parameter space, as determined with the emulator or directly with the model and show the model vector accuracy in Fig.\,\ref{fig:chi_acc}. The fractional error is better than $2\%$ ($95\%$ confidence level).

\subsection*{Reporting parameter constraints and goodness of fit}
\label{sec:parameter_estimation}
In this work, we followed the approach of \citet{Joachimi2021} to report our parameter constraints. In particular, we seek to report the global best fit to the data, that is the set of parameter values that provide the maximum a posteriori (MAP) distribution, computed as
\begin{equation}
    \boldsymbol{\Theta}_\mathrm{MAP} = \underset{\boldsymbol{\Theta}}{\mathrm{argmax}} \left[\boldsymbol{P}(\vb{m}(\boldsymbol{\Theta})|\vb{d},\Tilde{C})\right]\,.
\end{equation}
where we found the maximum by running several minimisation processes. To estimate the resulting uncertainties around the MAP, we use the suggested projected-joint-highest-posterior-density (PJ-HPD) method, which calculates the parameter ranges that encompass the $68\%$ and $95\%$ credible intervals. 

Furthermore, with the degrees of freedom (d.o.f.), we also report the reduced $\chi^2/\mathrm{d.o.f.}$ to quantify the goodness of fit, where the $\chi^2$ results from the point in the high-dimensional parameter space that has the highest posterior probability. To unbias the covariance matrix $\Tilde{C}$, which is used to estimate the $\chi^2$-values, we instead of inverting $\Tilde{C}\,\mathrm{h}$ with the known Hartlap factor \citep{Hartlap2007} defined as ${\mathrm{h}=(n_\mathrm{r}-1)/(n_\mathrm{r}-n_\mathrm{d}-2)}$, but rather use
\begin{equation}
    \Tilde{C}' = \frac{(n_\mathrm{r}-1)\left[1+B(n_\mathrm{d}-n_{\theta})\right]}{n_\mathrm{r}-n_\mathrm{d}+n_\theta-1}\Tilde{C} \, .
\end{equation}

To estimate the d.o.f. we measure for 1000 mock data vectors the best $\chi^2$ and fit a $\chi^2$ distribution to it. The 1000 mock data vectors are drawn from a multivariate Gaussian distribution, where the mean is the model prediction at the MAP values, and the covariance is the corresponding covariance matrix for that particular model. As we have only four free parameters and the rest are fixed by prior knowledge, we expect that the resulting d.o.f. is only slightly less than the raw number of elements in the data vector. Lastly, we report the $p$-value in each case, which provides the probability of finding a $\chi^2$ that is more extreme for the given d.o.f., and therefore indicates the goodness of fit. For our analysis, we choose a significance level of 0.01 to be a reliable fit. We have verified with some selected cosmo-SLICS nodes that the theoretical model for the KiDS-bright sample is valid and accurate for $\sigma_8 < 1.0$, with indications that it loses accuracy for larger values of $\sigma_8$. This does not affect our results, given that the preferred values of $\sigma_8$ are well below this limit.

\section{Validating the model on simulations}
\label{sec:validation}

In this section, we validate our model on simulated measurements, several of which are infused with known and controlled systematic effects. The first (fiducial) test establishes that our model is unbiased in the simplest setup, where lens galaxies linearly trace the pure dark matter density maps, while source galaxies are given by the noise-free pure gravitational shear. The second test verifies that our results are unchanged in the presence of IA, as described in Sect.~\ref{sect:cosmo-SLICS+IA}, while, lastly, we investigate the impact of baryonic physics on our statistics. The fiducial and the IA tests use 20 light cones, which have an unmasked area close to that of the KiDS-1000 footprint\footnote{After removing the outer strip, each light cone has an effective area of $36\,\mathrm{deg}^2$, which roughly matches the unmasked area of the KiDS-1000 data if 20 light cones are added.}. The measurements on the Magneticum mocks use the ten available light cones. Furthermore, we measured the shear profiles from KiDS-like mocks with shape noise to validate our model on a realistic data vector.

\begin{figure}[h!]
\includegraphics[width=\columnwidth]{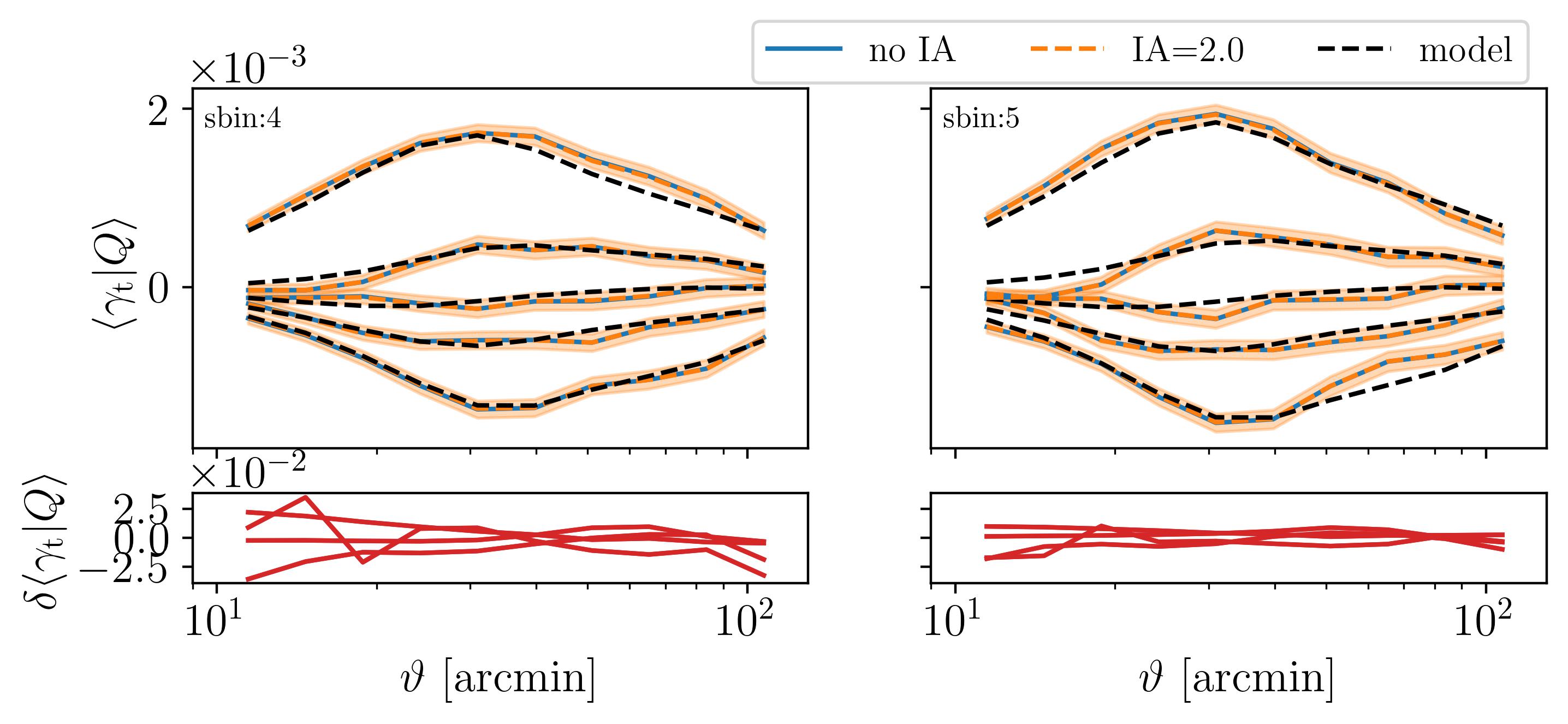}
\caption{Shear profiles measured with the adapted filter for the KiDS-bright-like lenses and sources from the cosmo-SLICS+IA simulations for two different IA amplitudes (see the legend). The orange regions are estimated from the covariance matrix, while the black dashed lines are obtained from our IA-free analytical predictions at the input cosmology and using the $n(z)$ shown in Fig.\,\ref{fig:nofz}. As the differences between the shear profiles with varying IA amplitude can barely be seen, we display the relative difference between them in the bottom panels for the highest and lowest two quantiles.}
\label{fig:shear_SLICS}
\end{figure}

\begin{figure}[h!]
\centering
\includegraphics[width=\columnwidth]{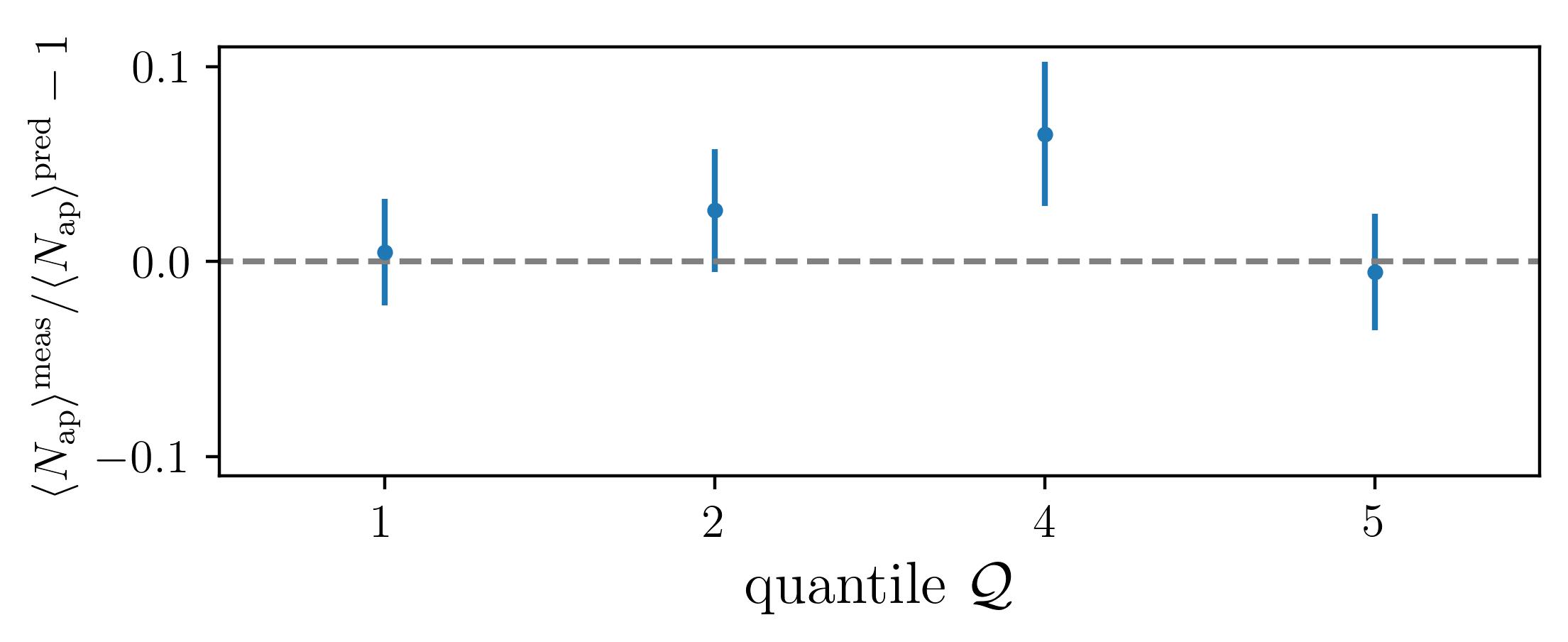}
    \caption{Relative difference between the mean $\langle N_\mathrm{ap} \rangle$ measured in the cosmo-SLICS+IA bright mocks for four quantiles, and the value of $\langle N_\mathrm{ap} \rangle$ predicted by the model at the same cosmology. The blue error bars show the KiDS-1000 statistical uncertainty measured from \texttt{FLASK}.}
    \label{fig:mean_Nap_SLICS}
\end{figure}

\subsection{Validation on intrinsic alignment}

To quantify the influence of IA on our results, we performed an MCMC analysis on the mock infused with a strong IA amplitude ($A_{\mathrm{IA}}=2.0$) and compared it to our fiducial model ($A_\mathrm{IA}=0$). For this, we used the mocks described in Sect.~\ref{sect:cosmo-SLICS+IA}, excluding the third source tomographic bin due to the lens-source coupling. The resulting profiles and mean relative number counts that are used for our pipeline validation are shown in Figs.\,\ref{fig:shear_SLICS} and \ref{fig:mean_Nap_SLICS}, respectively. Although the aperture number is not affected by IA, it has a slight effect on the profiles, and hence we verify how it impacts the full data analysis. Although the fourth quantile in $N_\mathrm{ap}$ has $2\,\sigma$ deviation, the $p$-value is approximately 0.2, which shows that this is consistent with being a statistical fluctuation. We performed this validation test for the adapted and top-hat filters but show the resulting shear signals, and the corresponding mean aperture number values only for the adapted filter since those of the top-hat filter are very similar and would not yield more insights.

To quantify our decision to discard the third source tomographic bin from our analysis, we show in Fig.~\ref{fig:pure_shear_cosmoSLICS} the same shear profiles as in Fig.\,\ref{fig:shear_SLICS} but also the ones resulting from the third redshift bin, where we clearly see the third tomographic bin is heavily affected by IA, due to a significant overlap in redshift between the lens and source populations, and therefore would need additional modelling of IA, which we disregard for this work, and thus we exclude the third bin.

\begin{figure}
\centering
\includegraphics[width=\columnwidth]{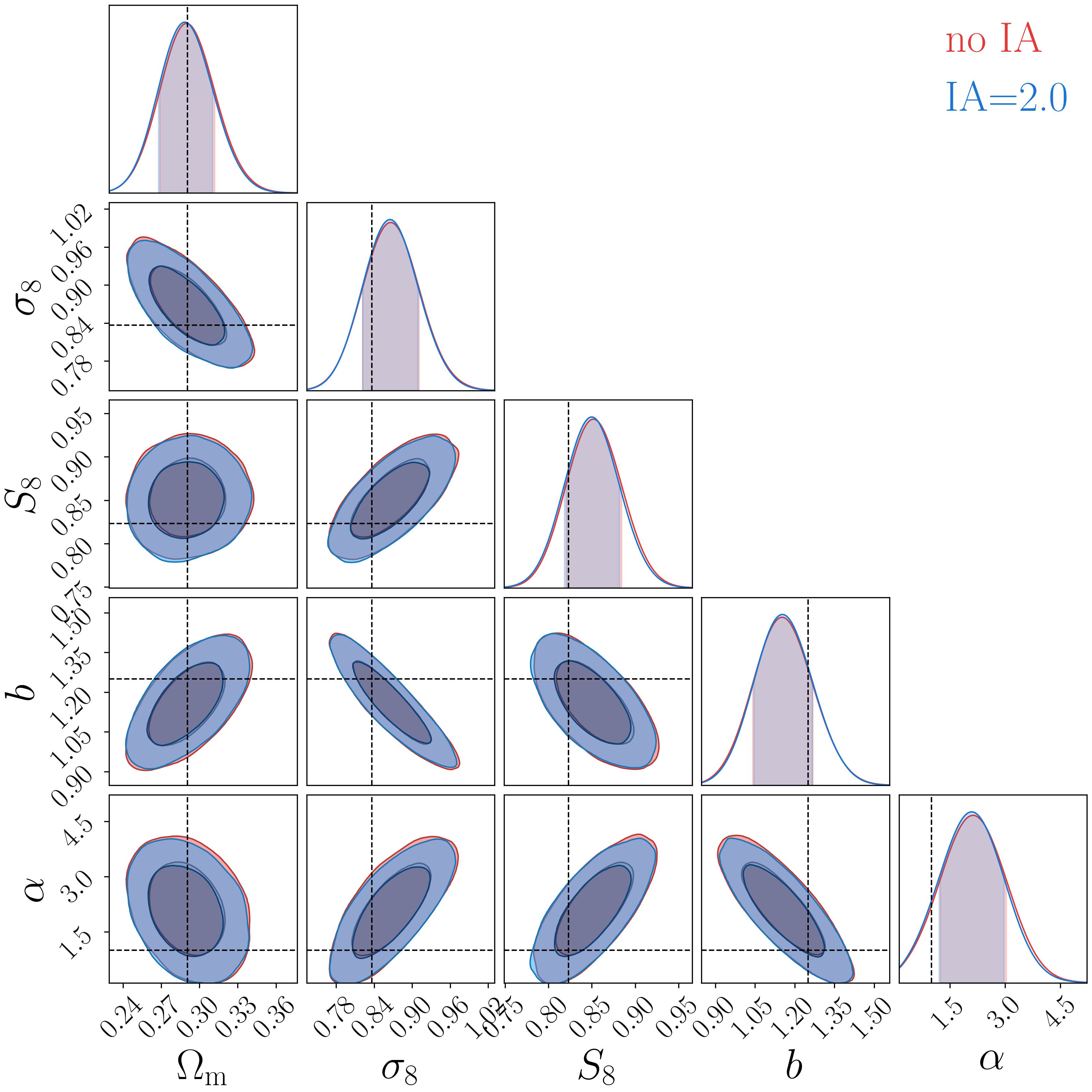}
    \caption{Pipeline validation: Cosmological inference with the adapted filter using the cosmo-SLICS simulations with and without IA infusion, analysed with our model that ignores IA. The posteriors are almost indistinguishable from each other.}
    \label{fig:MCMC_SLICS_adapted}
\end{figure}

The MCMC results for the two IA amplitudes (no $\mathrm{IA}$ and $\mathrm{IA}=2.0$) are shown in Fig.\,\ref{fig:MCMC_SLICS_adapted}. First, it is clearly shown that changing the IA amplitude does not affect the posterior at all. Second, and very importantly, the input cosmology is recovered. We observe a small offset on the parameter $\alpha$, but the other parameters are all recovered within the $1\,\sigma$ region. This confirms that the $<6\%$ deviations seen on the aperture number count presented in Fig.\,\ref{fig:MCMC_SLICS_adapted} do not impact our cosmological inference.
Finally, we observe an anti-correlation between the $S_8$ or $\sigma_8$ parameter and the galaxy bias $b$ parameter, which is expected since all three parameters are directly correlated with the amplitude of the shear signal. This correlation and the correlation of $S_8$ and $\sigma_8$ to the $\alpha$ parameter could potentially impair the robustness of the later constrained parameters. However, these parameters are particularly important for our model and, therefore, cannot be ignored. The same validation is done for the top-hat filter in Appendix \ref{sec:top-hat}.

\subsection{Validation on baryonic feedback}
\begin{figure}[h!]
\includegraphics[width=\columnwidth]{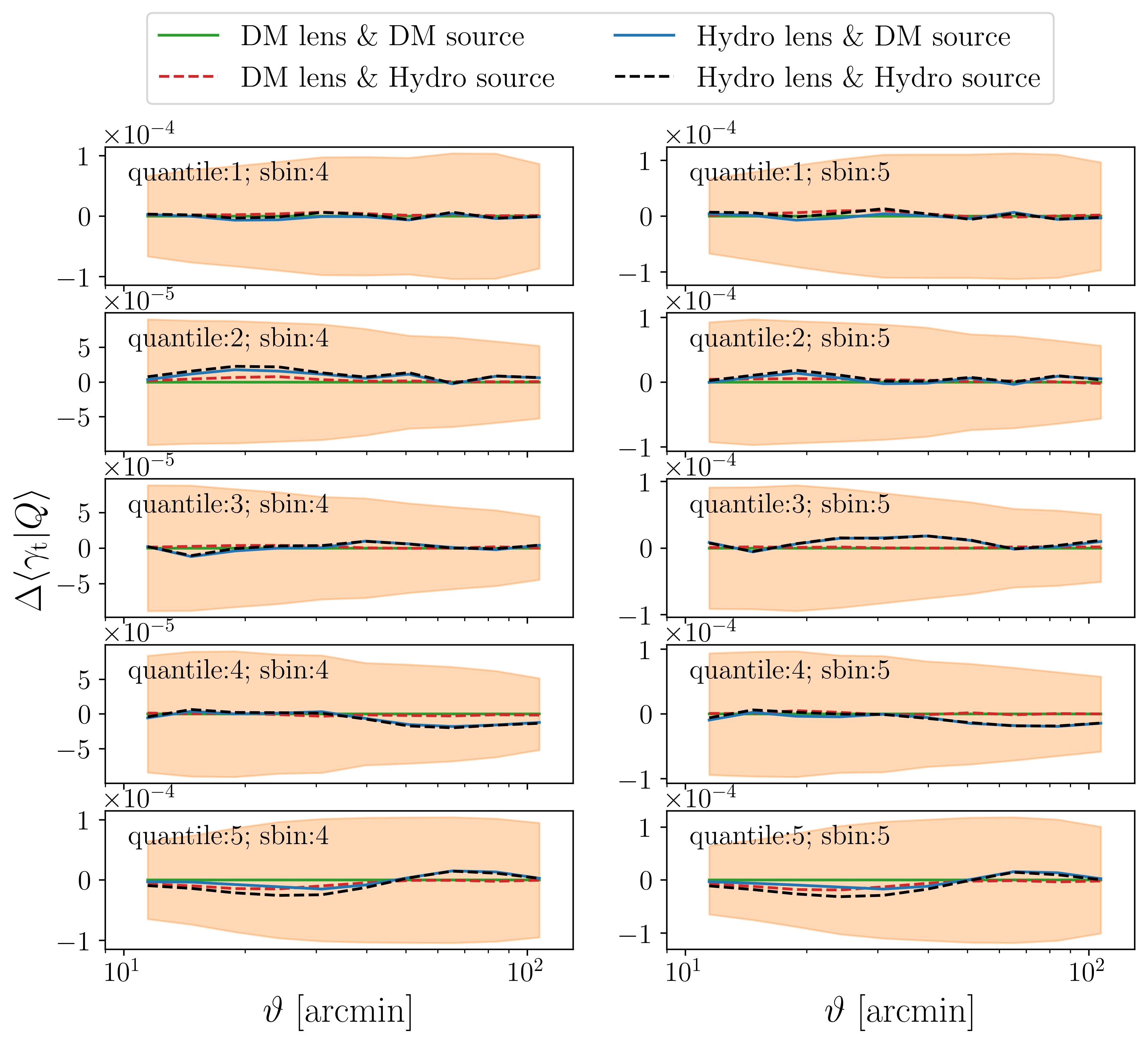}
\caption{Absolute differences between the mean shear profiles for all quantiles using either dark matter-only or full hydrodynamical Magneticum simulations. The residuals are with respect to the dark matter-only mocks for the lenses and sources and are always within the expected statistical uncertainty shown as the orange bands.}
\label{fig:shear_magneticum}
\end{figure}

\begin{figure}[h!]
\centering
\includegraphics[width=\columnwidth]{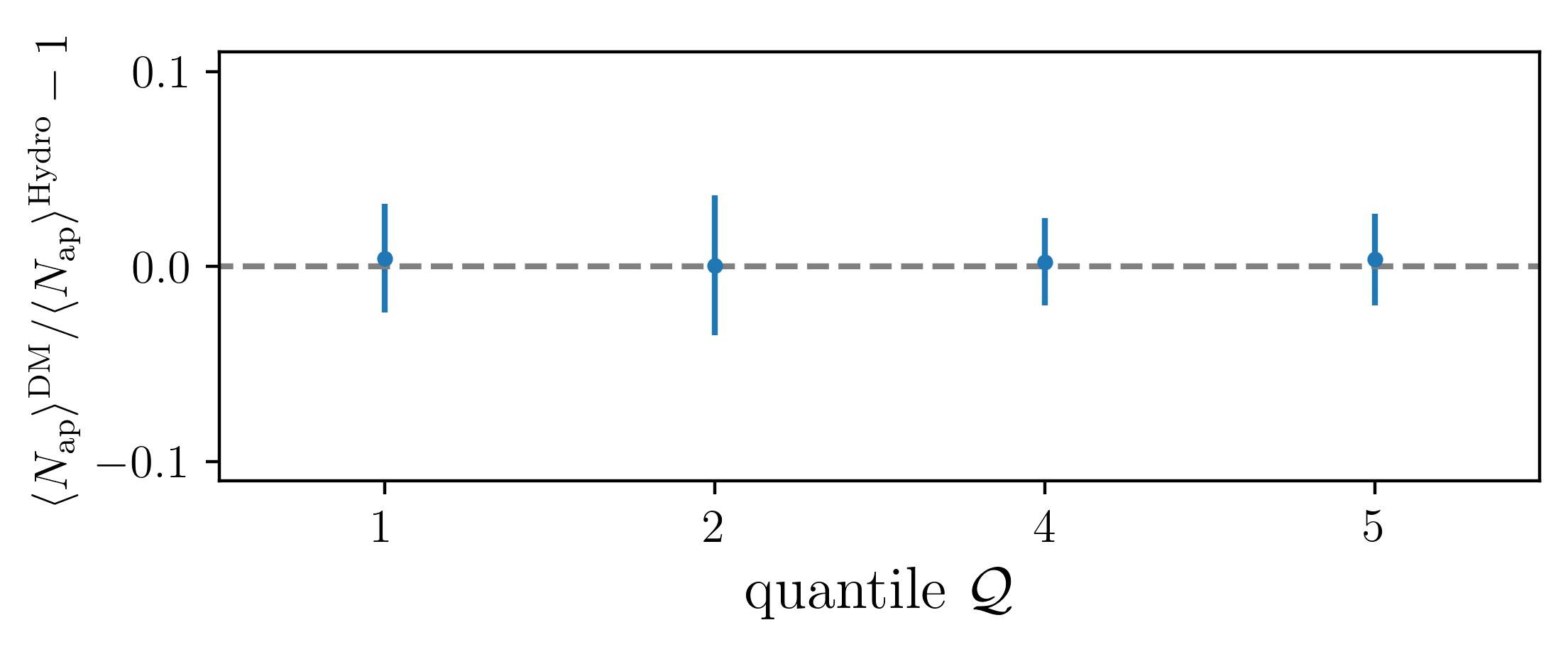}
\caption{Relative difference between the mean $\langle N_\mathrm{ap} \rangle$ to compare measurements from the hydro and dark matter-only Magneticum simulations. The relative difference is always well inside the expected statistical uncertainty of KiDS-1000.}
\label{fig:mean_Nap_magneticum}
\end{figure}

As a last important verification, we investigate for the first time the impact of baryons on the DSS with the Magneticum simulations described in Sect. \ref{sec:magneticum}. By combining the different DM-only and Hydro mock data for the lenses and sources, we end up with four scenarios (lens-source = DM-DM, DM-Hydro, Hydro-DM and Hydro-Hydro). Figure\,\ref{fig:shear_magneticum} shows the residuals between the shear profiles measured from dark matter-only mocks (DM-DM) to the other three combinations for each quantile. Clearly, the deviations are well inside the expected KiDS-1000 uncertainty. The biggest differences are seen if baryonic feedback processes are included in the lens mocks: some pixels, close to the $N_\mathrm{ap}$ threshold between two quantiles, are shifted to another quantile by the presence of baryons. This is in concordance with the mean aperture numbers reported in Fig.\,\ref{fig:mean_Nap_magneticum}, which shows that the mean aperture numbers with baryons are slightly lower. Different to our studies of baryonic feedback such as \citet{Heydenreich2022a} or \citet{Harnois-Deraps2021}, the inclusion of baryons in the sources has only a minor impact on the DSS, but as expected, becoming more important at small scales. In light of this, we can safely neglect the impact of baryons in our real data analysis. 


\section{Results and discussion}
\label{sec:results}
After validating our model to simulations in B22 and our additional testing on the impact of IA and baryonic physics, we are well equipped to analyse real lensing data accurately. As described in Sect.~\ref{Sect:Obs_Data}, we use the KiDS-bright sample as our foreground lenses and the fourth and fifth KiDS-1000 tomographic bins as our cosmic shear data. 
\begin{figure}
\includegraphics[width=\columnwidth]{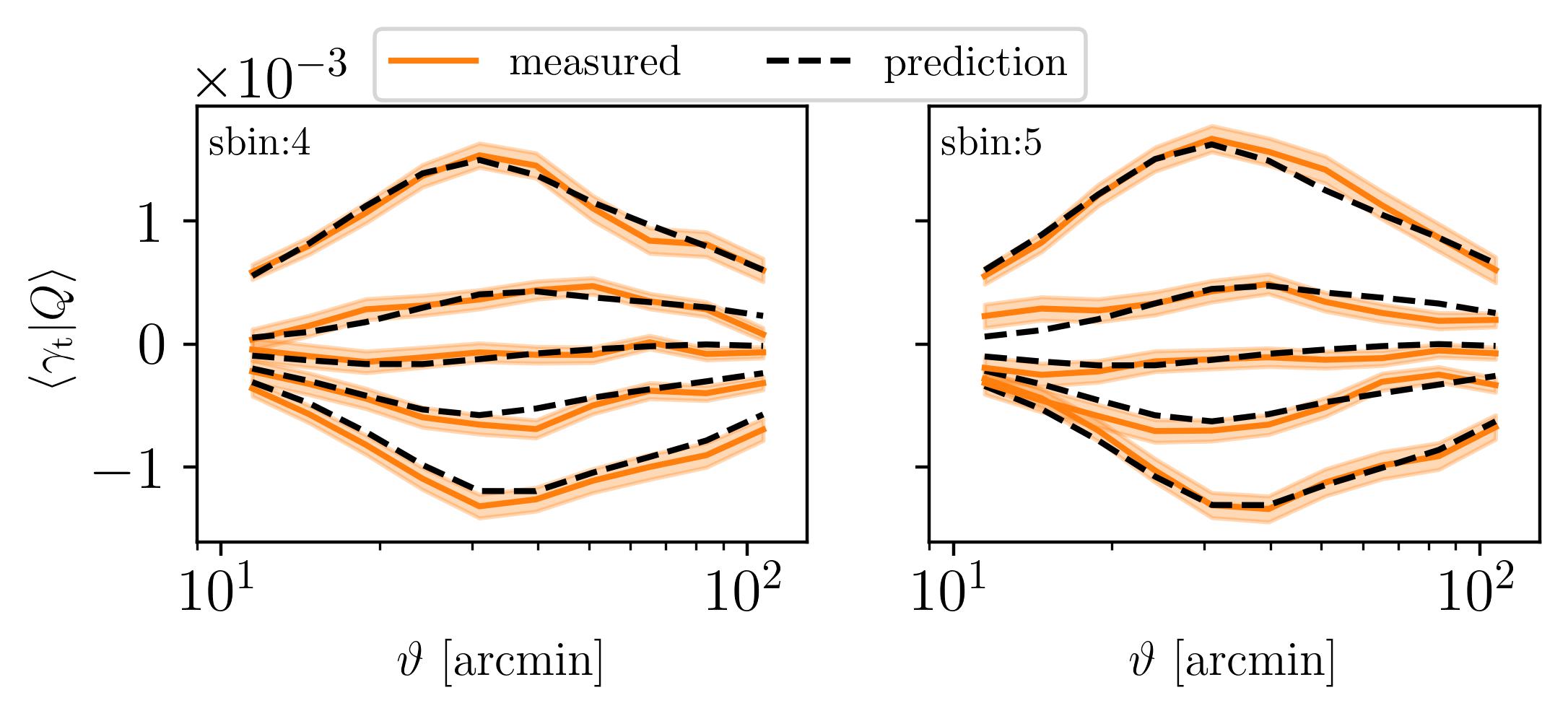}
\caption{Shear profiles measured in the data, compared to the best-fit predictions (MAP) obtained with values listed in Table~\ref{table:MAP_values}, for the adapted filter. The shaded region shows the statistical uncertainty estimated from 1000 \texttt{FLASK} realisations.}
\label{fig:shear_bright_full}
\end{figure}
\begin{figure}
\centering
\includegraphics[width=\columnwidth]{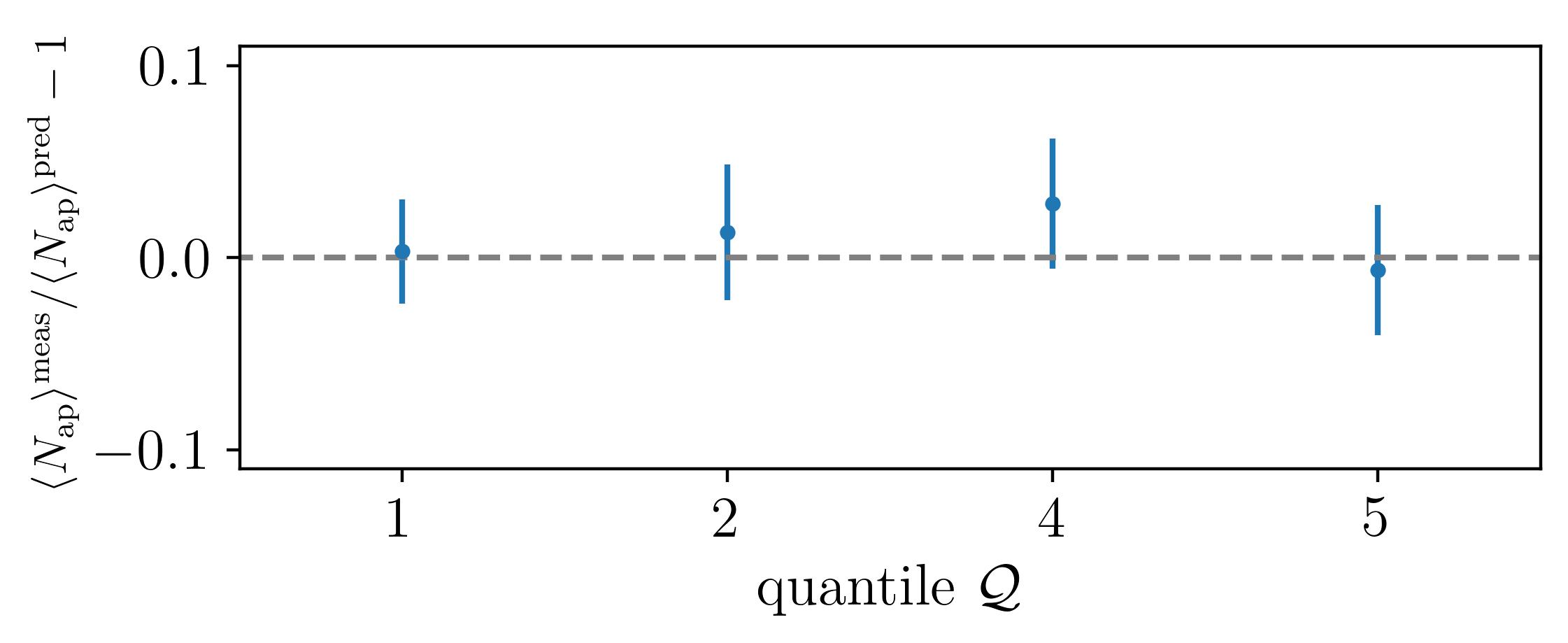}
\caption{Relative difference between the mean $\langle N_\mathrm{ap} \rangle$ to compare measurements from the KiDS-bright sample and our model, evaluated at the MAP values shown in Table~\ref{table:MAP_values}. The measured $\langle N_\mathrm{ap} \rangle$ are all greater than the predicted $\langle N_\mathrm{ap} \rangle$ at MAP just indicating that the measured $p(N_\mathrm{ap})$ is broader than the predicted one.}
\label{fig:mean_Nap_bright_full}
\end{figure}
To recap, in our fiducial analysis, we used the $n(z)$ shown as the black solid line in Fig.\ref{fig:nofz}, we varied the two cosmological parameters $\Omega_\mathrm{m}$ and $\sigma_8$ as well as the two astrophysical parameters $b$ and $\alpha$. We marginalised over the systematic effects parameters describing the $\delta \langle z \rangle$ and $m$-bias uncertainty. In Figs.~\ref{fig:shear_bright_full} and 
\ref{fig:mean_Nap_bright_full} we display the resulting shear profiles and mean aperture number. The model shown in these figures is computed at the best-fit MAP values, listed in the first column of Table \ref{table:MAP_values}. In that table, the $p$-values indicate that the data are well fitted by the model, being all well above our threshold value fixed at $p=0.01$. The d.o.f. are estimated as described in Sect.~\ref{sec:parameter_estimation}, where we show in Fig.~\ref{fig:chi2_around_MAP} the distribution of $\chi^2$ values and for which a $\chi^2$-distribution with 81 d.o.f. fits well. As expected, the resulting d.o.f is slightly lower than the raw number of elements. The reduced $\chi^2$ values are slightly below the expectation of 1.0, potentially indicating that the uncertainties could be slightly overestimated, although they are well inside the expected reduced $\chi^2$ scatter of $\pm\sqrt{2}$, hence do not warrant further investigation.

\begin{figure}
\includegraphics[width=\columnwidth]{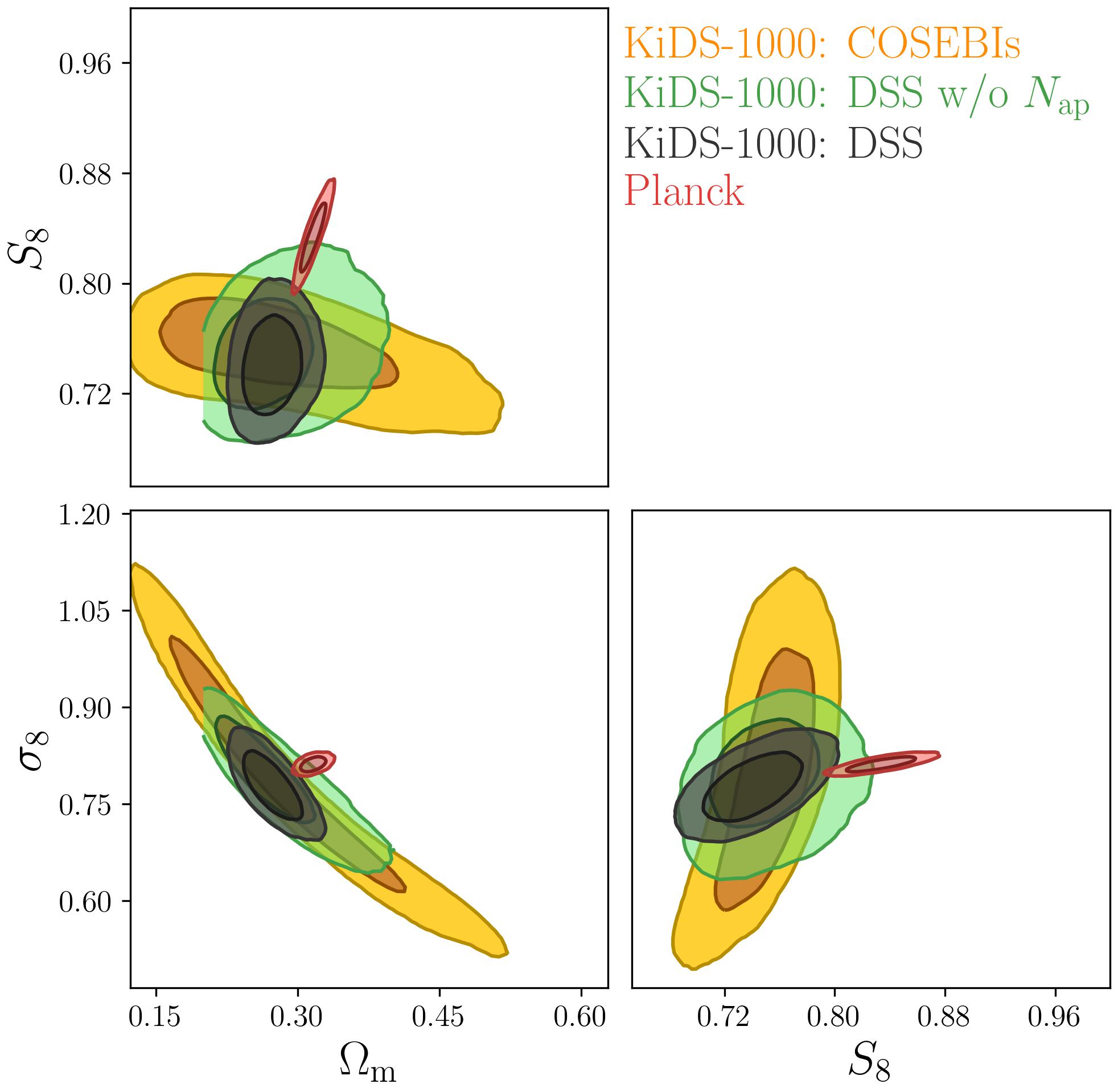}
\caption{Cosmological posteriors for the adapted filter for the best-estimated $n_\mathrm{be}(z)$ in black using the full data vector and in green using only shear information compared to the COSEBIs posteriors in orange presented in \citet{Asgari2021} and to the (TT, TE, EE+lowE) results of \citet{Aghanim:2020} in red. The sharp cut of the green posterior is due to the conservative prior of $\Omega_\mathrm{m}>0.2$, as the model does not agree with the cosmo-SLICS for smaller $\Omega_\mathrm{m}$ values. The shear-only posterior is shown in this figure to support the assumption of using a linear galaxy bias model.}
\label{fig:MCMC_vs_others}
\end{figure}

Using the approach described in Sect.~\ref{sec:parameter_estimation} to estimate the uncertainty around the MAP, we find 
\begin{eqnarray}
S_8^{\mathrm{DSS}}= 0.731^{+0.030}_{-0.018} \,,
\end{eqnarray} 
which is consistent with and competitive to the KiDS-1000 cosmic shear constraints from \citet{Asgari2021},
\begin{eqnarray}
S_8^{\mathrm{COSEBIs}}=0.759^{+0.024}_{-0.021} \, .
\end{eqnarray}
We present the posterior of these two analyses in Fig.~\ref{fig:MCMC_vs_others}, where the consistency between the two probes is obvious. The DSS has a slightly lower constraining power on $S_8$; however, the $\Omega_{\rm m}$-$\sigma_8$ degeneracy is broken, thanks to the additional information provided by the foreground data. But even for the shear-only case shown in green in Fig.~\ref{fig:MCMC_vs_others}, the DSS has better constraining power for the $\Omega_{\rm m}$ and $\sigma_8$ parameters, although the lower bound should be taken with caution as we excluded all $\Omega_{\rm m}<0.2$, as the model does not agree with the cosmo-SLICS for smaller $\Omega_\mathrm{m}$ values. Furthermore, although the results of Fig. \ref{fig:MCMC_bias} show that the inferred $S_8$ might be smaller compared to the truth if a linear galaxy bias model is not sufficient, the consistency between the shear-only DSS analysis to the complete DSS analysis supports the robustness of our inferred parameters with respect to the galaxy bias model (see the discussion at the end of Sect.~\ref{subsec:DSS_model}). The comparison to the COSEBIs analysis reveals competitive S/N for the $S_8$ parameter while using only a fraction of the lensing sources (tomographic bins four and five). Caution should be taken when comparing their respective constraining power, as the COSEBIs analysis marginalises over more cosmological parameters and samples the parameter space differently. Nevertheless, it seems that a joint COSEBIs-DSS analysis could further improve the constraints, which we leave for future work. Lastly, as the DSS estimates of $S_8$ are slightly lower but with higher uncertainty, we measure a similar tension to the CMB results as the COSEBIs analysis.

In the next section, we investigate the robustness of our results with respect to the lens redshift distribution $n(z)$, of varying the covariance matrix. We further present our galaxy colour-split analysis and additionally discuss the galaxy bias $b$ and $\alpha$ results.

\subsection{Impact of lens redshift distribution}
\label{sec:n_of_z_posterior}

\begin{figure}
\includegraphics[width=\columnwidth]{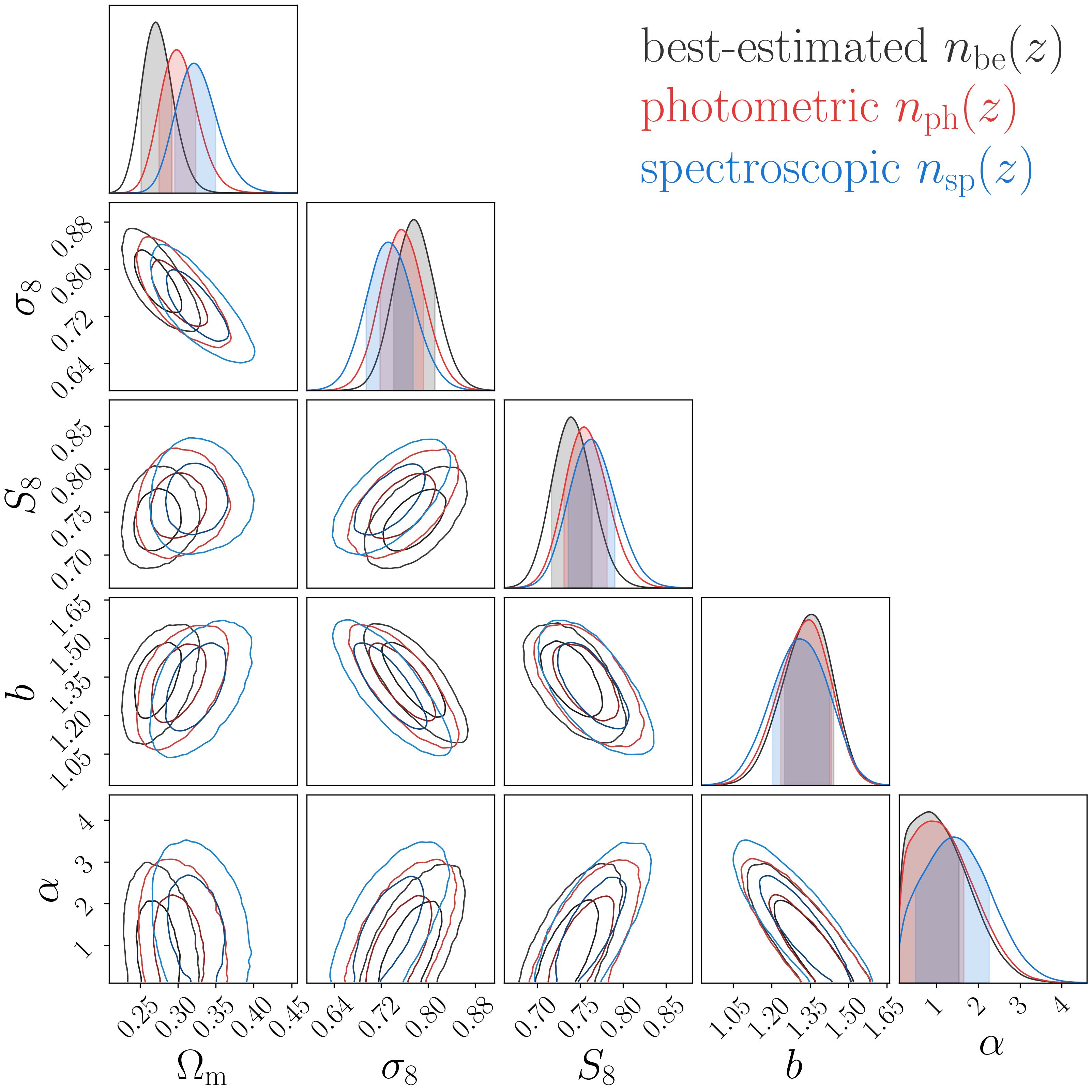}
\caption{Posteriors of the full DSS vector resulting from using the different lens $n(z)$ shown in Fig.~\ref{fig:nofz}. The $n(z_\mathrm{phot})$ and $n(z)$ have, by construction, the same mean redshift, while the mean redshift of the spectroscopic redshift estimate is $\sim 0.015$ lower. The main DSS results include a marginalisation over our uncertainty in the mean redshifts of both the lens and source samples, which partially compensates for some of these differences. The posterior obtained using the GAMA spectroscopic $n(z)$, shown in blue, should be taken with caution as it is estimated from a smaller sky coverage and, as such, contains larger statistical fluctuations.}
\label{fig:MCMC_redshift}
\end{figure}

To estimate the impact of the shape of the lens galaxy redshift distribution (on top of shifting the mean by $\delta \langle z \rangle$ in the sampling), we repeat the analysis for the three $n(z)$ shown in Fig.~\ref{fig:nofz}. These are the smoothed version of the photometric redshift distribution $n_\mathrm{be}(z)$, the photometric redshift distribution $n_\mathrm{ph}(z)$ itself without any smoothing, and the spectroscopic redshift $n_\mathrm{sp}(z)$ from those GAMA galaxies that are also in the KiDS-bright sample. Although Bi21 showed that GAMA is representative and that mismatches should be rare, the results from the GAMA spectroscopic $n_\mathrm{sp}(z)$ should be taken with caution because the equatorial fields have a relatively small sky coverage, leading to features in the $n_\mathrm{sp}(z)$ that are caused by the large-scale structure present in these fields. For this investigation, we use the same setup as for the fiducial analysis, varying the two cosmological parameters $\Omega_\mathrm{m}$ and $\sigma_8$ together with the $\alpha$ and the linear galaxy bias $b$ parameter. We also marginalised over the nuisance parameters shown bottom half of Table \ref{table:parameter_overview}. In Fig.~\ref{fig:MCMC_redshift} we display the different posteriors following from the three alternatives $n(z)$. It is clearly seen that the posteriors are shifted along the $\Omega_\mathrm{m}$-$\sigma_8$ degeneracy axis, whereas these shifts partially cancel out for $S_8$. 
Due to the different lens $n(z)$, the amplitude and the slope of the shear signal predictions are slightly different. In particular, higher amplitude and steeper slope of the shear profiles result in larger $\sigma_8$ and smaller $\Omega_\mathrm{m}$, and vice-versa. Furthermore, we notice that the linear galaxy bias $b$ and the noise $\alpha$ are stable against changes in the $n(z)$.

Lastly, in order to investigate the impact of our photometric redshift cut at $z_\mathrm{ph}=0.1$ (see Sect. \ref{Sect:Obs_Data:lenses}), we perform two additional analyses, this time modifying the photometric redshift threshold to $z_\mathrm{ph}>0.15$ and  $z_\mathrm{ph}>0.2$. We find that the posteriors shifts are smaller than the $68\%$ credibility region, indicating that removing low-redshift galaxies does not result in systematically different $\Omega_\mathrm{m}$ or $\sigma_8$.

\begin{table}[h!]
\centering
\caption{Marginalised MAP values and their $68\%$ confidence intervals for the different lens $n(z)$ of the full sample.}
\begin{tabular}{c|ccc}
\hline
\hline
 & $n_\mathrm{be}(z)$ & $n_\mathrm{ph}(z)$ & $n_\mathrm{sp}(z)$  \\
\hline
$\Omega_\nt{m}$ & $0.27^{+0.02}_{-0.02}$ & $0.30^{+0.03}_{-0.02}$ & $0.32^{+0.03}_{-0.02}$ \\
$\sigma_8$ &  $0.77^{+0.04}_{-0.03}$ & $0.75^{+0.04}_{-0.03}$ & $0.74^{+0.04}_{-0.04}$ \\
$S_8$ & $0.73^{+0.03}_{-0.02}$ & $0.75^{+0.03}_{-0.02}$ & $0.76^{+0.03}_{-0.02}$ \\
$b$ &  $1.37^{+0.10}_{-0.10}$ & $1.36^{+0.10}_{-0.11}$ & $1.32^{+0.13}_{-0.09}$  \\
$\alpha$ & $0.75^{+0.95}_{-0.44}$ & $0.83^{+0.79}_{-0.66}$ & $1.39^{+0.71}_{-0.95}$ \\
$\chi^2/\mathrm{d.o.f.}$ & $0.81$ & $0.81$ & $0.83$ \\
$p$-value & $0.90$ & $0.90$ & $0.87$ \\
\hline
\hline
\end{tabular}
\tablefoot{The $68\%$ confidence intervals result from the MCMC chains shown in Fig.~\ref{fig:MCMC_redshift}. Here $\Omega_\mathrm{m}$, $\sigma_8$, $\alpha$, and the linear galaxy bias parameter are varied. We fixed $h=0.6898$, $w_0=-1$ and $n_{\rm s}= 0.969$ but marginalised over the $\delta \langle z \rangle$ and $m$-bias uncertainties.}
\label{table:MAP_values}
\end{table}

\subsection{Cosmology scaling of the covariance matrix}
\label{sec:scaling_covariance}
The covariance matrix used in the main analysis is determined at a specific point in the parameter space (see Sect.~\ref{sec:FLASK}), which is not identical to the MAP. However, assuming the MAP values are the true parameters, the most robust likelihood analysis would be achieved with a data covariance matrix estimated at the MAP values. This section, therefore, explores the impact of considering a cosmology-dependent covariance matrix. 

In the related literature, there are two common approaches on whether the cosmology should be kept fixed in the covariance matrix \citep{vanUitert2018} or varied at each point sampled by the MCMC, as in \citet{Eifler2009}. It is argued in \citet{Carron2013} that the latter could result in over-constraints, whereas \citet{Kodwani2019....2E...3K} argues that the effect is small. We, therefore, explore both methods here.

We achieve our cosmology rescaling by assuming that the covariance matrix scales quadratically with the signal. This is only strictly true in the Gaussian regime; nevertheless, it is a good first approximation, even though the impact on the non-linear mode coupling is neglected in this approach.  

To achieve the rescaling, we compute at a new cosmology $\boldsymbol{\Theta}$ the ratio between the predicted model $\boldsymbol{m}(\boldsymbol{\Theta})$ and the model predicted at the \texttt{FLASK} cosmology $\boldsymbol{m}\left(\boldsymbol{\Theta}^\mathrm{F}\right)$,
\begin{equation}
    r_i(\boldsymbol{\Theta})=\frac{m_i(\boldsymbol{\Theta})}{m_i\left(\boldsymbol{\Theta}^\mathrm{F}\right)} \, .
\end{equation}
We then  multiply each element of the fiducial covariance matrix, $C^\mathrm{F}_{ij}$, by the scaling factors,
\begin{equation}
   C_{ij}(\boldsymbol{\Theta}) = C^\mathrm{F}_{ij}\, r_{i}(\boldsymbol{\Theta})\,r_{j}(\boldsymbol{\Theta}) \, ,
\end{equation}
and obtain a cosmology-rescaled covariance matrix.

As explained in \cite{Eifler2009}, this method is only valid for a noise-free covariance matrix since it wrongly rescales the shape-noise component and possibly over-estimates the cosmology changes. Finally, we note that in Eq.~\eqref{eq:t_distribution} the determinant of the covariance matrix changes with cosmology as well and needs to be recalculated.

\begin{figure}
\includegraphics[width=\columnwidth]{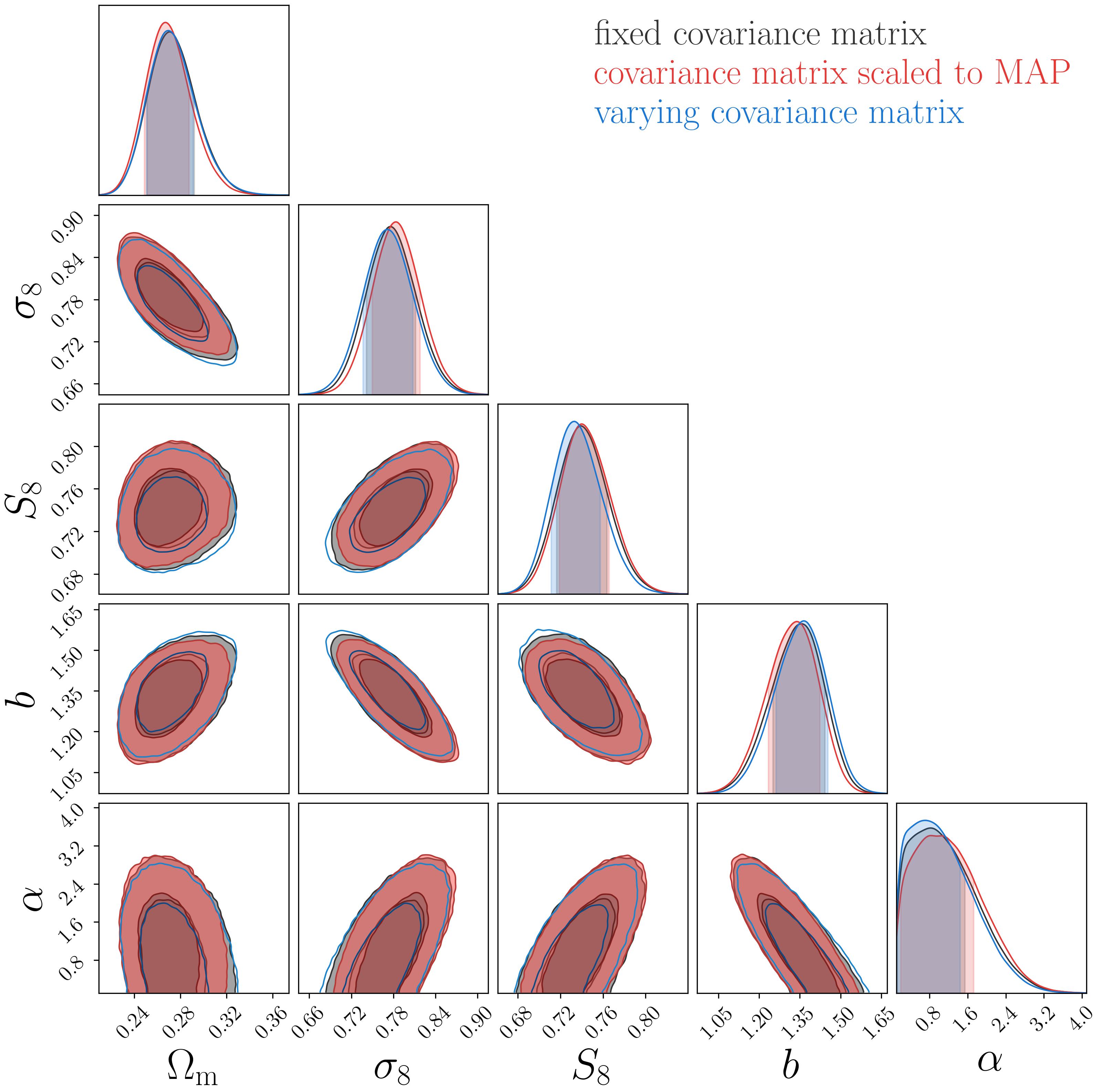}
\caption{Comparison between the posteriors obtained from our fiducial setup (with the covariance matrix calculated at the initial cosmology, see the black contours), with those obtained after scaling the covariance at the best-fit parameters (red) and to those obtained by varying the covariance with the parameters (blue).}
\label{fig:MCMC_covcos_adapted}
\end{figure}

Using our fiducial setup, we determine the posterior distribution in two distinct ways: first, by varying the covariance matrix alongside the model vector at each step of the MCMC, and second, by scaling the covariance matrix to the MAP value. For the latter approach, we use an iterative process, where we first estimate the MAP with the fiducial covariance matrix, then use the MAP parameters to scale the covariance matrix and find new MAP parameters; we repeated that process 100 times. As seen in Fig.~\ref{fig:cos_iterative} after approximately 20 iterations the MAP values converged. The results are shown in Fig.~\ref{fig:MCMC_covcos_adapted}, where the red posteriors used a  covariance rescaled to the converged MAP value; the blue posteriors are for a full parameter-dependent covariance matrix varied in the MCMC; the black posteriors show the fiducial covariance. The red and black posteriors are almost identical, which is not surprising given the fact that the MAP values are very close to the parameters used to determine the covariance matrix. However, the blue contours slightly differ from the other two but are still within half $1\,\sigma$; we are therefore not concerned about the impact on our constraints of this analysis choice. In Stage IV surveys, this is even less important due to the tighter posterior and the, therefore, smaller cosmology variation.

\subsection{Red and blue split}
\label{sec:red_blue}

\begin{figure}
\includegraphics[width=\columnwidth]{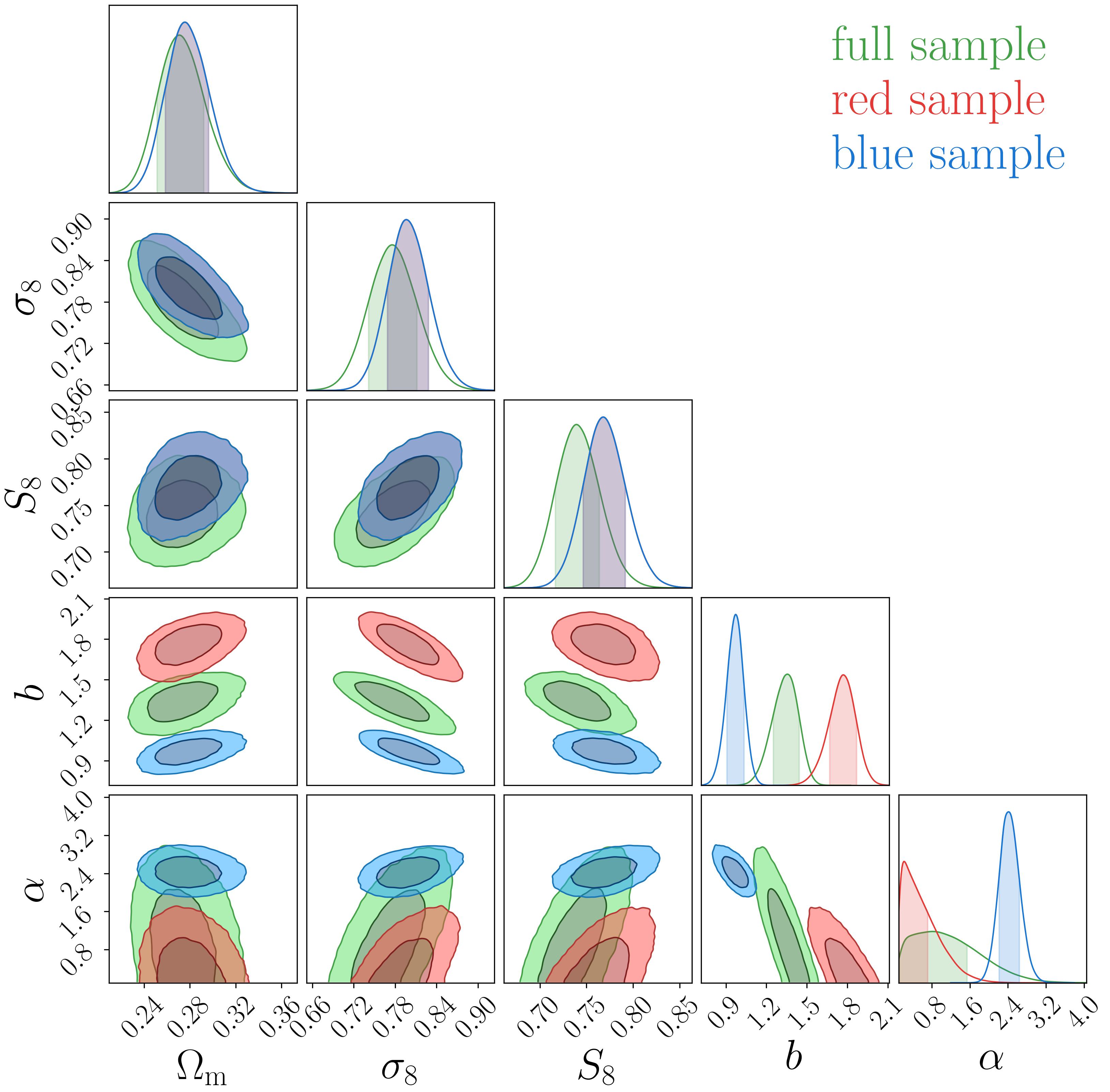}
\caption{Posterior for the full KiDS-bright sample shown in green, while the joint red+blue posteriors represent the results of the colour-selected samples. In the latter case, the red and blue samples share the same cosmology (the dark blue contours) by construction. The resulting measured and best-fit predicted shear profiles for the red and blue samples are displayed in Fig.\,\ref{fig:shear_bright_redblue} and the corresponding mean aperture number values are seen in Fig.\,\ref{fig:mean_Nap_redblue}.}
\label{fig:MCMC_bright_redblue_adapted}
\end{figure}

In this section, we present our final investigation, dividing the KiDS-bright sample into red and blue galaxies according to their colour as described in Sect.~\ref{Sect:Obs_Data:lenses}, and carry out a joint analysis. The motivation for this is to learn more about the behaviour of different galaxy types and as a cosmological robustness check. As for the main analysis, we use the best-estimated $n_\mathrm{be}(z)$, resulting from smoothing the photometric redshifts after applying a $z_\mathrm{ph}>0.1$ cut. In this setup, our data vector has 168 elements. But given the likelihood modelling described in Sect.~\ref{sec:inference_description}, we are still confident in our results with respect to the remaining noise in the covariance matrix. As shown in Table \ref{table:MAP_redblue}, the reduced $\chi^2$/d.o.f. = 1.00  indicates a valid fit. The resulting posteriors are shown in Fig.~\ref{fig:MCMC_bright_redblue_adapted} and the MAP values are stated in Table \ref{table:MAP_redblue}. 

First, we notice that the cosmological parameters of the full KiDS-bright sample analysis and the joint red and blue analysis are consistent and within $0.48\,\sigma$ in $\sigma_8$ and almost identical in $\Omega_\mathrm{m}$. Of particular interest in this investigation are the results obtained for the two astrophysical parameters, where we see that, as expected, the blue (red) sample prefers a smaller (larger) linear galaxy bias $b$ compared to the full sample. This is in line with the fact that red galaxies are known to be more strongly clustered than blue galaxies and therefore have a larger galaxy bias \citep{Mo2010}. Furthermore, we find that the $\alpha$ parameter for the blue sample is significantly larger than unity, whereas the red sample tends to be below one. This shows that blue galaxies follow a super-Poisson distribution and red galaxies a sub-Poisson distribution. Following the results of \citet{Friedrich2022} who found that a higher satellite fraction leads to a higher $\alpha$ value, the blue sample has more satellites. The full sample overlaps with the blue and red posteriors and is consistent with a normal-Poisson distribution ($\alpha=1.0$).  

\begin{table}[h!]
\centering
\caption{Marginalised MAP values and their $68\%$ confidence intervals for the different lens samples.}
\begin{tabular}{c|ccc}
\hline
\hline
& full & \multicolumn{2}{c}{red + blue }  \\
\hline
$\Omega_\mathrm{m}$ & $0.27^{+0.02}_{-0.02}$ & \multicolumn{2}{c}{$0.27^{+0.02}_{-0.02}$}\\
$\sigma_8$ & $0.77^{+0.04}_{-0.03}$ & \multicolumn{2}{c}{$0.79^{+0.04}_{-0.02}$}\\
$S_8$ & $0.73^{+0.03}_{-0.02}$ & \multicolumn{2}{c}{$0.75^{+0.03}_{-0.02}$}\\
$b$ & $1.37^{+0.10}_{-0.10}$ & $1.83^{+0.08}_{-0.14}$  & $1.02^{+0.04}_{-0.10}$ \\
$\alpha$ & $0.75^{+0.95}_{-0.44}$ & $0.10^{+0.62}$  & $2.25^{+0.33}_{-0.12}$ \\
$\chi^2/\mathrm{d.o.f.}$ & $0.81$ & \multicolumn{2}{c}{$1.01$} \\
$p$-value & $0.89$ & \multicolumn{2}{c}{$0.44$} \\
\hline
\hline
\end{tabular}
\tablefoot{The $68\%$ confidence intervals result from the MCMC chains shown in the right panel Fig.~\ref{fig:MCMC_bright_redblue_adapted}. The first column is the same as in Table \ref{table:MAP_values}. We fixed $h=0.6898$, $w_0=-1$ and $n_{\rm s}=0.969$ but marginalised over the $\delta \langle z \rangle$ and $m$-bias uncertainties. If limits are not given they are dominated by priors.}
\label{table:MAP_redblue}
\end{table}

\section{Summary and conclusions}
\label{sec:Conclusion}
In this work, we present an unblinded density split statistic analysis of the fourth data release of the Kilo-Degree survey (KiDS-1000). The analytical model used to infer cosmological and astrophysical parameters was first developed in \cite{Friedrich:Gruen:2018} and then modified in \cite{Burger2022}, and we further validated it on realistic simulated data. The lenses used to construct the foreground density map are taken from the KiDS-bright sample described in Bi21, while for our sources, we used the fourth and fifth tomographic redshift bins of the KiDS-1000 data described in H21.

We investigated for the first time the impact of baryons and IA on the DSS. While the effect of the former is suppressed due to the implied smoothing of the density map, IA can have an important role if the redshift distributions of the lenses and sources overlap. We carried out a full analysis on contaminated mock data without overlapping redshift distributions and found that for our selected data, we are immune to both of the systematic effects at the level of the inferred posteriors.

We explored the uncertainty of the redshift distribution of the lenses by investigating the impact on the posterior of changing the mean and the shape of the $n_\mathrm{ph}(z)$. In particular, for this, we used the photometric redshift distribution $n_\mathrm{ph}(z)$ with and without smoothing, as well as the distribution obtained directly from the overlapping spectroscopic GAMA galaxies. We found that the posteriors vary by less than $\sim 0.5\,\sigma$. Notably, we observed that of all parameters, $\Omega_\mathrm{m}$ is the most affected and is generally lower for broader $n(z)$; in contrast, $\sigma_8$ increased, leaving $S_8$ values changed by $\sim 0.5\,\sigma$. We assigned an extra error term to this uncertainty, resulting in $S_8=0.73^{+0.03}_{-0.02} \pm 0.01$ for the $n(z)$ shape after marginalising over the other systematic effects. These constraints are competitive and consistent with the KiDS-1000 cosmic shear constraints from \citet{Asgari2021}.

Furthermore, we investigated the impact of varying the covariance matrix with cosmological parameters, where we used an iterative process once to scale the covariance matrix to the MAP best-fit parameters, and we varied them alongside the parameters in the MCMC process once as well; for all cases, we recorded no significant deviations.

As our final result, we divided the full KiDS-bright sample into red and blue galaxies as a cosmology robustness check and to learn more about different galaxy types. For this, we performed a joint analysis of the red and blue samples with a joint covariance matrix with the smoothed version of $n_\mathrm{ph}(z)$ as our best redshift estimate. The resulting posteriors of the full and joint red+blue analyses agree as to the cosmological parameters within $0.35\,\sigma$. Furthermore, this shows the expected behaviour of the linear galaxy bias, where blue (red) galaxies have a lower (higher) bias than the full sample. The $\alpha$ parameter, which accounts for super-Poisson or sub-Poisson shot noise, also revealed interesting results. Whereas red galaxies have an $\alpha$ value that tends to be smaller than one ($\approx 2 \, \sigma$), blue galaxies have an $\alpha$ value significantly larger than one ($\approx 6 \, \sigma$), meaning that blue galaxies are super-Poisson distributed and red galaxies are sub-Poisson distributed. According to \citet{Friedrich2020}, this reveals information about the halo occupation distribution, with samples with a larger fraction of satellite galaxies tending to have larger $\alpha$ values.

We conclude from our results that the density split statistic is a valuable tool with a major advantage in the $\Omega_\mathrm{m}$-$\sigma_8$ degeneracy breaking. In addition to this, it also yields a new way to measure the galaxy bias on linear scales and the Poissanity of different galaxy types. We save the inference of the dark-energy-equation of state $w$ for future analysis; this can be achieved by a tomographic analysis of high-precision lensing and clustering data. Other aspects for future analysis include the modelling of a more complex galaxy bias model, and lastly, the impact of the filter size, where we expect smaller filter sizes to be more constraining yet also more sensitive to non-linear scales and baryonic analysis. 

\begin{acknowledgement}
We thank the anonymous referee for the very constructive and fruitful comments.
This paper went through the KiDS review process, where we especially want to thank the KiDS-internal referee Benjamin Joachimi for his fruitful comments to improve this work. Further, we would like to thank Mike Jarvis for maintaining \texttt{treecorr}, and Alessio Mancini to develop the \texttt{CosmoPower} emulator, which improved our work significantly. Lastly, we thank Sven Heydenreich, Laila Linke, Patrick Simon and Jan Luca van den Busch for very valuable discussions. PAB acknowledges support by the German Academic Scholarship Foundation. JHD acknowledges support from an STFC Ernest Rutherford Fellowship (project reference ST/S004858/1). OF gratefully acknowledges support by the Kavli Foundation and the International Newton Trust through a Newton-Kavli-Junior Fellowship and by Churchill College Cambridge through a postdoctoral By-Fellowship. MB is supported by the Polish National Science Center through grants no. 2020/38/E/ST9/00395, 2018/30/E/ST9/00698, 2018/31/G/ST9/03388 and 2020/39/B/ST9/03494, and by the Polish Ministry of Science and Higher Education through grant DIR/WK/2018/12. HH is supported by a Heisenberg grant of the Deutsche Forschungsgemeinschaft (Hi 1495/5-1) as well as an ERC Consolidator Grant (No. 770935). AHW is supported by a European Research Council Consolidator Grant (No. 770935). TC is supported by the INFN INDARK PD51 grant and the FARE MIUR grant ‘ClustersXEuclid’ R165SBKTMA. KD acknowledges support by the COMPLEX project from the European Research Council (ERC) under the European Union’s Horizon 2020 research and innovation program grant agreement ERC-2019-AdG 882679 and by the Deutsche Forschungsgemeinschaft (DFG, German Research Foundation) under Germany’s Excellence
Strategy - EXC-2094 - 390783311. The calculations for the Magneticum simulations were carried out at the Leibniz Supercomputer Center (LRZ) under the project pr83li and with the support by M. Petkova through the Computational Center for Particle and Astrophysics (C2PAP). CH acknowledges support from the European Research Council under grant number 647112, and support from the Max Planck Society and the Alexander von Humboldt Foundation in the framework of the Max Planck-Humboldt Research Award endowed by the Federal Ministry of Education and Research. BJ acknowledges support by STFC Consolidated Grant ST/V000780/1. KK acknowledges support from the Royal Society, and Imperial College NM acknowledges support from the Centre National d’Etudes Spatiales (CNES) fellowship. HYS acknowledges the support from CMS-CSST-2021-A01 and CMS-CSST-2021-B01, NSFC of China under grant 11973070, the Shanghai Committee of Science and Technology grant No.19ZR1466600 and Key Research Program of Frontier Sciences, CAS, Grant No. ZDBS-LY-7013. TT acknowledges support from the Leverhulme Trust.\\

Author contributions: all authors contributed to the development and writing of this paper. The authorship list is given in three groups: the lead authors (PAB, OF, JHD, PS), where we ordered (OF, JHD, PS) alphabetically. PAB led the paper, JHD provided all necessary simulations, OF and PS contributed to the modelling, and all four helped in the development of the analysis. The first author group is followed by two further alphabetical groups. The first alphabetical group includes those who are key contributors to both the scientific analysis and the data products. The second group covers those who have either made a significant contribution to the data products or to the scientific analysis.
\end{acknowledgement}

\bibliographystyle{aa}
\bibliography{cite}

\begin{thebibliography}{87}
\expandafter\ifx\csname natexlab\endcsname\relax\def\natexlab#1{#1}\fi

\bibitem[{{Abbott} {et~al.}(2018){Abbott}, {Abdalla}, {Alarcon}, {Aleksi{\'c}},
  {Allam}, {Allen}, {Amara}, {Annis}, {Asorey}, {Avila}, {Bacon}, {Balbinot},
  {Banerji}, {Banik}, {Barkhouse}, {Baumer}, {Baxter}, {Bechtol}, {Becker},
  {Benoit-L{\'e}vy}, {Benson}, {Bernstein}, {Bertin}, {Blazek}, {Bridle},
  {Brooks}, {Brout}, {Buckley-Geer}, {Burke}, {Busha}, {Campos}, {Capozzi},
  {Carnero Rosell}, {Carrasco Kind}, {Carretero}, {Castander}, {Cawthon},
  {Chang}, {Chen}, {Childress}, {Choi}, {Conselice}, {Crittenden}, {Crocce},
  {Cunha}, {D'Andrea}, {da Costa}, {Das}, {Davis}, {Davis}, {De Vicente},
  {DePoy}, {DeRose}, {Desai}, {Diehl}, {Dietrich}, {Dodelson}, {Doel},
  {Drlica-Wagner}, {Eifler}, {Elliott}, {Elsner}, {Elvin-Poole}, {Estrada},
  {Evrard}, {Fang}, {Fernandez}, {Fert{\'e}}, {Finley}, {Flaugher}, {Fosalba},
  {Friedrich}, {Frieman}, {Garc{\'\i}a-Bellido}, {Garcia-Fernandez}, {Gatti},
  {Gaztanaga}, {Gerdes}, {Giannantonio}, {Gill}, {Glazebrook}, {Goldstein},
  {Gruen}, {Gruendl}, {Gschwend}, {Gutierrez}, {Hamilton}, {Hartley}, {Hinton},
  {Honscheid}, {Hoyle}, {Huterer}, {Jain}, {James}, {Jarvis}, {Jeltema},
  {Johnson}, {Johnson}, {Kacprzak}, {Kent}, {Kim}, {King}, {Kirk}, {Kokron},
  {Kovacs}, {Krause}, {Krawiec}, {Kremin}, {Kuehn}, {Kuhlmann}, {Kuropatkin},
  {Lacasa}, {Lahav}, {Li}, {Liddle}, {Lidman}, {Lima}, {Lin}, {MacCrann},
  {Maia}, {Makler}, {Manera}, {March}, {Marshall}, {Martini}, {McMahon},
  {Melchior}, {Menanteau}, {Miquel}, {Miranda}, {Mudd}, {Muir}, {M{\"o}ller},
  {Neilsen}, {Nichol}, {Nord}, {Nugent}, {Ogando}, {Palmese}, {Peacock},
  {Peiris}, {Peoples}, {Percival}, {Petravick}, {Plazas}, {Porredon}, {Prat},
  {Pujol}, {Rau}, {Refregier}, {Ricker}, {Roe}, {Rollins}, {Romer}, {Roodman},
  {Rosenfeld}, {Ross}, {Rozo}, {Rykoff}, {Sako}, {Salvador}, {Samuroff},
  {S{\'a}nchez}, {Sanchez}, {Santiago}, {Scarpine}, {Schindler}, {Scolnic},
  {Secco}, {Serrano}, {Sevilla-Noarbe}, {Sheldon}, {Smith}, {Smith}, {Smith},
  {Soares-Santos}, {Sobreira}, {Suchyta}, {Tarle}, {Thomas}, {Troxel},
  {Tucker}, {Tucker}, {Uddin}, {Varga}, {Vielzeuf}, {Vikram}, {Vivas},
  {Walker}, {Wang}, {Wechsler}, {Weller}, {Wester}, {Wolf}, {Yanny}, {Yuan},
  {Zenteno}, {Zhang}, {Zhang}, {Zuntz}, \& {Dark Energy Survey
  Collaboration}}]{DES:2018}
{Abbott}, T.~M.~C., {Abdalla}, F.~B., {Alarcon}, A., {et~al.} 2018, \prd, 98,
  043526

\bibitem[{{Amon} {et~al.}(2022){Amon}, {Gruen}, {Troxel}, {MacCrann},
  {Dodelson}, {Choi}, {Doux}, {Secco}, {Samuroff}, {Krause}, {Cordero},
  {Myles}, {DeRose}, {Wechsler}, {Gatti}, {Navarro-Alsina}, {Bernstein},
  {Jain}, {Blazek}, {Alarcon}, {Fert{\'e}}, {Lemos}, {Raveri}, {Campos},
  {Prat}, {S{\'a}nchez}, {Jarvis}, {Alves}, {Andrade-Oliveira}, {Baxter},
  {Bechtol}, {Becker}, {Bridle}, {Camacho}, {Carnero Rosell}, {Carrasco Kind},
  {Cawthon}, {Chang}, {Chen}, {Chintalapati}, {Crocce}, {Davis}, {Diehl},
  {Drlica-Wagner}, {Eckert}, {Eifler}, {Elvin-Poole}, {Everett}, {Fang},
  {Fosalba}, {Friedrich}, {Gaztanaga}, {Giannini}, {Gruendl}, {Harrison},
  {Hartley}, {Herner}, {Huang}, {Huff}, {Huterer}, {Kuropatkin}, {Leget},
  {Liddle}, {McCullough}, {Muir}, {Pandey}, {Park}, {Porredon}, {Refregier},
  {Rollins}, {Roodman}, {Rosenfeld}, {Ross}, {Rykoff}, {Sanchez},
  {Sevilla-Noarbe}, {Sheldon}, {Shin}, {Troja}, {Tutusaus}, {Tutusaus},
  {Varga}, {Weaverdyck}, {Yanny}, {Yin}, {Zhang}, {Zuntz}, {Aguena}, {Allam},
  {Annis}, {Bacon}, {Bertin}, {Bhargava}, {Brooks}, {Buckley-Geer}, {Burke},
  {Carretero}, {Costanzi}, {da Costa}, {Pereira}, {De Vicente}, {Desai},
  {Dietrich}, {Doel}, {Ferrero}, {Flaugher}, {Frieman}, {Garc{\'\i}a-Bellido},
  {Gaztanaga}, {Gerdes}, {Giannantonio}, {Gschwend}, {Gutierrez}, {Hinton},
  {Hollowood}, {Honscheid}, {Hoyle}, {James}, {Kron}, {Kuehn}, {Lahav}, {Lima},
  {Lin}, {Maia}, {Marshall}, {Martini}, {Melchior}, {Menanteau}, {Miquel},
  {Mohr}, {Morgan}, {Ogando}, {Palmese}, {Paz-Chinch{\'o}n}, {Petravick},
  {Pieres}, {Romer}, {Sanchez}, {Scarpine}, {Schubnell}, {Serrano}, {Smith},
  {Soares-Santos}, {Tarle}, {Thomas}, {To}, {Weller}, \& {DES
  Collaboration}}]{Amon2022}
{Amon}, A., {Gruen}, D., {Troxel}, M.~A., {et~al.} 2022, \prd, 105, 023514

\bibitem[{{Arnouts} {et~al.}(1999){Arnouts}, {Cristiani}, {Moscardini},
  {Matarrese}, {Lucchin}, {Fontana}, \& {Giallongo}}]{Arnouts1999}
{Arnouts}, S., {Cristiani}, S., {Moscardini}, L., {et~al.} 1999, \mnras, 310,
  540

\bibitem[{{Asgari} {et~al.}(2021){Asgari}, {Lin}, {Joachimi}, {Giblin},
  {Heymans}, {Hildebrandt}, {Kannawadi}, {St{\"o}lzner}, {Tr{\"o}ster}, {van
  den Busch}, {Wright}, {Bilicki}, {Blake}, {de Jong}, {Dvornik}, {Erben},
  {Getman}, {Hoekstra}, {K{\"o}hlinger}, {Kuijken}, {Miller}, {Radovich},
  {Schneider}, {Shan}, \& {Valentijn}}]{Asgari2021}
{Asgari}, M., {Lin}, C.-A., {Joachimi}, B., {et~al.} 2021, \aap, 645, A104

\bibitem[{{Asgari} {et~al.}(2020){Asgari}, {Tr{\"o}ster}, {Heymans},
  {Hildebrandt}, {van den Busch}, {Wright}, {Choi}, {Erben}, {Joachimi},
  {Joudaki}, {Kannawadi}, {Kuijken}, {Lin}, {Schneider}, \&
  {Zuntz}}]{Asgari2020}
{Asgari}, M., {Tr{\"o}ster}, T., {Heymans}, C., {et~al.} 2020, \aap, 634, A127

\bibitem[{{Barthelemy} {et~al.}(2021){Barthelemy}, {Codis}, \&
  {Bernardeau}}]{Barthelemy2020}
{Barthelemy}, A., {Codis}, S., \& {Bernardeau}, F. 2021, \mnras, 503, 5204

\bibitem[{{Berg{\'e}} {et~al.}(2010){Berg{\'e}}, {Amara}, \&
  {R{\'e}fr{\'e}gier}}]{Berge2010}
{Berg{\'e}}, J., {Amara}, A., \& {R{\'e}fr{\'e}gier}, A. 2010, \apj, 712, 992

\bibitem[{{Bilicki} {et~al.}(2021){Bilicki}, {Dvornik}, {Hoekstra}, {Wright},
  {Chisari}, {Vakili}, {Asgari}, {Giblin}, {Heymans}, {Hildebrandt},
  {Holwerda}, {Hopkins}, {Johnston}, {Kannawadi}, {Kuijken}, {Nakoneczny},
  {Shan}, {Sonnenfeld}, \& {Valentijn}}]{Bilicki2021}
{Bilicki}, M., {Dvornik}, A., {Hoekstra}, H., {et~al.} 2021, \aap, 653, A82

\bibitem[{{Bilicki} {et~al.}(2014){Bilicki}, {Jarrett}, {Peacock}, {Cluver}, \&
  {Steward}}]{Bilicki2014}
{Bilicki}, M., {Jarrett}, T.~H., {Peacock}, J.~A., {Cluver}, M.~E., \&
  {Steward}, L. 2014, \apjs, 210, 9

\bibitem[{{Bridle} \& {King}(2007)}]{Bridle2007}
{Bridle}, S. \& {King}, L. 2007, New Journal of Physics, 9, 444

\bibitem[{{Brouwer} {et~al.}(2018){Brouwer}, {Demchenko}, {Harnois-D{\'e}raps},
  {Bilicki}, {Heymans}, {Hoekstra}, {Kuijken}, {Alpaslan}, {Brough}, {Cai},
  {Costa-Duarte}, {Dvornik}, {Erben}, {Hildebrandt}, {Holwerda}, {Schneider},
  {Sif{\'o}n}, \& {van Uitert}}]{Brouwer2018}
{Brouwer}, M.~M., {Demchenko}, V., {Harnois-D{\'e}raps}, J., {et~al.} 2018,
  \mnras, 481, 5189

\bibitem[{{Brown} {et~al.}(2002){Brown}, {Taylor}, {Hambly}, \&
  {Dye}}]{Brown2002}
{Brown}, M.~L., {Taylor}, A.~N., {Hambly}, N.~C., \& {Dye}, S. 2002, \mnras,
  333, 501

\bibitem[{{Burger} {et~al.}(2022){Burger}, {Friedrich}, {Harnois-D{\'e}raps},
  \& {Schneider}}]{Burger2022}
{Burger}, P., {Friedrich}, O., {Harnois-D{\'e}raps}, J., \& {Schneider}, P.
  2022, \aap, 661, A137

\bibitem[{{Burger} {et~al.}(2020){Burger}, {Schneider}, {Demchenko},
  {Harnois-Deraps}, {Heymans}, {Hildebrandt}, \& {Unruh}}]{Burger:2020}
{Burger}, P., {Schneider}, P., {Demchenko}, V., {et~al.} 2020, \aap, 642, A161

\bibitem[{{Carron}(2013)}]{Carron2013}
{Carron}, J. 2013, \aap, 551, A88

\bibitem[{{Castro} {et~al.}(2021){Castro}, {Borgani}, {Dolag}, {Marra},
  {Quartin}, {Saro}, \& {Sefusatti}}]{Castro2021}
{Castro}, T., {Borgani}, S., {Dolag}, K., {et~al.} 2021, \mnras, 500, 2316

\bibitem[{{Chisari} {et~al.}(2015){Chisari}, {Codis}, {Laigle}, {Dubois},
  {Pichon}, {Devriendt}, {Slyz}, {Miller}, {Gavazzi}, \&
  {Benabed}}]{Chisari2015}
{Chisari}, N., {Codis}, S., {Laigle}, C., {et~al.} 2015, \mnras, 454, 2736

\bibitem[{{Chisari} {et~al.}(2019){Chisari}, {Alonso}, {Krause}, {Leonard},
  {Bull}, {Neveu}, {Villarreal}, {Singh}, {McClintock}, {Ellison}, {Du},
  {Zuntz}, {Mead}, {Joudaki}, {Lorenz}, {Tr{\"o}ster}, {Sanchez}, {Lanusse},
  {Ishak}, {Hlozek}, {Blazek}, {Campagne}, {Almoubayyed}, {Eifler}, {Kirby},
  {Kirkby}, {Plaszczynski}, {Slosar}, {Vrastil}, {Wagoner}, \& {LSST Dark
  Energy Science Collaboration}}]{pyccl}
{Chisari}, N.~E., {Alonso}, D., {Krause}, E., {et~al.} 2019, \apjs, 242, 2

\bibitem[{{de Jong} {et~al.}(2015){de Jong}, {Verdoes Kleijn}, {Boxhoorn},
  {Buddelmeijer}, {Capaccioli}, {Getman}, {Grado}, {Helmich}, {Huang},
  {Irisarri}, {Kuijken}, {La Barbera}, {McFarland}, {Napolitano}, {Radovich},
  {Sikkema}, {Valentijn}, {Begeman}, {Brescia}, {Cavuoti}, {Choi}, {Cordes},
  {Covone}, {Dall'Ora}, {Hildebrandt}, {Longo}, {Nakajima}, {Paolillo},
  {Puddu}, {Rifatto}, {Tortora}, {van Uitert}, {Buddendiek},
  {Harnois-D{\'e}raps}, {Erben}, {Eriksen}, {Heymans}, {Hoekstra}, {Joachimi},
  {Kitching}, {Klaes}, {Koopmans}, {K{\"o}hlinger}, {Roy}, {Sif{\'o}n},
  {Schneider}, {Sutherland}, {Viola}, \& {Vriend}}]{deJong:2015}
{de Jong}, J. T.~A., {Verdoes Kleijn}, G.~A., {Boxhoorn}, D.~R., {et~al.} 2015,
  \aap, 582, A62

\bibitem[{{de Jong} {et~al.}(2017){de Jong}, {Verdoes Kleijn}, {Erben},
  {Hildebrandt}, {Kuijken}, {Sikkema}, {Brescia}, {Bilicki}, {Napolitano}, \&
  {Amaro}}]{deJong:2017}
{de Jong}, J. T.~A., {Verdoes Kleijn}, G.~A., {Erben}, T., {et~al.} 2017, \aap,
  604, A134

\bibitem[{{DES Collaboration: Abbott} {et~al.}(2022){DES Collaboration:
  Abbott}, {Aguena}, {Alarcon}, {Allam}, {Alves}, {Amon}, {Andrade-Oliveira},
  {Annis}, {Avila}, {Bacon}, {Baxter}, {Bechtol}, {Becker}, {Bernstein},
  {Bhargava}, {Birrer}, {Blazek}, {Brandao-Souza}, {Bridle}, {Brooks},
  {Buckley-Geer}, {Burke}, {Camacho}, {Campos}, {Carnero Rosell}, {Carrasco
  Kind}, {Carretero}, {Castander}, {Cawthon}, {Chang}, {Chen}, {Chen}, {Choi},
  {Conselice}, {Cordero}, {Costanzi}, {Crocce}, {da Costa}, {da Silva Pereira},
  {Davis}, {Davis}, {De Vicente}, {DeRose}, {Desai}, {Di Valentino}, {Diehl},
  {Dietrich}, {Dodelson}, {Doel}, {Doux}, {Drlica-Wagner}, {Eckert}, {Eifler},
  {Elsner}, {Elvin-Poole}, {Everett}, {Evrard}, {Fang}, {Farahi}, {Fernandez},
  {Ferrero}, {Fert{\'e}}, {Fosalba}, {Friedrich}, {Frieman},
  {Garc{\'\i}a-Bellido}, {Gatti}, {Gaztanaga}, {Gerdes}, {Giannantonio},
  {Giannini}, {Gruen}, {Gruendl}, {Gschwend}, {Gutierrez}, {Harrison},
  {Hartley}, {Herner}, {Hinton}, {Hollowood}, {Honscheid}, {Hoyle}, {Huff},
  {Huterer}, {Jain}, {James}, {Jarvis}, {Jeffrey}, {Jeltema}, {Kovacs},
  {Krause}, {Kron}, {Kuehn}, {Kuropatkin}, {Lahav}, {Leget}, {Lemos}, {Liddle},
  {Lidman}, {Lima}, {Lin}, {MacCrann}, {Maia}, {Marshall}, {Martini},
  {McCullough}, {Melchior}, {Mena-Fern{\'a}ndez}, {Menanteau}, {Miquel},
  {Mohr}, {Morgan}, {Muir}, {Myles}, {Nadathur}, {Navarro-Alsina}, {Nichol},
  {Ogando}, {Omori}, {Palmese}, {Pandey}, {Park}, {Paz-Chinch{\'o}n},
  {Petravick}, {Pieres}, {Plazas Malag{\'o}n}, {Porredon}, {Prat}, {Raveri},
  {Rodriguez-Monroy}, {Rollins}, {Romer}, {Roodman}, {Rosenfeld}, {Ross},
  {Rykoff}, {Samuroff}, {S{\'a}nchez}, {Sanchez}, {Sanchez}, {Sanchez Cid},
  {Scarpine}, {Schubnell}, {Scolnic}, {Secco}, {Serrano}, {Sevilla-Noarbe},
  {Sheldon}, {Shin}, {Smith}, {Soares-Santos}, {Suchyta}, {Swanson}, {Tabbutt},
  {Tarle}, {Thomas}, {To}, {Troja}, {Troxel}, {Tucker}, {Tutusaus}, {Varga},
  {Walker}, {Weaverdyck}, {Wechsler}, {Weller}, {Yanny}, {Yin}, {Zhang},
  {Zuntz}, \& {DES Collaboration}}]{DES2021}
{DES Collaboration: Abbott}, T.~M.~C., {Aguena}, M., {Alarcon}, A., {et~al.}
  2022, \prd, 105, 023520

\bibitem[{{Di Valentino} {et~al.}(2021{\natexlab{a}}){Di Valentino},
  {Anchordoqui}, {Akarsu}, {Ali-Haimoud}, {Amendola}, {Arendse}, {Asgari},
  {Ballardini}, {Basilakos}, {Battistelli}, {Benetti}, {Birrer}, {Bouchet},
  {Bruni}, {Calabrese}, {Camarena}, {Capozziello}, {Chen}, {Chluba},
  {Chudaykin}, {Colg{\'a}in}, {Cyr-Racine}, {de Bernardis}, {de Cruz
  P{\'e}rez}, {Delabrouille}, {Dunkley}, {Escamilla-Rivera}, {Fert{\'e}},
  {Finelli}, {Freedman}, {Frusciante}, {Giusarma}, {G{\'o}mez-Valent}, {Guy},
  {Handley}, {Harrison}, {Hart}, {Heavens}, {Hildebrandt}, {Holz}, {Huterer},
  {Ivanov}, {Joudaki}, {Kamionkowski}, {Karwal}, {Knox}, {Kumar}, {Lamagna},
  {Lesgourgues}, {Lucca}, {Marra}, {Masi}, {Matarrese}, {Mazumdar},
  {Melchiorri}, {Mena}, {Mersini-Houghton}, {Miranda}, {Moreno-Pulido}, {Mota},
  {Muir}, {Mukherjee}, {Niedermann}, {Notari}, {Nunes}, {Pace},
  {Paliathanasis}, {Palmese}, {Pan}, {Paoletti}, {Pettorino}, {Piacentini},
  {Poulin}, {Raveri}, {Riess}, {Salzano}, {Saridakis}, {Sen}, {Shafieloo},
  {Shajib}, {Silk}, {Silvestri}, {Sloth}, {Smith}, {Sol{\`a} Peracaula}, {van
  de Bruck}, {Verde}, {Visinelli}, {Wandelt}, {Wang}, {Wang}, {Yadav}, \&
  {Yang}}]{diValentino:2021h}
{Di Valentino}, E., {Anchordoqui}, L.~A., {Akarsu}, {\"O}., {et~al.}
  2021{\natexlab{a}}, Astroparticle Physics, 131, 102605

\bibitem[{{Di Valentino} {et~al.}(2021{\natexlab{b}}){Di Valentino},
  {Anchordoqui}, {Akarsu}, {Ali-Haimoud}, {Amendola}, {Arendse}, {Asgari},
  {Ballardini}, {Basilakos}, {Battistelli}, {Benetti}, {Birrer}, {Bouchet},
  {Bruni}, {Calabrese}, {Camarena}, {Capozziello}, {Chen}, {Chluba},
  {Chudaykin}, {Colg{\'a}in}, {Cyr-Racine}, {de Bernardis}, {de Cruz
  P{\'e}rez}, {Delabrouille}, {Dunkley}, {Escamilla-Rivera}, {Fert{\'e}},
  {Finelli}, {Freedman}, {Frusciante}, {Giusarma}, {G{\'o}mez-Valent},
  {Handley}, {Harrison}, {Hart}, {Heavens}, {Hildebrandt}, {Holz}, {Huterer},
  {Ivanov}, {Joudaki}, {Kamionkowski}, {Karwal}, {Knox}, {Kumar}, {Lamagna},
  {Lesgourgues}, {Lucca}, {Marra}, {Masi}, {Matarrese}, {Mazumdar},
  {Melchiorri}, {Mena}, {Mersini-Houghton}, {Miranda}, {Moreno-Pulido}, {Mota},
  {Muir}, {Mukherjee}, {Niedermann}, {Notari}, {Nunes}, {Pace},
  {Paliathanasis}, {Palmese}, {Pan}, {Paoletti}, {Pettorino}, {Piacentini},
  {Poulin}, {Raveri}, {Riess}, {Salzano}, {Saridakis}, {Sen}, {Shafieloo},
  {Shajib}, {Silk}, {Silvestri}, {Sloth}, {Smith}, {Sol{\`a} Peracaula}, {van
  de Bruck}, {Verde}, {Visinelli}, {Wandelt}, {Wang}, {Wang}, {Yadav}, \&
  {Yang}}]{diValentino:2021s}
{Di Valentino}, E., {Anchordoqui}, L.~A., {Akarsu}, {\"O}., {et~al.}
  2021{\natexlab{b}}, Astroparticle Physics, 131, 102604

\bibitem[{Driver {et~al.}(2011)Driver, Hill, Kelvin, Robotham, Liske, Norberg,
  Baldry, Bamford, Hopkins, Loveday, Peacock, Andrae, Bland-Hawthorn, Brough,
  Brown, Cameron, Ching, Colless, Conselice, Croom, Cross, De~Propris, Dye,
  Drinkwater, Ellis, Graham, Grootes, Gunawardhana, Jones, van Kampen,
  Maraston, Nichol, Parkinson, Phillipps, Pimbblet, Popescu, Prescott,
  Roseboom, Sadler, Sansom, Sharp, Smith, Taylor, Thomas, Tuffs, Wijesinghe,
  Dunne, Frenk, Jarvis, Madore, Meyer, Seibert, Staveley-Smith, Sutherland, \&
  Warren}]{Driver:2011}
Driver, S.~P., Hill, D.~T., Kelvin, L.~S., {et~al.} 2011, Monthly Notices of
  the Royal Astronomical Society, 413, 971

\bibitem[{{Edge} {et~al.}(2013){Edge}, {Sutherland}, {Kuijken}, {Driver},
  {McMahon}, {Eales}, \& {Emerson}}]{Viking2013}
{Edge}, A., {Sutherland}, W., {Kuijken}, K., {et~al.} 2013, The Messenger, 154,
  32

\bibitem[{{Eifler} {et~al.}(2009){Eifler}, {Schneider}, \&
  {Hartlap}}]{Eifler2009}
{Eifler}, T., {Schneider}, P., \& {Hartlap}, J. 2009, \aap, 502, 721

\bibitem[{{Fenech Conti} {et~al.}(2017){Fenech Conti}, {Herbonnet}, {Hoekstra},
  {Merten}, {Miller}, \& {Viola}}]{Fenech2017}
{Fenech Conti}, I., {Herbonnet}, R., {Hoekstra}, H., {et~al.} 2017, \mnras,
  467, 1627

\bibitem[{{Friedrich} {et~al.}(2018){Friedrich}, {Gruen}, {DeRose}, {Kirk},
  {Krause}, {McClintock}, {Rykoff}, {Seitz}, {Wechsler}, \&
  {Bernstein}}]{Friedrich:Gruen:2018}
{Friedrich}, O., {Gruen}, D., {DeRose}, J., {et~al.} 2018, \prd, 98, 023508

\bibitem[{{Friedrich} {et~al.}(2022){Friedrich}, {Halder}, {Boyle}, {Uhlemann},
  {Britt}, {Codis}, {Gruen}, \& {Hahn}}]{Friedrich2022}
{Friedrich}, O., {Halder}, A., {Boyle}, A., {et~al.} 2022, \mnras, 510, 5069

\bibitem[{{Friedrich} {et~al.}(2020){Friedrich}, {Uhlemann},
  {Villaescusa-Navarro}, {Baldauf}, {Manera}, \& {Nishimichi}}]{Friedrich2020}
{Friedrich}, O., {Uhlemann}, C., {Villaescusa-Navarro}, F., {et~al.} 2020,
  \mnras, 498, 464

\bibitem[{{Fu} {et~al.}(2014){Fu}, {Kilbinger}, {Erben}, {Heymans},
  {Hildebrandt}, {Hoekstra}, {Kitching}, {Mellier}, {Miller}, {Semboloni},
  {Simon}, {Van Waerbeke}, {Coupon}, {Harnois-D{\'e}raps}, {Hudson}, {Kuijken},
  {Rowe}, {Schrabback}, {Vafaei}, \& {Velander}}]{Fu:2014}
{Fu}, L., {Kilbinger}, M., {Erben}, T., {et~al.} 2014, \mnras, 441, 2725

\bibitem[{{Giblin} {et~al.}(2021){Giblin}, {Heymans}, {Asgari}, {Hildebrandt},
  {Hoekstra}, {Joachimi}, {Kannawadi}, {Kuijken}, {Lin}, {Miller},
  {Tr{\"o}ster}, {van den Busch}, {Wright}, {Bilicki}, {Blake}, {de Jong},
  {Dvornik}, {Erben}, {Getman}, {Napolitano}, {Schneider}, {Shan}, \&
  {Valentijn}}]{Giblin:2020}
{Giblin}, B., {Heymans}, C., {Asgari}, M., {et~al.} 2021, \aap, 645, A105

\bibitem[{{G{\'o}rski} {et~al.}(2005){G{\'o}rski}, {Hivon}, {Banday},
  {Wandelt}, {Hansen}, {Reinecke}, \& {Bartelmann}}]{HEALPix2005}
{G{\'o}rski}, K.~M., {Hivon}, E., {Banday}, A.~J., {et~al.} 2005, \apj, 622,
  759

\bibitem[{{Gruen} {et~al.}(2016){Gruen}, {Friedrich}, {Amara}, {Bacon},
  {Bonnett}, {Hartley}, {Jain}, {Jarvis}, {Kacprzak}, {Krause}, {Mana}, {Rozo},
  {Rykoff}, {Seitz}, {Sheldon}, {Troxel}, {Vikram}, {Abbott}, {Abdalla},
  {Allam}, {Armstrong}, {Banerji}, {Bauer}, {Becker}, {Benoit-L{\'e}vy},
  {Bernstein}, {Bernstein}, {Bertin}, {Bridle}, {Brooks}, {Buckley-Geer},
  {Burke}, {Capozzi}, {Carnero Rosell}, {Carrasco Kind}, {Carretero}, {Crocce},
  {Cunha}, {D'Andrea}, {da Costa}, {DePoy}, {Desai}, {Diehl}, {Dietrich},
  {Doel}, {Eifler}, {Neto}, {Fernandez}, {Flaugher}, {Fosalba}, {Frieman},
  {Gerdes}, {Gruendl}, {Gutierrez}, {Honscheid}, {James}, {Kuehn},
  {Kuropatkin}, {Lahav}, {Li}, {Lima}, {Maia}, {March}, {Martini}, {Melchior},
  {Miller}, {Miquel}, {Mohr}, {Nord}, {Ogando}, {Plazas}, {Reil}, {Romer},
  {Roodman}, {Sako}, {Sanchez}, {Scarpine}, {Schubnell}, {Sevilla-Noarbe},
  {Smith}, {Soares-Santos}, {Sobreira}, {Suchyta}, {Swanson}, {Tarle},
  {Thaler}, {Thomas}, {Walker}, {Wechsler}, {Weller}, {Zhang}, \&
  {Zuntz}}]{Gruen:2015}
{Gruen}, D., {Friedrich}, O., {Amara}, A., {et~al.} 2016, \mnras, 455, 3367

\bibitem[{{Gruen} {et~al.}(2018){Gruen}, {Friedrich}, {Krause}, {DeRose},
  {Cawthon}, {Davis}, {Elvin-Poole}, {Rykoff}, {Wechsler}, {Alarcon},
  {Bernstein}, {Blazek}, {Chang}, {Clampitt}, {Crocce}, {De Vicente}, {Gatti},
  {Gill}, {Hartley}, {Hilbert}, {Hoyle}, {Jain}, {Jarvis}, {Lahav}, {MacCrann},
  {McClintock}, {Prat}, {Rollins}, {Ross}, {Rozo}, {Samuroff}, {S{\'a}nchez},
  {Sheldon}, {Troxel}, {Zuntz}, {Abbott}, {Abdalla}, {Allam}, {Annis},
  {Bechtol}, {Benoit-L{\'e}vy}, {Bertin}, {Bridle}, {Brooks}, {Buckley-Geer},
  {Carnero Rosell}, {Carrasco Kind}, {Carretero}, {Cunha}, {D'Andrea}, {da
  Costa}, {Desai}, {Diehl}, {Dietrich}, {Doel}, {Drlica-Wagner}, {Fernandez},
  {Flaugher}, {Fosalba}, {Frieman}, {Garc{\'\i}a-Bellido}, {Gaztanaga},
  {Giannantonio}, {Gruendl}, {Gschwend}, {Gutierrez}, {Honscheid}, {James},
  {Jeltema}, {Kuehn}, {Kuropatkin}, {Lima}, {March}, {Marshall}, {Martini},
  {Melchior}, {Menanteau}, {Miquel}, {Mohr}, {Plazas}, {Roodman}, {Sanchez},
  {Scarpine}, {Schubnell}, {Sevilla-Noarbe}, {Smith}, {Smith}, {Soares-Santos},
  {Sobreira}, {Swanson}, {Tarle}, {Thomas}, {Vikram}, {Walker}, {Weller},
  {Zhang}, \& {DES Collaboration}}]{Gruen:Friedrich:2018}
{Gruen}, D., {Friedrich}, O., {Krause}, E., {et~al.} 2018, \prd, 98, 023507

\bibitem[{{Halder} \& {Barreira}(2022)}]{Halder2022}
{Halder}, A. \& {Barreira}, A. 2022, \mnras, 515, 4639

\bibitem[{{Halder} {et~al.}(2021){Halder}, {Friedrich}, {Seitz}, \&
  {Varga}}]{Halder2021}
{Halder}, A., {Friedrich}, O., {Seitz}, S., \& {Varga}, T.~N. 2021, \mnras,
  506, 2780

\bibitem[{{Hamana} {et~al.}(2020){Hamana}, {Shirasaki}, {Miyazaki}, {Hikage},
  {Oguri}, {More}, {Armstrong}, {Leauthaud}, {Mandelbaum}, {Miyatake},
  {Nishizawa}, {Simet}, {Takada}, {Aihara}, {Bosch}, {Komiyama}, {Lupton},
  {Murayama}, {Strauss}, \& {Tanaka}}]{Hamana:2020}
{Hamana}, T., {Shirasaki}, M., {Miyazaki}, S., {et~al.} 2020, \pasj, 72, 16

\bibitem[{{Harnois-D{\'e}raps} {et~al.}(2018){Harnois-D{\'e}raps}, {Amon},
  {Choi}, {Demchenko}, {Heymans}, {Kannawadi}, {Nakajima}, {Sirks}, {van
  Waerbeke}, \& {Cai}}]{Harnois-Deraps2018}
{Harnois-D{\'e}raps}, J., {Amon}, A., {Choi}, A., {et~al.} 2018, \mnras, 481,
  1337

\bibitem[{{Harnois-D{\'e}raps} {et~al.}(2019){Harnois-D{\'e}raps}, {Giblin}, \&
  {Joachimi}}]{Harnois-Deraps2019}
{Harnois-D{\'e}raps}, J., {Giblin}, B., \& {Joachimi}, B. 2019, \aap, 631, A160

\bibitem[{{Harnois-D{\'e}raps} {et~al.}(2021){Harnois-D{\'e}raps}, {Martinet},
  {Castro}, {Dolag}, {Giblin}, {Heymans}, {Hildebrandt}, \&
  {Xia}}]{Harnois-Deraps2021}
{Harnois-D{\'e}raps}, J., {Martinet}, N., {Castro}, T., {et~al.} 2021, \mnras,
  506, 1623

\bibitem[{{Harnois-D{\'e}raps} {et~al.}(2022){Harnois-D{\'e}raps}, {Martinet},
  \& {Reischke}}]{Harnois-Deraps2022}
{Harnois-D{\'e}raps}, J., {Martinet}, N., \& {Reischke}, R. 2022, \mnras, 509,
  3868

\bibitem[{{Harnois-D{\'e}raps} {et~al.}(2013){Harnois-D{\'e}raps}, {Pen},
  {Iliev}, {Merz}, {Emberson}, \& {Desjacques}}]{Harnois-Deraps2013}
{Harnois-D{\'e}raps}, J., {Pen}, U.-L., {Iliev}, I.~T., {et~al.} 2013, \mnras,
  436, 540

\bibitem[{{Hartlap} {et~al.}(2007){Hartlap}, {Simon}, \&
  {Schneider}}]{Hartlap2007}
{Hartlap}, J., {Simon}, P., \& {Schneider}, P. 2007, \aap, 464, 399

\bibitem[{{Heitmann} {et~al.}(2014){Heitmann}, {Lawrence}, {Kwan}, {Habib}, \&
  {Higdon}}]{Heitmann:Lawrence:2014}
{Heitmann}, K., {Lawrence}, E., {Kwan}, J., {Habib}, S., \& {Higdon}, D. 2014,
  \apj, 780, 111

\bibitem[{{Heydenreich} {et~al.}(2022{\natexlab{a}}){Heydenreich}, {Br{\"u}ck},
  {Burger}, {Harnois-D{\'e}raps}, {Unruh}, {Castro}, {Dolag}, \&
  {Martinet}}]{Heydenreich2022a}
{Heydenreich}, S., {Br{\"u}ck}, B., {Burger}, P., {et~al.} 2022{\natexlab{a}},
  \aap, 667, A125

\bibitem[{{Heydenreich} {et~al.}(2022{\natexlab{b}}){Heydenreich}, {Linke},
  {Burger}, \& {Schneider}}]{Heydenreich2022b}
{Heydenreich}, S., {Linke}, L., {Burger}, P., \& {Schneider}, P.
  2022{\natexlab{b}}, arXiv:2208.11686

\bibitem[{{Heymans} {et~al.}(2021){Heymans}, {Tr{\"o}ster}, {Asgari}, {Blake},
  {Hildebrandt}, {Joachimi}, {Kuijken}, {Lin}, {S{\'a}nchez}, {van den Busch},
  {Wright}, {Amon}, {Bilicki}, {de Jong}, {Crocce}, {Dvornik}, {Erben},
  {Fortuna}, {Getman}, {Giblin}, {Glazebrook}, {Hoekstra}, {Joudaki},
  {Kannawadi}, {K{\"o}hlinger}, {Lidman}, {Miller}, {Napolitano}, {Parkinson},
  {Schneider}, {Shan}, {Valentijn}, {Verdoes Kleijn}, \& {Wolf}}]{Heymans:2021}
{Heymans}, C., {Tr{\"o}ster}, T., {Asgari}, M., {et~al.} 2021, \aap, 646, A140

\bibitem[{{Hilbert} {et~al.}(2011){Hilbert}, {Hartlap}, \&
  {Schneider}}]{Hilbert:2011}
{Hilbert}, S., {Hartlap}, J., \& {Schneider}, P. 2011, \aap, 536, A85

\bibitem[{{Hildebrandt} {et~al.}(2021){Hildebrandt}, {van den Busch}, {Wright},
  {Blake}, {Joachimi}, {Kuijken}, {Tr{\"o}ster}, {Asgari}, {Bilicki}, {de
  Jong}, {Dvornik}, {Erben}, {Getman}, {Giblin}, {Heymans}, {Kannawadi}, {Lin},
  \& {Shan}}]{Hildebrandt2021}
{Hildebrandt}, H., {van den Busch}, J.~L., {Wright}, A.~H., {et~al.} 2021,
  \aap, 647, A124

\bibitem[{{Hildebrandt} {et~al.}(2017){Hildebrandt}, {Viola}, {Heymans},
  {Joudaki}, {Kuijken}, {Blake}, {Erben}, {Joachimi}, {Klaes}, {Miller},
  {Morrison}, {Nakajima}, {Verdoes Kleijn}, {Amon}, {Choi}, {Covone}, {de
  Jong}, {Dvornik}, {Fenech Conti}, {Grado}, {Harnois-D{\'e}raps}, {Herbonnet},
  {Hoekstra}, {K{\"o}hlinger}, {McFarland}, {Mead}, {Merten}, {Napolitano},
  {Peacock}, {Radovich}, {Schneider}, {Simon}, {Valentijn}, {van den Busch},
  {van Uitert}, \& {Van Waerbeke}}]{Hildebrandt:2017}
{Hildebrandt}, H., {Viola}, M., {Heymans}, C., {et~al.} 2017, \mnras, 465, 1454

\bibitem[{{Hirschmann} {et~al.}(2014){Hirschmann}, {Dolag}, {Saro}, {Bachmann},
  {Borgani}, \& {Burkert}}]{Hirschmann2014}
{Hirschmann}, M., {Dolag}, K., {Saro}, A., {et~al.} 2014, \mnras, 442, 2304

\bibitem[{{Jarvis} {et~al.}(2004){Jarvis}, {Bernstein}, \&
  {Jain}}]{Jarvis:Bernstein:2004}
{Jarvis}, M., {Bernstein}, G., \& {Jain}, B. 2004, \mnras, 352, 338

\bibitem[{{Joachimi} {et~al.}(2021){Joachimi}, {Lin}, {Asgari}, {Tr{\"o}ster},
  {Heymans}, {Hildebrandt}, {K{\"o}hlinger}, {S{\'a}nchez}, {Wright},
  {Bilicki}, {Blake}, {van den Busch}, {Crocce}, {Dvornik}, {Erben}, {Getman},
  {Giblin}, {Hoekstra}, {Kannawadi}, {Kuijken}, {Napolitano}, {Schneider},
  {Scoccimarro}, {Sellentin}, {Shan}, {von Wietersheim-Kramsta}, \&
  {Zuntz}}]{Joachimi2021}
{Joachimi}, B., {Lin}, C.~A., {Asgari}, M., {et~al.} 2021, \aap, 646, A129

\bibitem[{{Joudaki} {et~al.}(2018){Joudaki}, {Blake}, {Johnson}, {Amon},
  {Asgari}, {Choi}, {Erben}, {Glazebrook}, {Harnois-D{\'e}raps}, {Heymans},
  {Hildebrandt}, {Hoekstra}, {Klaes}, {Kuijken}, {Lidman}, {Mead}, {Miller},
  {Parkinson}, {Poole}, {Schneider}, {Viola}, \& {Wolf}}]{Joudaki2018}
{Joudaki}, S., {Blake}, C., {Johnson}, A., {et~al.} 2018, \mnras, 474, 4894

\bibitem[{{Joudaki} {et~al.}(2020){Joudaki}, {Hildebrandt}, {Traykova},
  {Chisari}, {Heymans}, {Kannawadi}, {Kuijken}, {Wright}, {Asgari}, {Erben},
  {Hoekstra}, {Joachimi}, {Miller}, {Tr{\"o}ster}, \& {van den
  Busch}}]{joudaki2020}
{Joudaki}, S., {Hildebrandt}, H., {Traykova}, D., {et~al.} 2020, \aap, 638, L1

\bibitem[{{Kaiser}(1992)}]{Kaiser1992}
{Kaiser}, N. 1992, \apj, 388, 272

\bibitem[{{Kilbinger} \& {Schneider}(2005)}]{Kilbinger2005}
{Kilbinger}, M. \& {Schneider}, P. 2005, \aap, 442, 69

\bibitem[{{Kodwani} {et~al.}(2019){Kodwani}, {Alonso}, \&
  {Ferreira}}]{Kodwani2019....2E...3K}
{Kodwani}, D., {Alonso}, D., \& {Ferreira}, P. 2019, The Open Journal of
  Astrophysics, 2, 3

\bibitem[{{Kuijken} {et~al.}(2019){Kuijken}, {Heymans}, {Dvornik},
  {Hildebrandt}, {de Jong}, {Wright}, {Erben}, {Bilicki}, {Giblin}, {Shan},
  {Getman}, {Grado}, {Hoekstra}, {Miller}, {Napolitano}, {Paolilo}, {Radovich},
  {Schneider}, {Sutherland }, {Tewes}, {Tortora}, {Valentijn}, \& {Verdoes
  Kleijn}}]{Kuijken:2019}
{Kuijken}, K., {Heymans}, C., {Dvornik}, A., {et~al.} 2019, \aap, 625, A2

\bibitem[{{Kuijken} {et~al.}(2015){Kuijken}, {Heymans}, {Hildebrandt},
  {Nakajima}, {Erben}, {de Jong}, {Viola}, {Choi}, {Hoekstra}, {Miller}, {van
  Uitert}, {Amon}, {Blake}, {Brouwer}, {Buddendiek}, {Conti}, {Eriksen},
  {Grado}, {Harnois-D{\'e}raps}, {Helmich}, {Herbonnet}, {Irisarri},
  {Kitching}, {Klaes}, {La Barbera}, {Napolitano}, {Radovich}, {Schneider},
  {Sif{\'o}n}, {Sikkema}, {Simon}, {Tudorica}, {Valentijn}, {Verdoes Kleijn},
  \& {van Waerbeke}}]{Kuijken:2015}
{Kuijken}, K., {Heymans}, C., {Hildebrandt}, H., {et~al.} 2015, \mnras, 454,
  3500

\bibitem[{{Martinet} {et~al.}(2021){Martinet}, {Castro}, {Harnois-D{\'e}raps},
  {Jullo}, {Giocoli}, \& {Dolag}}]{Martinet2021b}
{Martinet}, N., {Castro}, T., {Harnois-D{\'e}raps}, J., {et~al.} 2021, \aap,
  648, A115

\bibitem[{{Miller} {et~al.}(2013){Miller}, {Heymans}, {Kitching}, {van
  Waerbeke}, {Erben}, {Hildebrandt}, {Hoekstra}, {Mellier}, {Rowe}, {Coupon},
  {Dietrich}, {Fu}, {Harnois-D{\'e}raps}, {Hudson}, {Kilbinger}, {Kuijken},
  {Schrabback}, {Semboloni}, {Vafaei}, \& {Velander}}]{Miller2013}
{Miller}, L., {Heymans}, C., {Kitching}, T.~D., {et~al.} 2013, \mnras, 429,
  2858

\bibitem[{{Mo} {et~al.}(2010){Mo}, {van den Bosch}, \& {White}}]{Mo2010}
{Mo}, H., {van den Bosch}, F.~C., \& {White}, S. 2010, {Galaxy Formation and
  Evolution}

\bibitem[{{Peacock} \& {Bilicki}(2018)}]{PeacockBilicki2018}
{Peacock}, J.~A. \& {Bilicki}, M. 2018, \mnras, 481, 1133

\bibitem[{{Percival} {et~al.}(2022){Percival}, {Friedrich}, {Sellentin}, \&
  {Heavens}}]{Percival2021}
{Percival}, W.~J., {Friedrich}, O., {Sellentin}, E., \& {Heavens}, A. 2022,
  \mnras, 510, 3207

\bibitem[{{Pires} {et~al.}(2012){Pires}, {Leonard}, \& {Starck}}]{Pires:2012}
{Pires}, S., {Leonard}, A., \& {Starck}, J.-L. 2012, \mnras, 423, 983

\bibitem[{{Planck Collaboration} {et~al.}(2020){Planck Collaboration},
  {Aghanim}, {Akrami}, {Ashdown}, {Aumont}, {Baccigalupi}, {Ballardini},
  {Banday}, {Barreiro}, {Bartolo}, {Basak}, {Benabed}, {Bernard}, {Bersanelli},
  {Bielewicz}, {Bock}, {Bond}, {Borrill}, {Bouchet}, {Boulanger}, {Bucher},
  {Burigana}, {Butler}, {Calabrese}, {Cardoso}, {Carron}, {Casaponsa},
  {Challinor}, {Chiang}, {Colombo}, {Combet}, {Crill}, {Cuttaia}, {de
  Bernardis}, {de Rosa}, {de Zotti}, {Delabrouille}, {Delouis}, {Di Valentino},
  {Diego}, {Dor{\'e}}, {Douspis}, {Ducout}, {Dupac}, {Dusini}, {Efstathiou},
  {Elsner}, {En{\ss}lin}, {Eriksen}, {Fantaye}, {Fernandez-Cobos}, {Finelli},
  {Frailis}, {Fraisse}, {Franceschi}, {Frolov}, {Galeotta}, {Galli}, {Ganga},
  {G{\'e}nova-Santos}, {Gerbino}, {Ghosh}, {Giraud-H{\'e}raud},
  {Gonz{\'a}lez-Nuevo}, {G{\'o}rski}, {Gratton}, {Gruppuso}, {Gudmundsson},
  {Hamann}, {Handley}, {Hansen}, {Herranz}, {Hivon}, {Huang}, {Jaffe}, {Jones},
  {Keih{\"a}nen}, {Keskitalo}, {Kiiveri}, {Kim}, {Kisner}, {Krachmalnicoff},
  {Kunz}, {Kurki-Suonio}, {Lagache}, {Lamarre}, {Lasenby}, {Lattanzi},
  {Lawrence}, {Le Jeune}, {Levrier}, {Lewis}, {Liguori}, {Lilje}, {Lilley},
  {Lindholm}, {L{\'o}pez-Caniego}, {Lubin}, {Ma}, {Mac{\'\i}as-P{\'e}rez},
  {Maggio}, {Maino}, {Mandolesi}, {Mangilli}, {Marcos-Caballero}, {Maris},
  {Martin}, {Mart{\'\i}nez-Gonz{\'a}lez}, {Matarrese}, {Mauri}, {McEwen},
  {Meinhold}, {Melchiorri}, {Mennella}, {Migliaccio}, {Millea},
  {Miville-Desch{\^e}nes}, {Molinari}, {Moneti}, {Montier}, {Morgante}, {Moss},
  {Natoli}, {N{\o}rgaard-Nielsen}, {Pagano}, {Paoletti}, {Partridge},
  {Patanchon}, {Peiris}, {Perrotta}, {Pettorino}, {Piacentini}, {Polenta},
  {Puget}, {Rachen}, {Reinecke}, {Remazeilles}, {Renzi}, {Rocha}, {Rosset},
  {Roudier}, {Rubi{\~n}o-Mart{\'\i}n}, {Ruiz-Granados}, {Salvati}, {Sandri},
  {Savelainen}, {Scott}, {Shellard}, {Sirignano}, {Sirri}, {Spencer},
  {Sunyaev}, {Suur-Uski}, {Tauber}, {Tavagnacco}, {Tenti}, {Toffolatti},
  {Tomasi}, {Trombetti}, {Valiviita}, {Van Tent}, {Vielva}, {Villa},
  {Vittorio}, {Wandelt}, {Wehus}, {Zacchei}, \& {Zonca}}]{Aghanim:2020}
{Planck Collaboration}, {Aghanim}, N., {Akrami}, Y., {et~al.} 2020, \aap, 641,
  A5

\bibitem[{{Pyne} \& {Joachimi}(2021)}]{Pyne2021}
{Pyne}, S. \& {Joachimi}, B. 2021, \mnras, 503, 2300

\bibitem[{{Ragagnin} {et~al.}(2017){Ragagnin}, {Dolag}, {Biffi}, {Cadolle Bel},
  {Hammer}, {Krukau}, {Petkova}, \& {Steinborn}}]{Ragagnin2017}
{Ragagnin}, A., {Dolag}, K., {Biffi}, V., {et~al.} 2017, Astronomy and
  Computing, 20, 52

\bibitem[{{Reimberg} \& {Bernardeau}(2018)}]{Reimberg2018}
{Reimberg}, P. \& {Bernardeau}, F. 2018, \prd, 97, 023524

\bibitem[{{Sadeh} {et~al.}(2016){Sadeh}, {Abdalla}, \& {Lahav}}]{Sadeh2016}
{Sadeh}, I., {Abdalla}, F.~B., \& {Lahav}, O. 2016, \pasp, 128, 104502

\bibitem[{{S{\'a}nchez} {et~al.}(2017){S{\'a}nchez}, {Scoccimarro}, {Crocce},
  {Grieb}, {Salazar-Albornoz}, {Dalla Vecchia}, {Lippich}, {Beutler},
  {Brownstein}, {Chuang}, {Eisenstein}, {Kitaura}, {Olmstead}, {Percival},
  {Prada}, {Rodr{\'\i}guez-Torres}, {Ross}, {Samushia}, {Seo}, {Tinker},
  {Tojeiro}, {Vargas-Maga{\~n}a}, {Wang}, \& {Zhao}}]{Sanchez2017}
{S{\'a}nchez}, A.~G., {Scoccimarro}, R., {Crocce}, M., {et~al.} 2017, \mnras,
  464, 1640

\bibitem[{{Schneider}(1996)}]{Schneider:1996}
{Schneider}, P. 1996, \mnras, 283, 837

\bibitem[{{Schneider}(1998)}]{Schneider:1998}
{Schneider}, P. 1998, \apj, 498, 43

\bibitem[{{Seitz} \& {Schneider}(1997)}]{Seitz1997}
{Seitz}, C. \& {Schneider}, P. 1997, \aap, 318, 687

\bibitem[{{Sellentin} \& {Heavens}(2016)}]{Sellentin2016}
{Sellentin}, E. \& {Heavens}, A.~F. 2016, \mnras, 456, L132

\bibitem[{{Smith} {et~al.}(2017){Smith}, {Cole}, {Baugh}, {Zheng}, {Angulo},
  {Norberg}, \& {Zehavi}}]{Smith:2017}
{Smith}, A., {Cole}, S., {Baugh}, C., {et~al.} 2017, \mnras, 470, 4646

\bibitem[{{Springel}(2005)}]{Springel2005}
{Springel}, V. 2005, \mnras, 364, 1105

\bibitem[{{Spurio Mancini} {et~al.}(2022){Spurio Mancini}, {Piras}, {Alsing},
  {Joachimi}, \& {Hobson}}]{COSMOPOWER2022}
{Spurio Mancini}, A., {Piras}, D., {Alsing}, J., {Joachimi}, B., \& {Hobson},
  M.~P. 2022, \mnras, 511, 1771

\bibitem[{{Takahashi} {et~al.}(2017){Takahashi}, {Hamana}, {Shirasaki},
  {Namikawa}, {Nishimichi}, {Osato}, \& {Shiroyama}}]{Takahashi2017}
{Takahashi}, R., {Hamana}, T., {Shirasaki}, M., {et~al.} 2017, \apj, 850, 24

\bibitem[{{van Daalen} {et~al.}(2011){van Daalen}, {Schaye}, {Booth}, \& {Dalla
  Vecchia}}]{vanDaalen2011}
{van Daalen}, M.~P., {Schaye}, J., {Booth}, C.~M., \& {Dalla Vecchia}, C. 2011,
  \mnras, 415, 3649

\bibitem[{{van den Busch} {et~al.}(2022){van den Busch}, {Wright},
  {Hildebrandt}, {Bilicki}, {Asgari}, {Joudaki}, {Blake}, {Heymans},
  {Kannawadi}, {Shan}, \& {Tr{\"o}ster}}]{JLB2022}
{van den Busch}, J.~L., {Wright}, A.~H., {Hildebrandt}, H., {et~al.} 2022,
  \aap, 664, A170

\bibitem[{{van Uitert} {et~al.}(2018){van Uitert}, {Joachimi}, {Joudaki},
  {Amon}, {Heymans}, {K{\"o}hlinger}, {Asgari}, {Blake}, {Choi}, {Erben},
  {Farrow}, {Harnois-D{\'e}raps}, {Hildebrandt}, {Hoekstra}, {Kitching},
  {Klaes}, {Kuijken}, {Merten}, {Miller}, {Nakajima}, {Schneider}, {Valentijn},
  \& {Viola}}]{vanUitert2018}
{van Uitert}, E., {Joachimi}, B., {Joudaki}, S., {et~al.} 2018, \mnras, 476,
  4662

\bibitem[{{Wright} {et~al.}(2020){Wright}, {Hildebrandt}, {van den Busch},
  {Heymans}, {Joachimi}, {Kannawadi}, \& {Kuijken}}]{Wright:2020}
{Wright}, A.~H., {Hildebrandt}, H., {van den Busch}, J.~L., {et~al.} 2020,
  \aap, 640, L14

\bibitem[{{Xavier} {et~al.}(2016){Xavier}, {Abdalla}, \&
  {Joachimi}}]{FLASK2016}
{Xavier}, H.~S., {Abdalla}, F.~B., \& {Joachimi}, B. 2016, \mnras, 459, 3693

\bibitem[{{Z{\"u}rcher} {et~al.}(2022){Z{\"u}rcher}, {Fluri}, {Sgier},
  {Kacprzak}, {Gatti}, {Doux}, {Whiteway}, {R{\'e}fr{\'e}gier}, {Chang},
  {Jeffrey}, {Jain}, {Lemos}, {Bacon}, {Alarcon}, {Amon}, {Bechtol}, {Becker},
  {Bernstein}, {Campos}, {Chen}, {Choi}, {Davis}, {Derose}, {Dodelson},
  {Elsner}, {Elvin-Poole}, {Everett}, {Ferte}, {Gruen}, {Harrison}, {Huterer},
  {Jarvis}, {Leget}, {Maccrann}, {Mccullough}, {Muir}, {Myles}, {Navarro
  Alsina}, {Pandey}, {Prat}, {Raveri}, {Rollins}, {Roodman}, {Sanchez},
  {Secco}, {Sheldon}, {Shin}, {Troxel}, {Tutusaus}, {Yin}, {Aguena}, {Allam},
  {Andrade-Oliveira}, {Annis}, {Bertin}, {Brooks}, {Burke}, {Carnero Rosell},
  {Carrasco Kind}, {Carretero}, {Castander}, {Cawthon}, {Conselice},
  {Costanzi}, {da Costa}, {da Silva Pereira}, {Davis}, {De Vicente}, {Desai},
  {Diehl}, {Dietrich}, {Doel}, {Eckert}, {Evrard}, {Ferrero}, {Flaugher},
  {Fosalba}, {Friedel}, {Frieman}, {Garcia-Bellido}, {Gaztanaga}, {Gerdes},
  {Giannantonio}, {Gruendl}, {Gschwend}, {Gutierrez}, {Hinton}, {Hollowood},
  {Honscheid}, {Hoyle}, {James}, {Kuehn}, {Kuropatkin}, {Lahav}, {Lidman},
  {Lima}, {Maia}, {Marshall}, {Melchior}, {Menanteau}, {Miquel}, {Morgan},
  {Palmese}, {Paz-Chinchon}, {Pieres}, {Plazas Malag{\'o}n}, {Reil}, {Rodriguez
  Monroy}, {Romer}, {Sanchez}, {Scarpine}, {Schubnell}, {Serrano}, {Sevilla},
  {Smith}, {Suchyta}, {Tarle}, {Thomas}, {To}, {Varga}, {Weller}, {Wilkinson},
  \& {DES Collaboration}}]{Zuercher2022}
{Z{\"u}rcher}, D., {Fluri}, J., {Sgier}, R., {et~al.} 2022, \mnras, 511, 2075

\end{thebibliography}

\clearpage
\begin{appendix}
\section{Additional plots}
\begin{figure}[ht]
\includegraphics[width=\columnwidth]{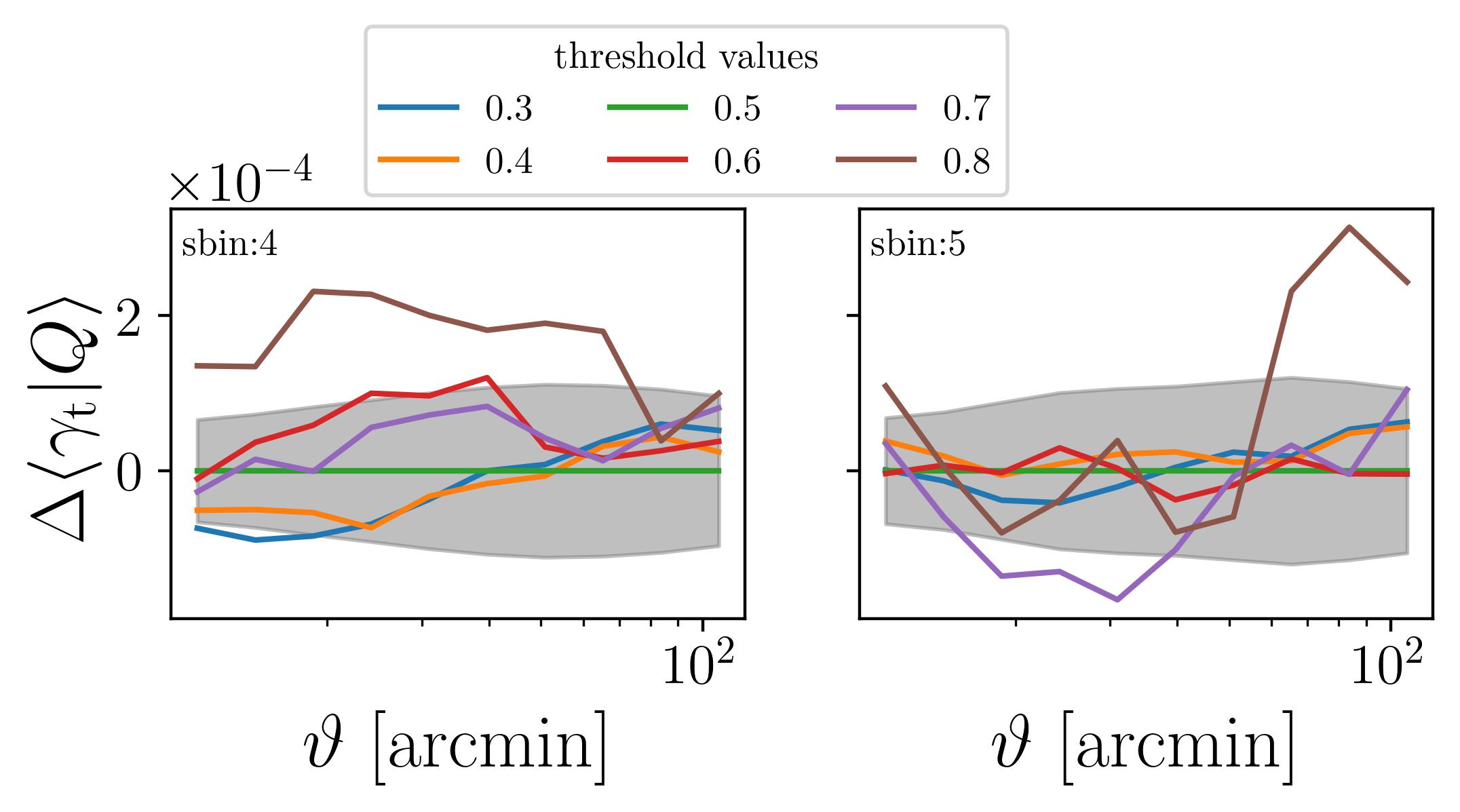}
\caption{Residual shear profiles of the highest quantile filter for different threshold values that the effective area must exceed to be used with respect to the shear profiles with a 0.5 threshold. It is seen that for threshold values above $0.6$ the shear profiles deviate more than one standard deviation (grey band) due to decreasing remaining area.}
\label{fig:shear_cutoff}
\end{figure}

In this section, we show additional plots which belong to the main text. We start by visualising the shear profiles that result in different threshold values for the effective area above which pixels are used for the analysis in Fig.~\ref{fig:shear_cutoff}. 

\begin{figure}[ht]
\begin{subfigure}{0.49\columnwidth}
\includegraphics[width=\columnwidth]{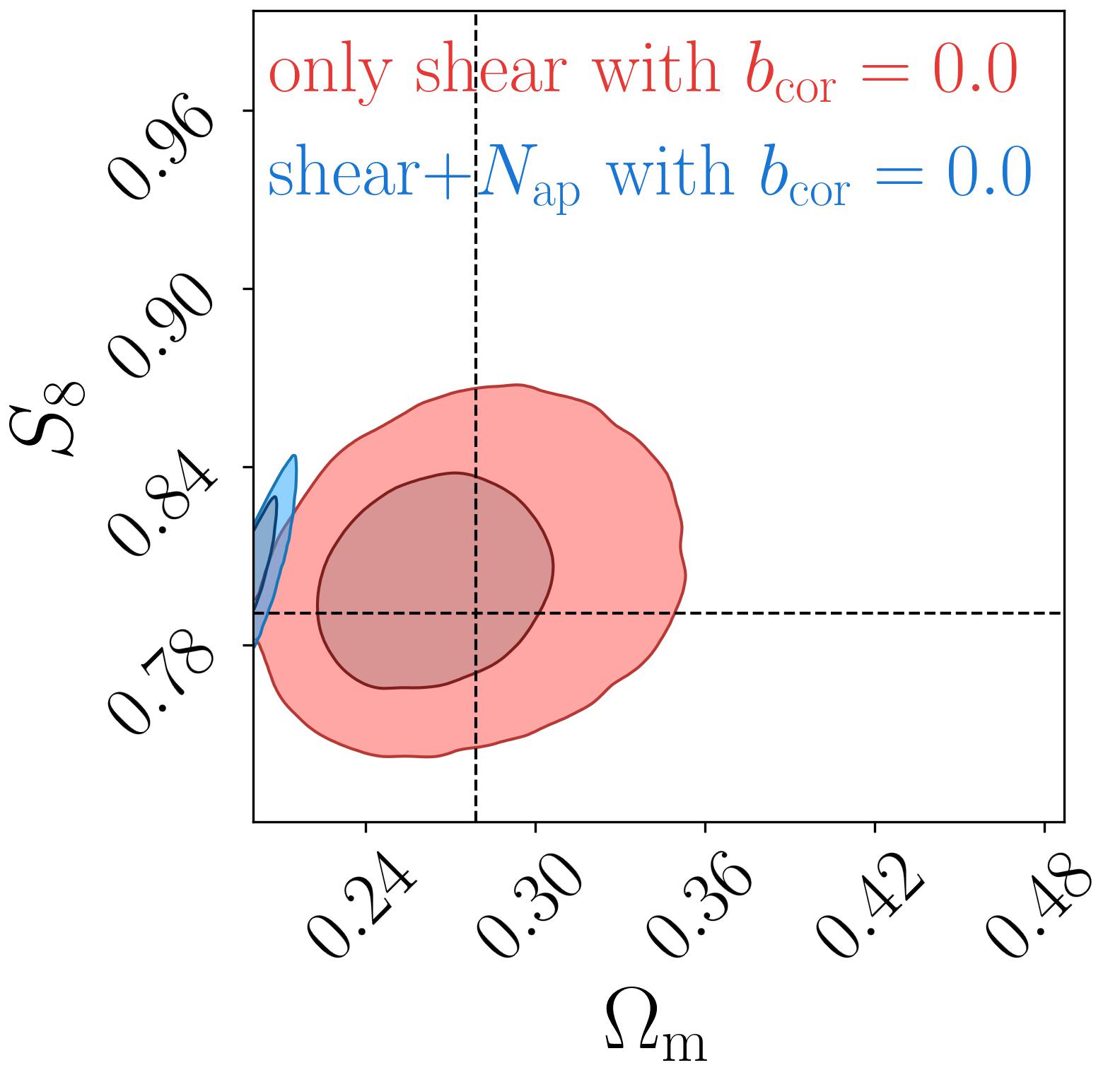}
\end{subfigure}
\begin{subfigure}{0.49\columnwidth}
\includegraphics[width=\columnwidth]{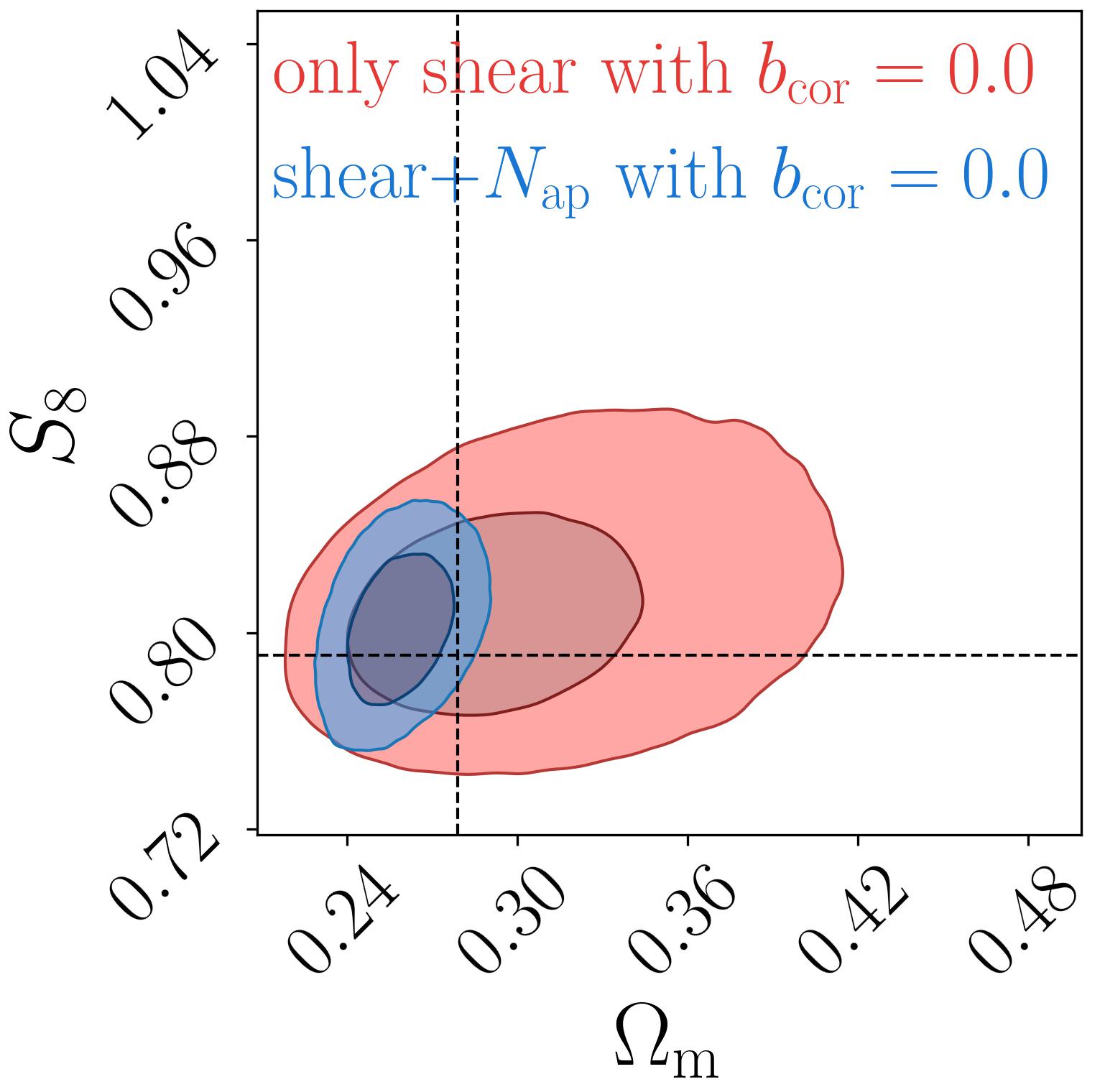}
\end{subfigure}
\begin{subfigure}{0.49\columnwidth}
\includegraphics[width=\columnwidth]{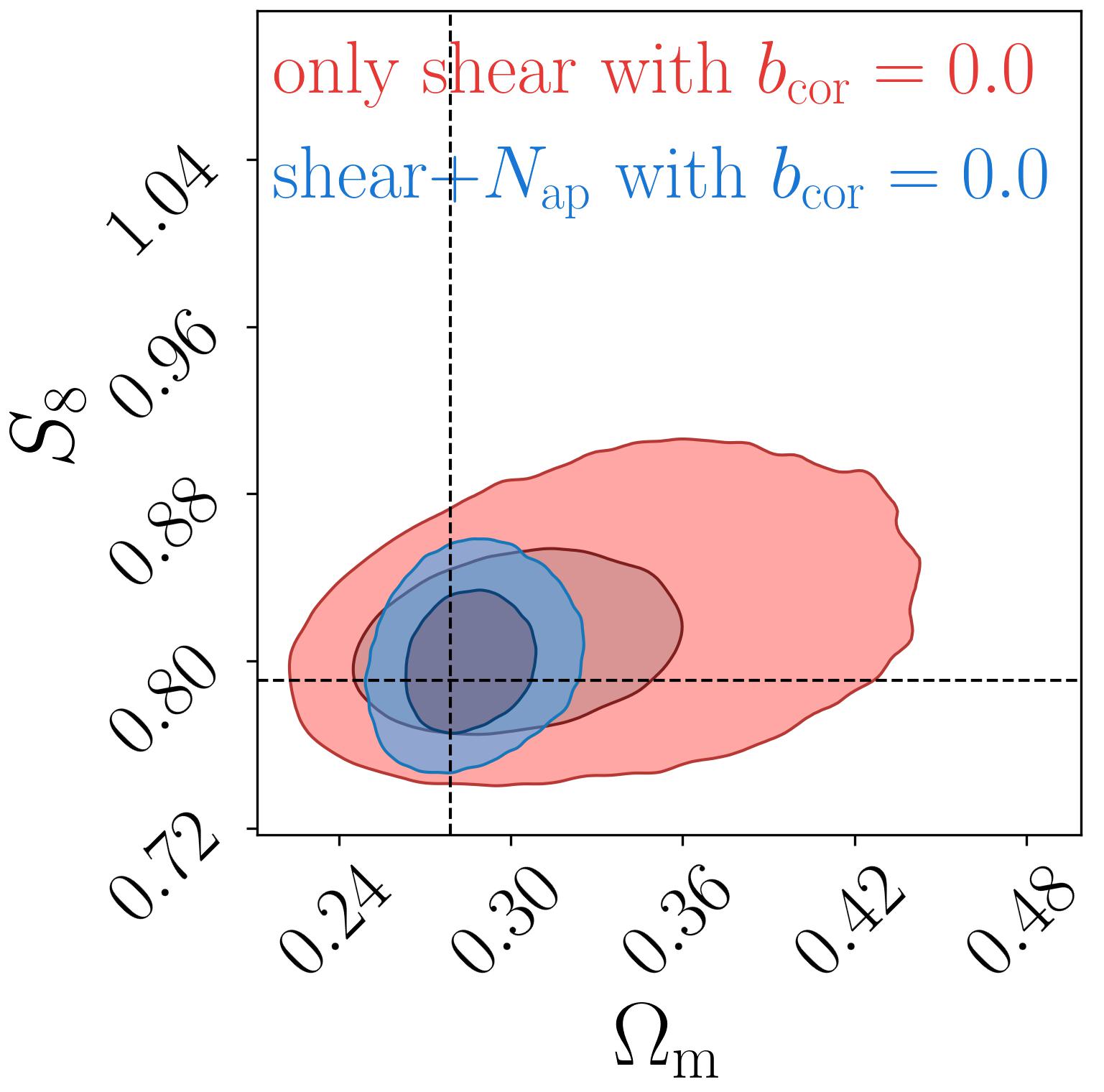}
\end{subfigure}
\begin{subfigure}{0.49\columnwidth}
\includegraphics[width=\columnwidth]{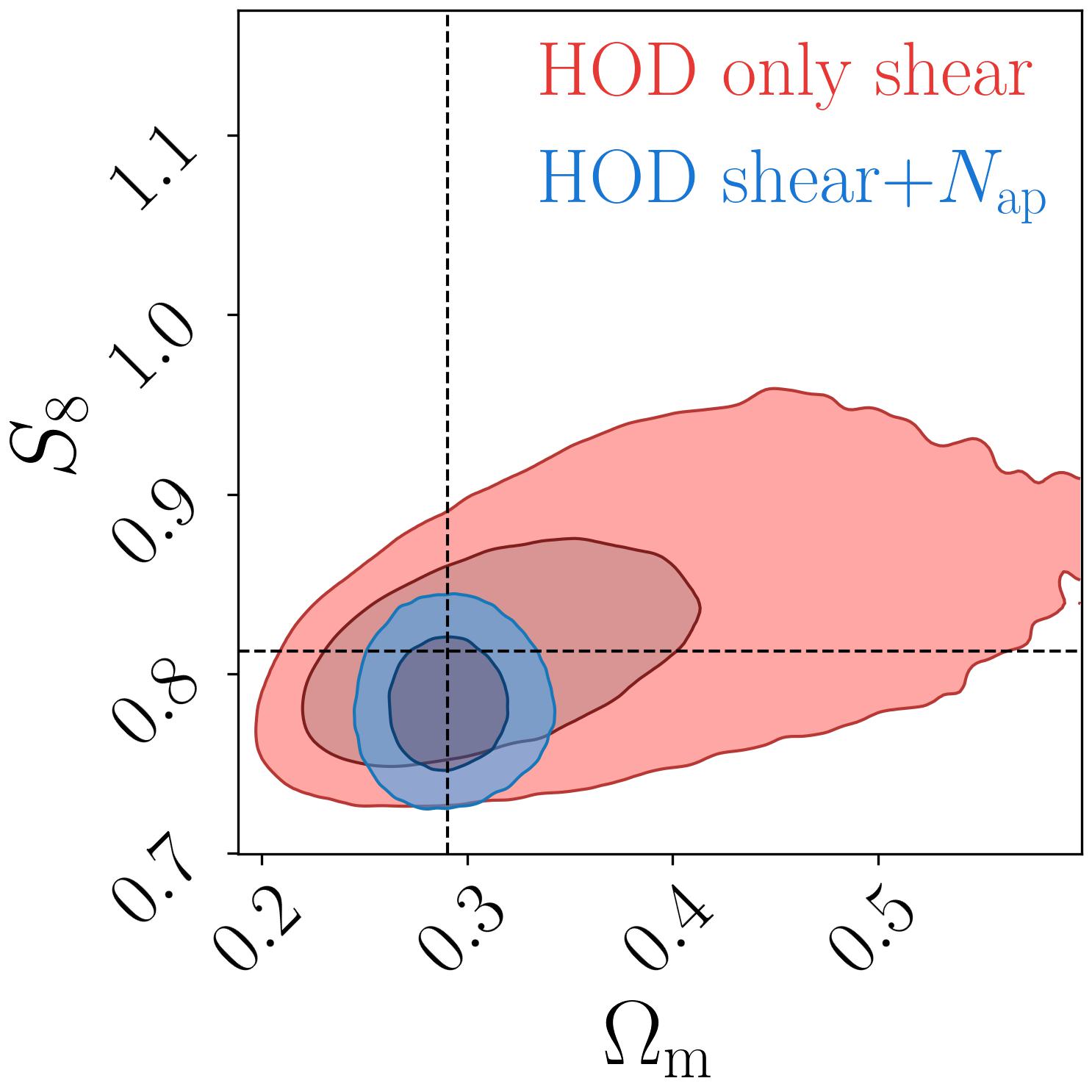}
\end{subfigure}
\caption{Posterior distributions, using different $b_\mathrm{2}$ values to distribute galaxies to the same density contrast, from which a data vector is measured. The lower right panel shows the posterior results from simulations, where galaxies are distributed with a HOD. The model vector uses for all four cases a linear galaxy bias model. }
\label{fig:MCMC_bias}
\end{figure}

In Fig.~\ref{fig:MCMC_bias} we show different posterior distributions, to test whether our analytical model with a linear galaxy bias model gives robust cosmological results even if galaxies are placed with a Poisson process with mean $\lambda=n_\mathrm{eff}\big[1+b_\mathrm{1}\,\delta_{\mathrm{m,2D}}({\boldsymbol \theta})+b_\mathrm{2}\,(\delta_{\mathrm{m,2D}}^2({\boldsymbol \theta})-\langle\delta_{\mathrm{m,2D}}^2)\rangle\big]$, where $\delta_{\mathrm{m,2D}}$ is the projected density contrast or with a halo occupation distribution (HOD) description \citep{Smith:2017}. For the Poisson process test, we use the simulation of \citet{Takahashi2017} with sources that mimic the fourth and fifth KiDS-1000 bins and lenses that mimic the KiDS-bright sample. For the HOD analysis, we measure the data vector, and the covariance from 614 SLICS realisations with noisy KiDS-1000 sources and GAMA lens mocks \citep[see][for a detailed description]{Harnois-Deraps2018}, where the covariance is scaled to approximately match the KIDS-1000 footprint. It is clearly seen that the posterior for $b_\mathrm{2}>0$ is strongly biased if shear and $N_\mathrm{ap}$ information are used. In contrast, using only shear information, the posterior is unbiased as they are almost insensitive to the galaxy bias model. Although the HOD analysis already shows that the linear galaxy bias assumption is sufficient, if the posterior using shear and $N_\mathrm{ap}$ information is consistent with the posterior using only shear information gives additional confidence that $b_\mathrm{2}\approx 0$. 

\begin{figure}[ht]
\includegraphics[width=\columnwidth]{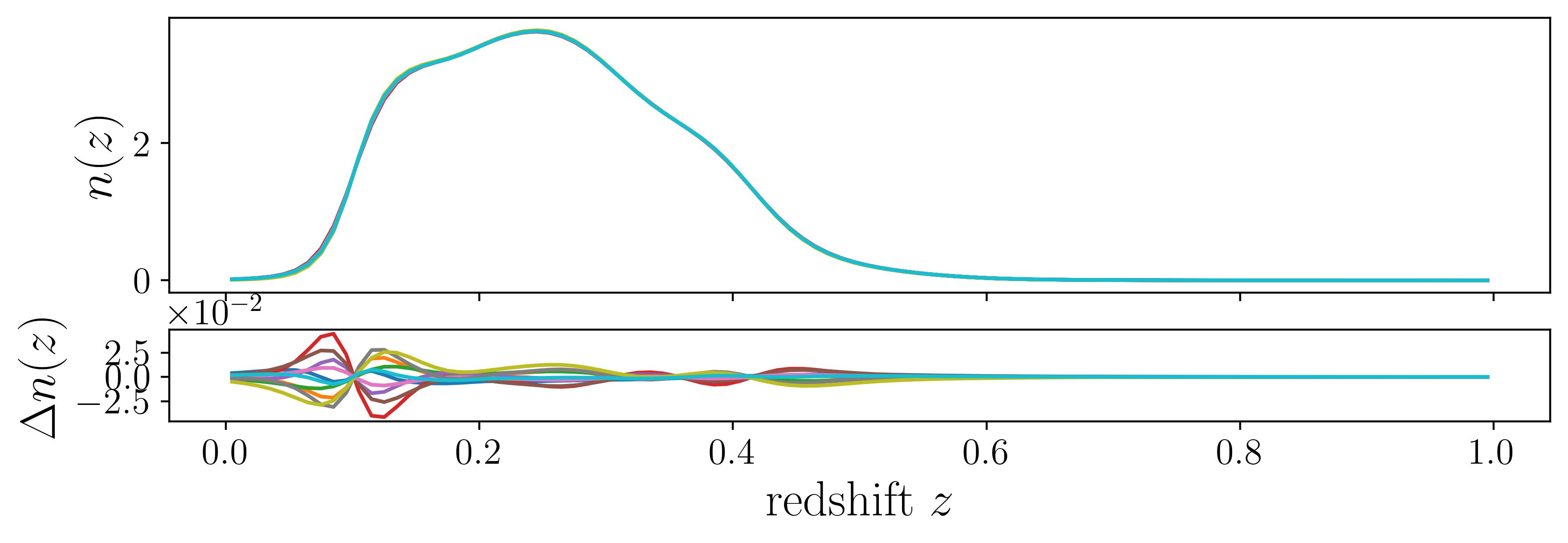}
\caption{Redshift distribution of the full KiDS-bright sample resulting from Lorentzian fitting parameters ($a$ and $s$), which in turn are determined from different patches on the sky. In the bottom panel, the absolute differences to the best-estimated $n_\mathrm{be}(z)$ are shown.}
\label{fig:nz_a_s}
\end{figure}

\begin{figure}[ht]
\includegraphics[width=\columnwidth]{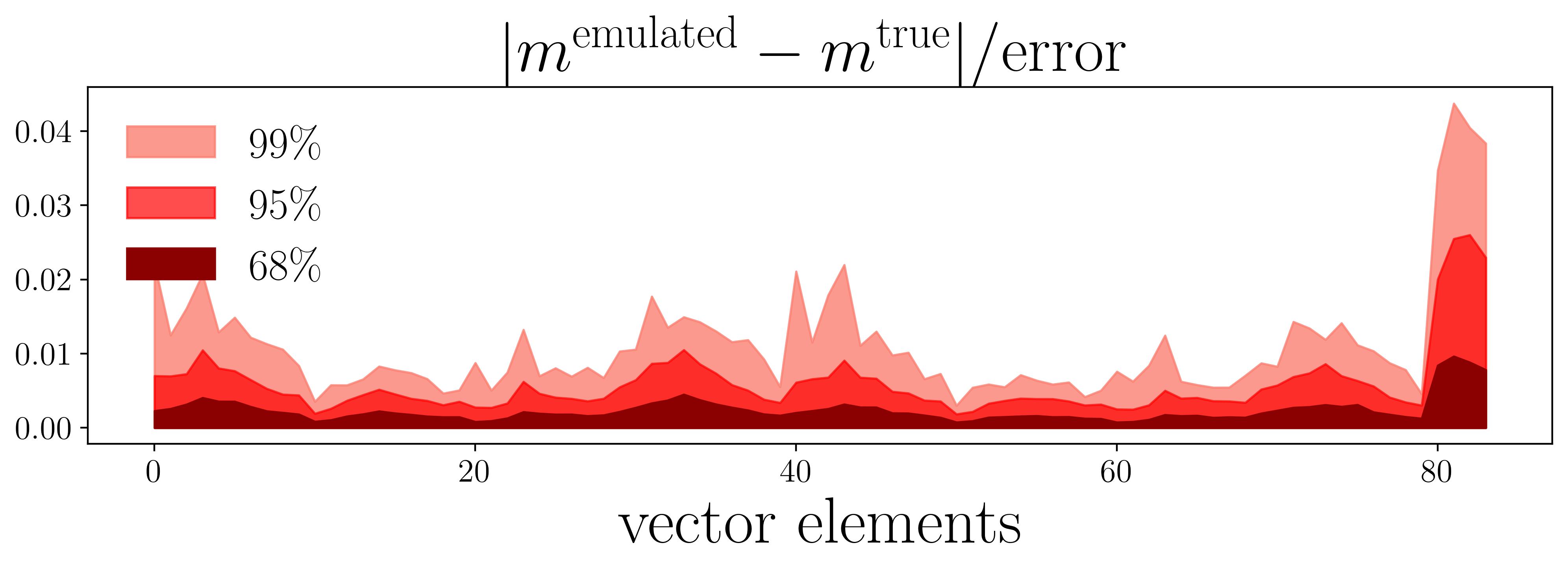}
\caption{Difference between the true and emulated model data vector $m(\boldsymbol{\Theta})$ scaled by the standard deviation of the measured data estimated with the \texttt{FLASK}.}
\label{fig:chi_acc}
\end{figure}

\begin{figure}[ht]
\includegraphics[width=\columnwidth]{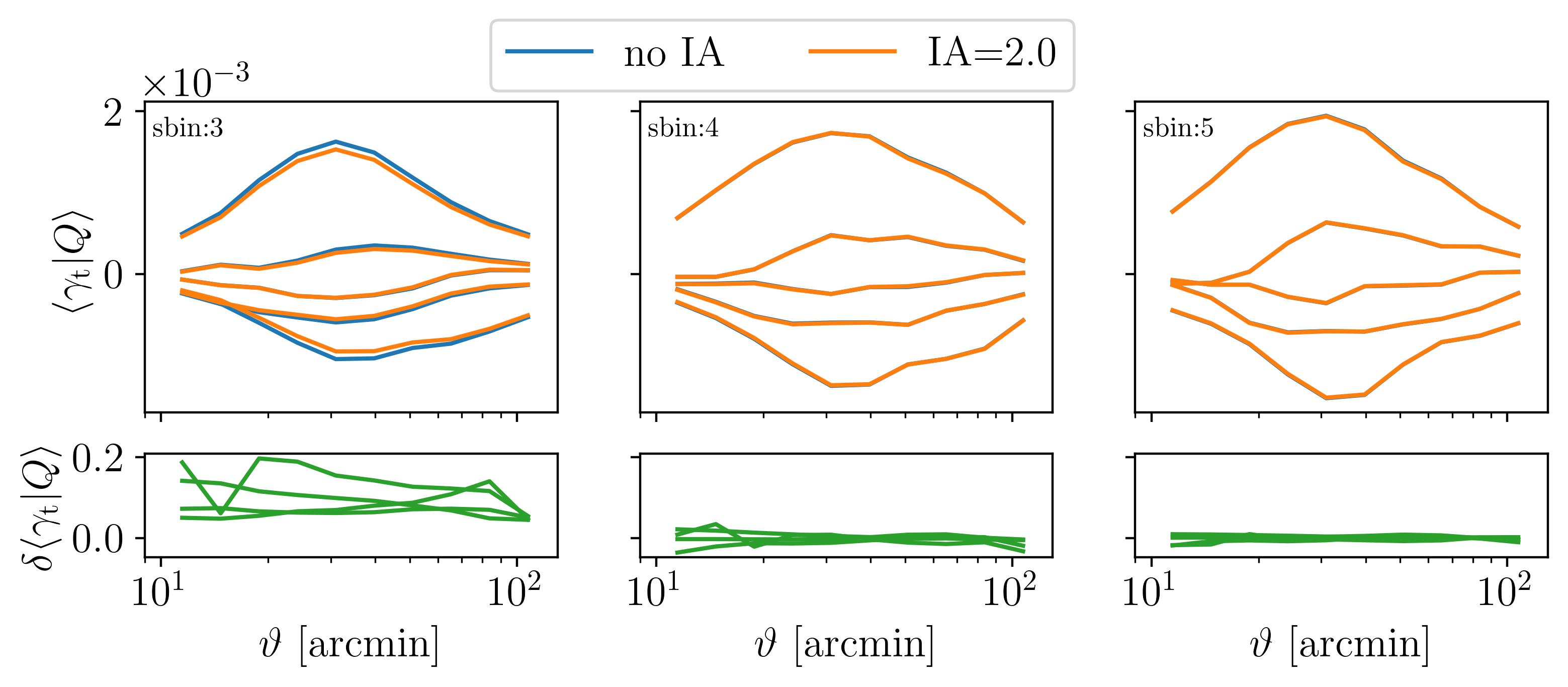}
\caption{Shear profiles of the cosmo-SLICS for the bright lens and all three source mocks with the $n(z)$ shown in Fig.\,\ref{fig:nofz} for two different IA amplitudes for the adapted filter. The third source tomographic bin is strongly contaminated by IA due to the significant overlap with the lens $n(z)$ and is therefore excluded from this analysis.}
\label{fig:pure_shear_cosmoSLICS}
\end{figure}

\begin{figure}[ht]
\includegraphics[width=\columnwidth]{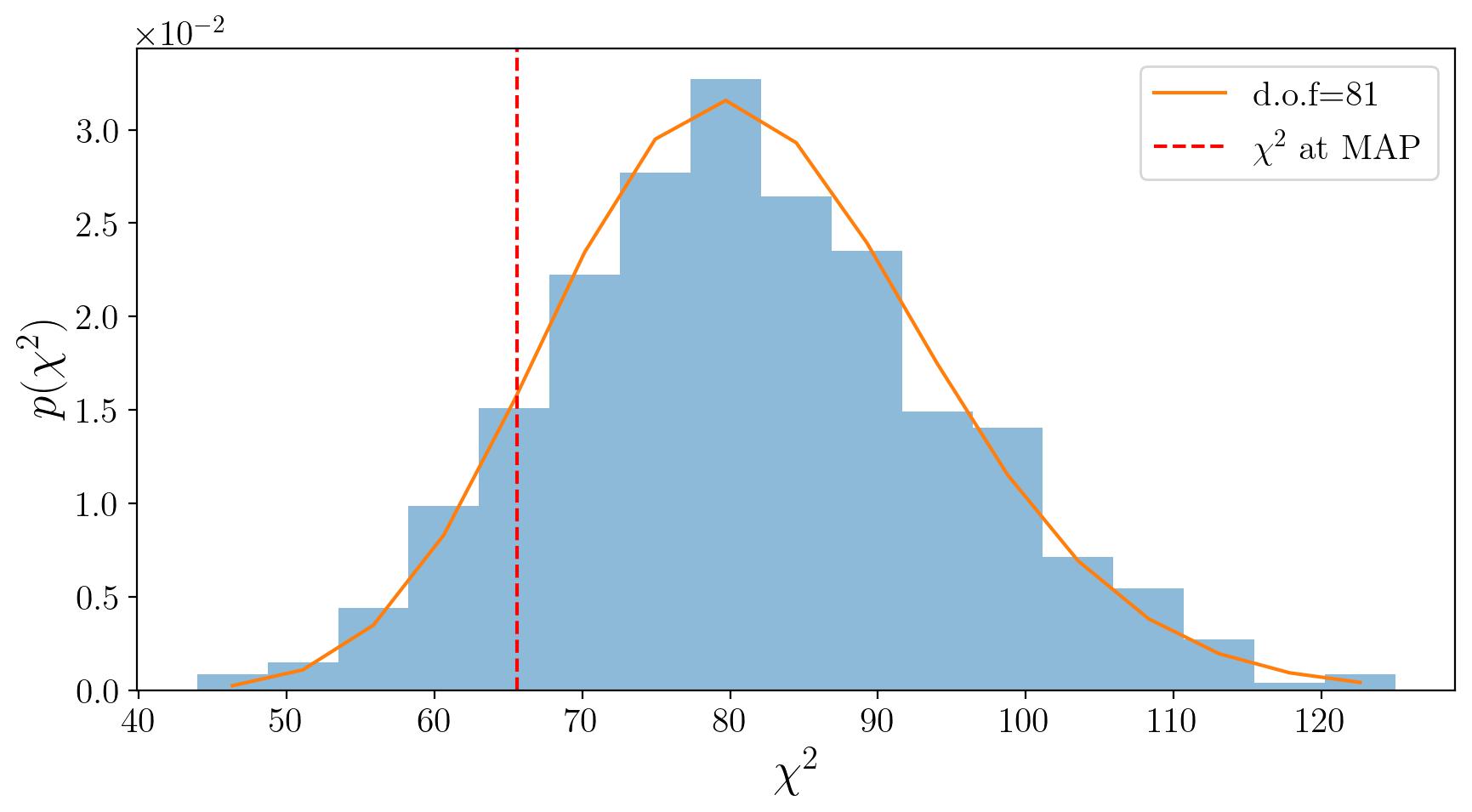}
\caption{Distribution of $\chi^2$ values of mock data vectors that follow from multivariate Gaussian distributions, where the mean is the model prediction at MAP and the covariance is the corresponding covariance for that particular model. The red line shows the $\chi^2$ values using the real data vector. The orange line is a $\chi^2$ distribution with $\mathrm{d.o.f.}=81$, which is slightly smaller than the 84 elements of the data vector.}
\label{fig:chi2_around_MAP}
\end{figure}

\begin{figure}[ht]
\centering
\includegraphics[width=\columnwidth]{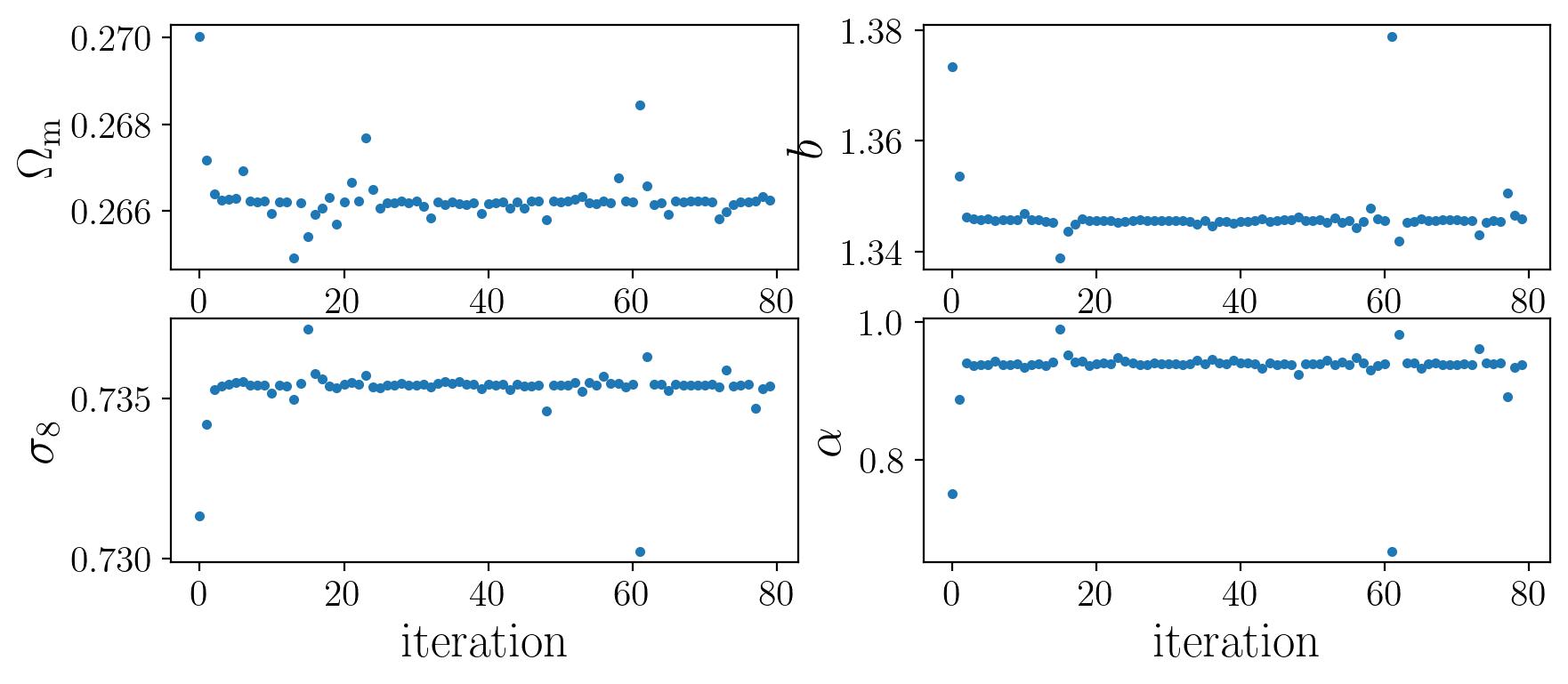}
    \caption{Change of MAP values due to scaling of the covariance to the previously measured MAP values. Roughly after ten iterations, the process converged, where the occasional outliers happened if the minimisation process stopped to early by coincidence.}
    \label{fig:cos_iterative}
\end{figure}

In Fig.~\ref{fig:nz_a_s} we show the negligible difference of the lens $n(z)$ if the Lorentzian fitting parameters ($a$ and $s$) are estimated from different patches of the sky. In Fig.~\ref{fig:chi_acc} we display the accuracy of the emulator, and in Fig.~\ref{fig:pure_shear_cosmoSLICS}, that the third tomographic bin is indeed contaminated by IA, in Fig.~\ref{fig:chi2_around_MAP} the $\chi^2$ distribution of mock data vectors around the MAP for the adapted filter with the best-estimated $n_\mathrm{be}(z)$, and lastly in Fig.~\ref{fig:cos_iterative} the iterative process to find the optimal MAP values by scaling the covariance matrix to the previously found MAP.

\section{Additional material for the red and blue analysis}

\begin{figure}[ht]
\centering
\includegraphics[width=\columnwidth]{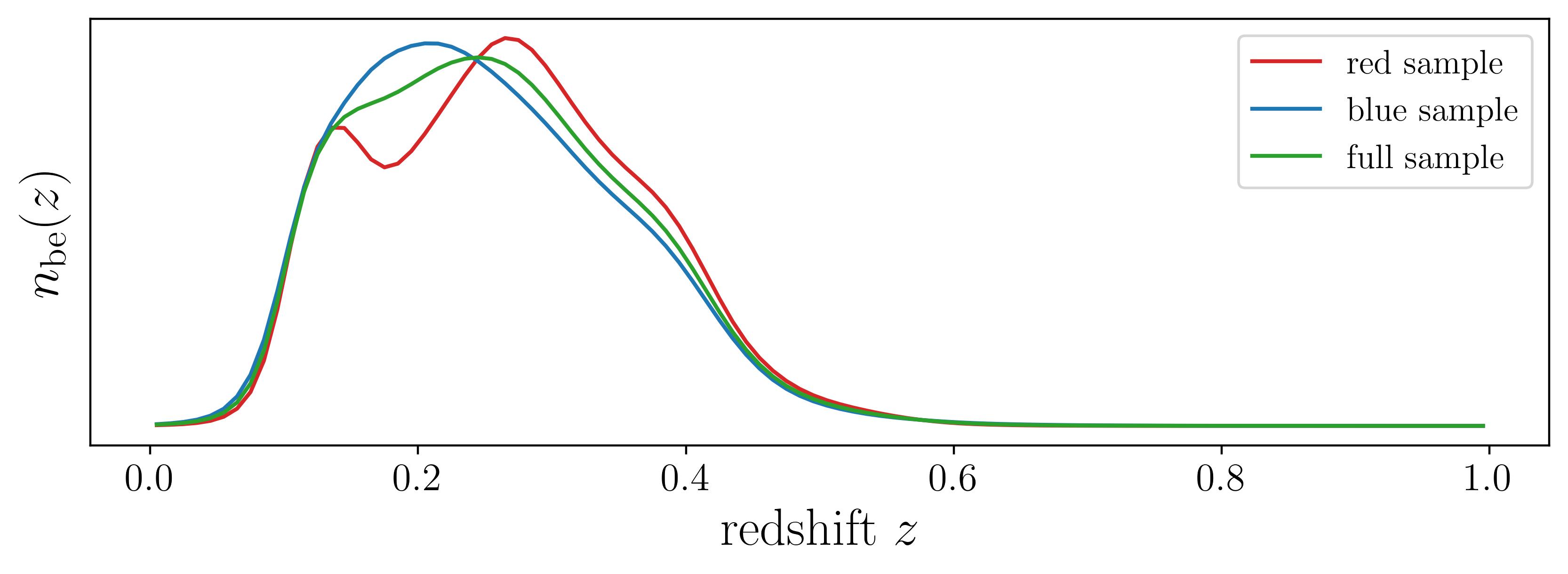}
    \caption{Best estimated redshift distributions of the red and blue KiDS-bright samples in red and blue, with the full KiDS-bright sample in green.}
    \label{fig:nofz_redblue}
\end{figure}

\begin{figure}[ht]
\begin{subfigure}{\columnwidth}
\includegraphics[width=\columnwidth]{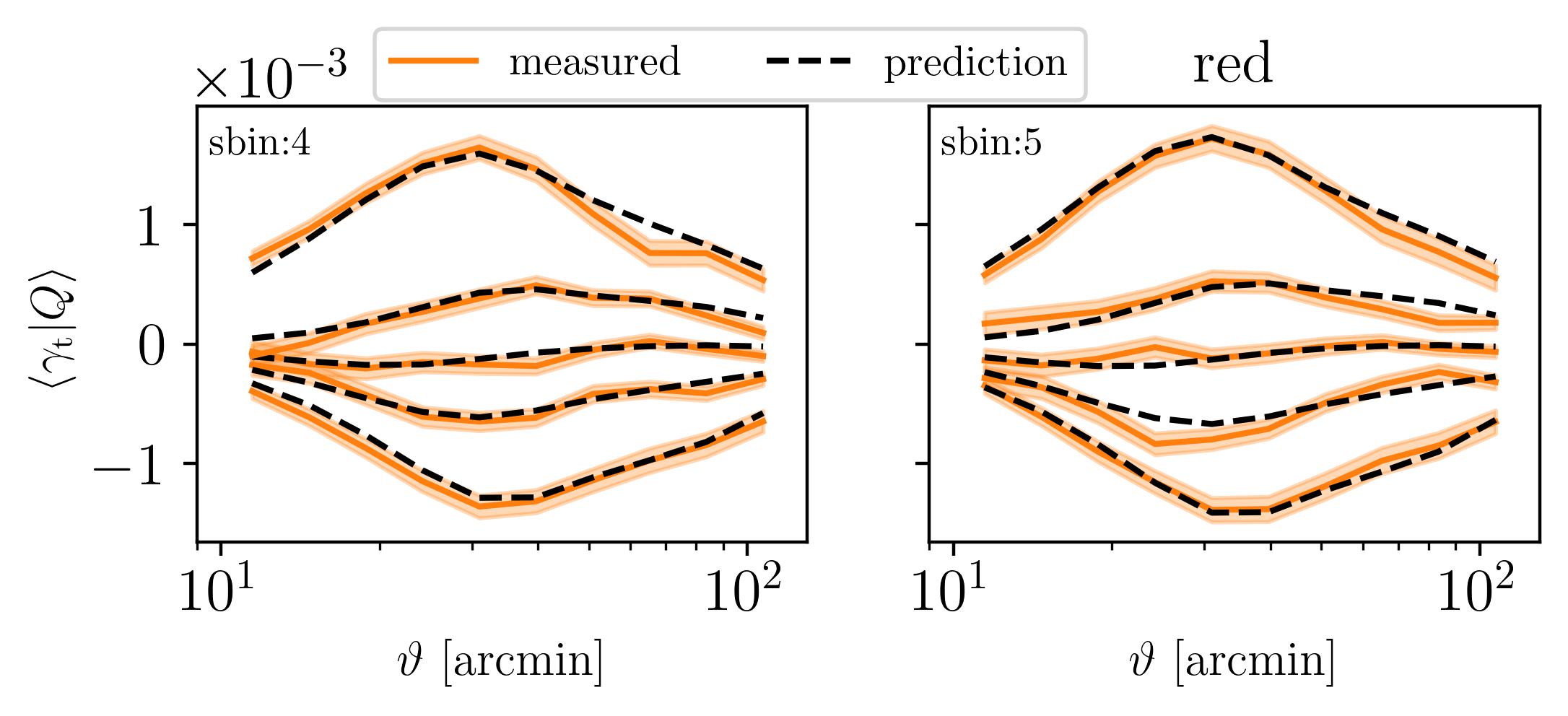}
\end{subfigure}
\begin{subfigure}{\columnwidth}
\includegraphics[width=\columnwidth]{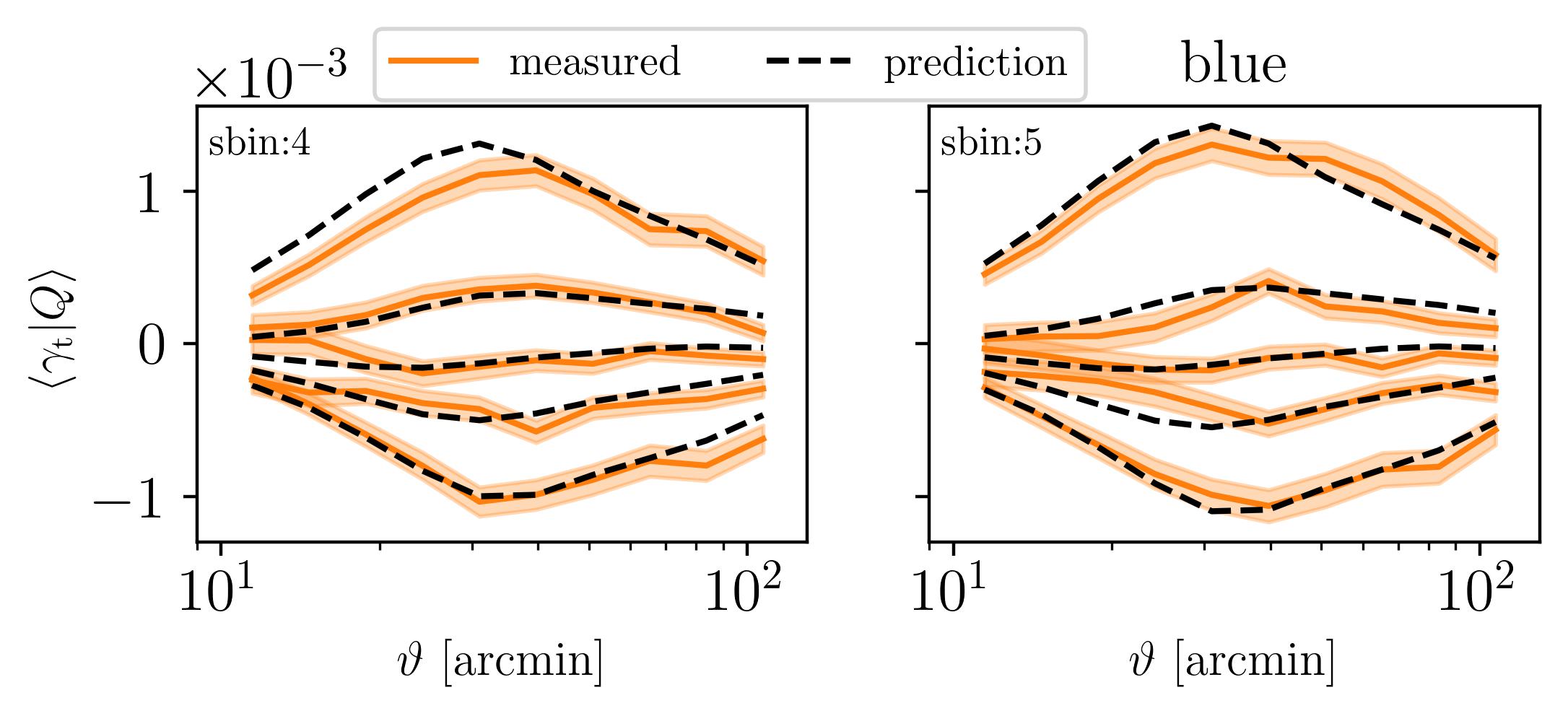}
\end{subfigure}
\caption{Measured and predicted shear profiles at the MAP values for the adapted filter for the red and blue samples. The shaded region is the expected KiDS-1000 uncertainty estimated from the 1000 \texttt{FLASK} realisations with shape noise.}
\label{fig:shear_bright_redblue}
\end{figure}

\begin{figure}[ht]
\includegraphics[width=\columnwidth]{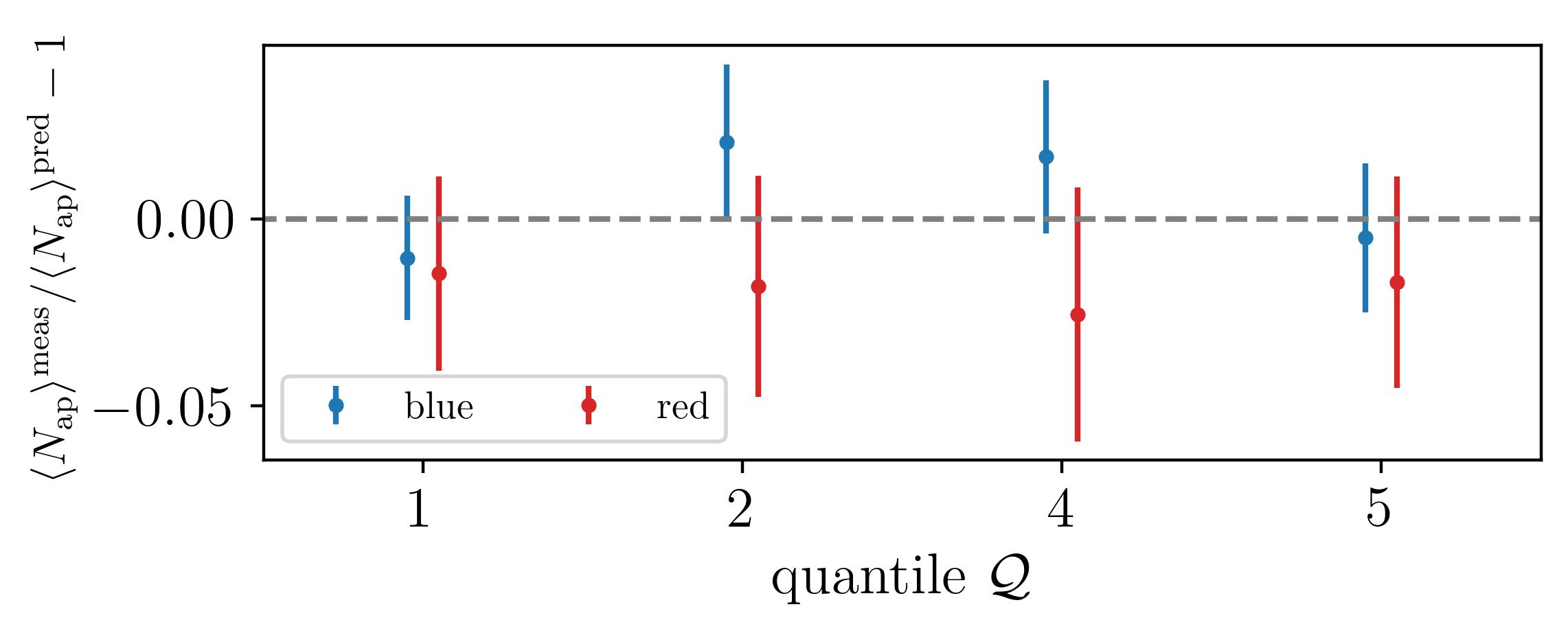}
\caption{Mean aperture number of the red and blue sample, where the predictions follow from a joint minimisation process. Different to the full bright sample the measured $p(N_\mathrm{ap})$ is smaller than the predicted Map, resulting in the measured $\langle N_\mathrm{ap} \rangle$ being smaller.}
\label{fig:mean_Nap_redblue}
\end{figure}

Here in this chapter, we show the complementary plots for the joint red+blue analysis. In Fig.~\ref{fig:nofz_redblue} the redshift distribution for the red and blue samples is shown, resulting from the smoothing method described in Sect.~\ref{Sect:Obs_Data:lenses}. In Fig.~\ref{fig:shear_bright_redblue} we show the shear profiles; in Fig.~\ref{fig:mean_Nap_redblue} the mean aperture number values for both samples which are used as the model and data vectors.

\clearpage
\section{Top-hat filter analysis}
\label{sec:top-hat}
\begin{figure}[ht]
\centering
\includegraphics[width=\columnwidth]{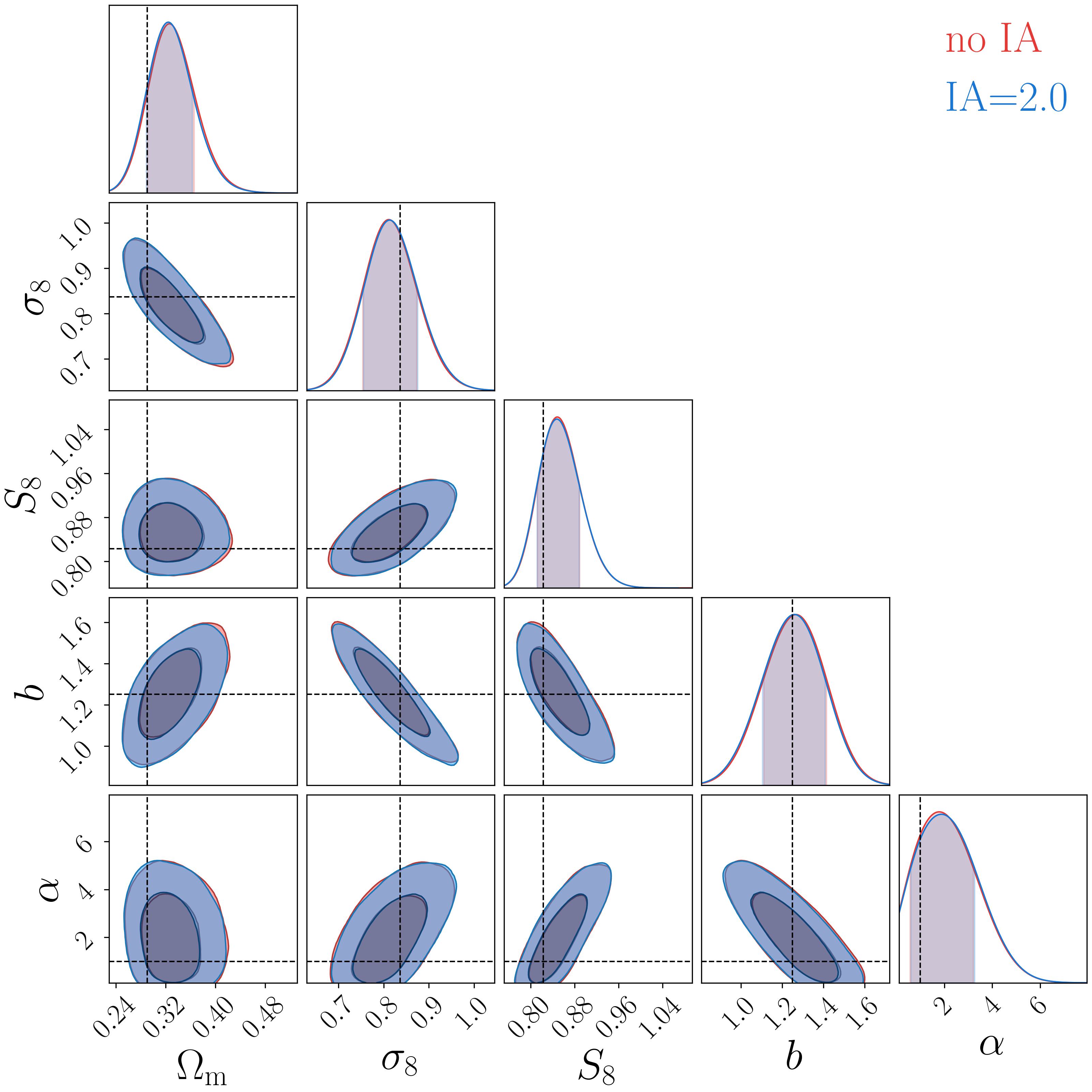}
    \caption{MCMC results for the top-hat filter using the model and simulations where we infused the pure shear signal with different amplitudes of IA.}
    \label{fig:MCMC_SLICS_tophat}
\end{figure}

This section shows the corresponding plots that belong to the analysis with the top-hat filter. We start by showing in Fig.~\ref{fig:MCMC_SLICS_tophat} the validation for the top-hat with the cosmo-SLICS, which shows that the true parameters are always inside $1\,\sigma$. The discussion from Sect.~\ref{sec:n_of_z_posterior} about the impact of different redshifts distributions is summarised for the top-hat filter in Table \ref{table:MAP_values_tophat}, which shows with the given reduced $\chi^2$ and corresponding $p$-value that the model is well fitted to the data given the covariance matrix, and result in parameters that are consistent to the ones constrained with the adapted filter.
 
\begin{table}[h!]
\centering
\caption{Marginalised MAP values and their $68\%$ confidence intervals for the different lens $n(z)$ of the full sample.}
\begin{tabular}{c|ccc}
\hline
\hline
 & $n_\mathrm{be}(z)$ & $n_\mathrm{ph}(z)$ & $n_\mathrm{sp}(z)$  \\
\hline
$\Omega_\nt{m}$ & $0.30^{+0.04}_{-0.04}$ & $0.33^{+0.05}_{-0.04}$ & $0.36^{+0.05}_{-0.04}$ \\
$\sigma_8$ &  $0.68^{+0.06}_{-0.05}$ & $0.65^{+0.06}_{-0.05}$ & $0.62^{+0.06}_{-0.04}$ \\
$S_8$ & $0.68^{+0.04}_{-0.02}$ & $0.68^{+0.04}_{-0.02}$ & $0.68^{+0.04}_{-0.02}$ \\
$b$ &  $1.66^{+0.11}_{-0.23}$ & $1.68^{+0.11}_{-0.24}$ & $1.72^{+0.11}_{-0.27}$  \\
$\alpha$ & $0.10^{+2.25}$ & $0.10^{+2.25}$ & $0.10^{+2.47}$ \\
$\chi^2/\mathrm{d.o.f.}$ & $1.03$ & $1.03$ & $1.05$ \\
$p$-value & $0.39$ & $0.40$ & $0.35$ \\
\hline
\hline
\end{tabular}
\tablefoot{The $68\%$ confidence intervals result from MCMC chains. Here $\Omega_\mathrm{m}$, $\sigma_8$, the $\alpha$ and the linear galaxy bias parameter are varied. We fixed $h=0.6898$, $w_0=-1$ and $n_{\rm s}= 0.969$ but marginalised over the $\delta \langle z \rangle$ and $m$-bias uncertainties. If limits are not given, they are dominated by priors.}
\label{table:MAP_values_tophat}
\end{table}

\begin{figure}
\includegraphics[width=\columnwidth]{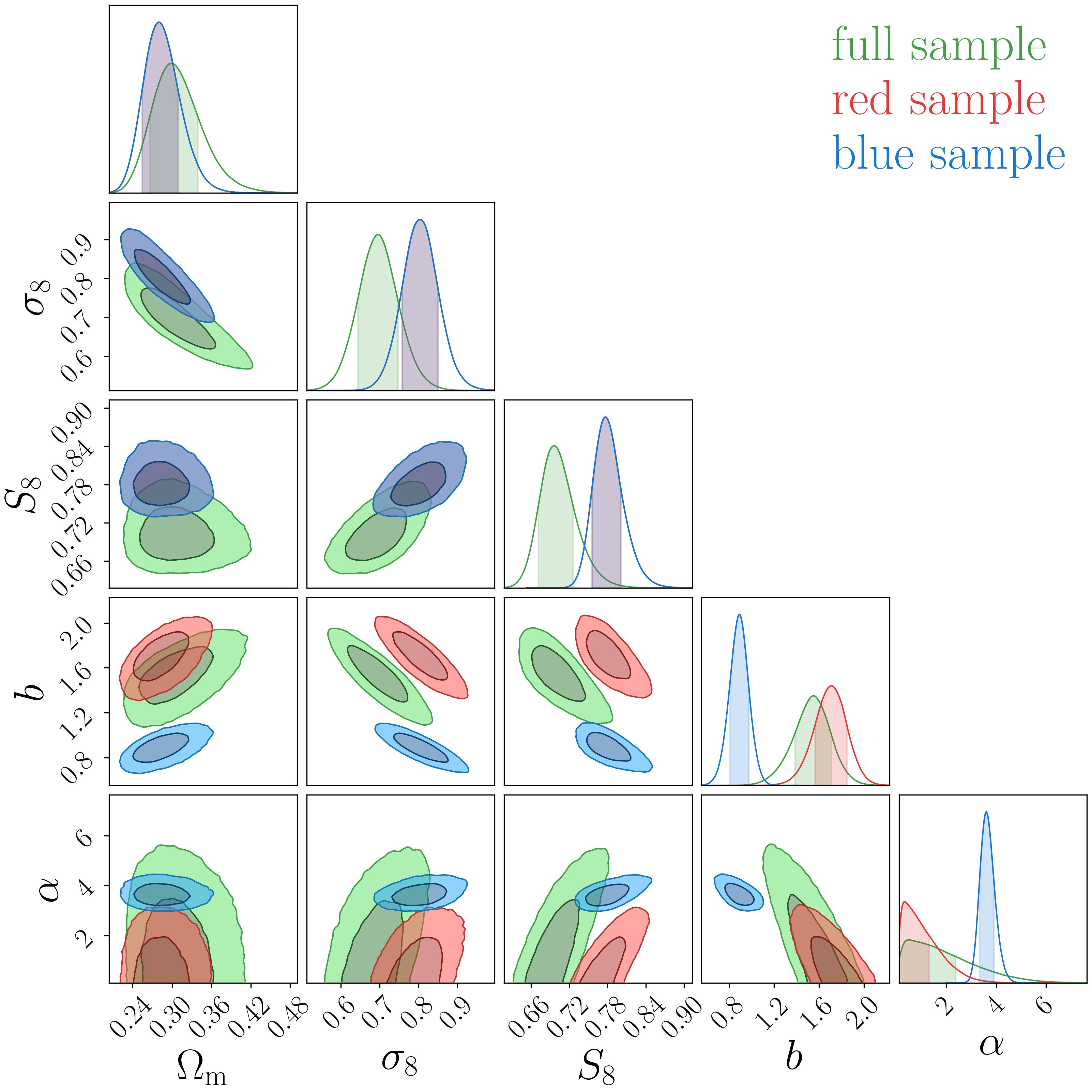}
\caption{Posterior for the full KiDS-bright sample shown in green, and the joint red+blue posteriors that represent the results of the individual colour-selected samples. In the latter case, by construction, the red and blue samples share the same cosmology (the dark blue contours).}
\label{fig:MCMC_bright_bias_tophat}
\end{figure}

Finally we show in Fig.~\ref{fig:MCMC_bright_bias_tophat} and Table \ref{table:MAP_redblue_tophat} the analogous results for the top-hat filter to the adapted filter as shown in Fig.~\ref{fig:MCMC_bright_redblue_adapted} and Table \ref{table:MAP_redblue}. The $p$-value for the joint red+blue indicates that the given d.o.f. has a significant tension between the measured data and the best fit model. Although this reduced $\chi^2$ is not ideal for the given d.o.f., we still perform the analysis, but the posteriors should be taken with caution, which is already true because we are uncertain about the true $n(z)$ of the sub-samples. Overall the results show the same trends as for the adapted filter, where the blue galaxies result in smaller bias $b$ and larger $\alpha$ than the red galaxies. The reason for the top-hat filter performing worse than the adapted filter is unclear. Nevertheless, besides the fact that the $n(z)$ of the red and blue sample is quite uncertain, both filters are probing on fundamentally different scales so that different behaviours are not surprising.

\begin{table}[h!]
\centering
\caption{Marginalised MAP values and their $68\%$ confidence intervals for the different lens samples.}
\begin{tabular}{c|ccc}
\hline
\hline
 & full & \multicolumn{2}{c}{red + blue}\\
\hline
$\Omega_\mathrm{m}$ & $0.30^{+0.04}_{-0.04}$ & \multicolumn{2}{c}{$0.28^{+0.03}_{-0.02}$}\\
$\sigma_8$ & $0.68^{+0.06}_{-0.05}$ & \multicolumn{2}{c}{$0.79^{+0.06}_{-0.03}$}\\
$S_8$ & $0.68^{+0.04}_{-0.02}$ & \multicolumn{2}{c}{$0.77^{+0.03}_{-0.01}$}\\
$b$ & $1.66^{+0.11}_{-0.23}$ & $1.78^{+0.11}_{-0.18}$  & $0.94^{+0.06}_{-0.12}$ \\
$\alpha$ & $0.10^{+2.25}$ & $0.10^{+1.21}$  & $3.40^{+0.42}_{-0.20}$ \\
$\chi^2/\mathrm{d.o.f.}$ & $1.03$ & \multicolumn{2}{c}{$1.37$} \\
$p$-value & $0.39$ & \multicolumn{2}{c}{$0.001$} \\
\hline
\hline
\end{tabular}
\tablefoot{The $68\%$ confidence intervals result from the MCMC chains shown 
in Fig.~\ref{fig:MCMC_bright_bias_tophat}. We fixed $h=0.6898$, $w_0=-1$ and $n_{\rm s}=0.969$ but marginalised over the $\delta \langle z \rangle$ and $m$-bias uncertainties. If limits are not given, they are dominated by priors.}
\label{table:MAP_redblue_tophat}
\end{table}

\end{appendix}
\end{document}